# Drag penalty during relaminarization and Kelvin–Helmholtz-promoted retransition in an accelerating turbulent boundary layer over initially drag-reducing riblets

**Benjamin S. Savino[1], Wen Wu[1]†**

[1] Department of Mechanical Engineering, University of Mississippi, Oxford, MS, 38677, USA

Direct numerical simulations of an accelerating turbulent boundary layer (TBL) over a smooth wall and a wall fully covered with streamwise-aligned riblets are performed to investigate drag modulation and its underlying mechanisms. The riblet-scale flow is resolved using an immersed boundary method. Starting from a zero-pressure-gradient (ZPG) TBL at $Re_{\delta,o} = 6,800$, the flow undergoes a threefold freestream acceleration over seventy-five boundary-layer thicknesses, matching the development reported by Warnack & Fernholz (1998), and consequently experiences relaminarization followed by retransition farther downstream. The riblets, defined by a sinusoidal spanwise profile with initial $s_o^+ = 15.2$ and $\ell_{g,o}^+ = 10.5$, correspond to near-optimal drag-reducing size in ZPG flows. However, even modest acceleration renders them drag-increasing, showing that the conventional ZPG interpretation based on total-drag viscous scaling does not apply directly in this non-equilibrium flow. During relaminarization, the drag penalty arises primarily from geometry-determined concentration of viscous shear near the riblet crest, with negligible direct Reynolds- and dispersive-stress contributions prior to retransition. Despite the drag increase, the overlying TBL remains statistically similar to the smooth-wall case when scaled with the total shear stress at the groove opening, demonstrating that this shear sets the relevant scaling for the TBL, while the additional drag generated within the grooves remains largely decoupled from the outer-layer turbulence dynamics. This partial decoupling persists until the onset of retransition, when spanwise Kelvin–Helmholtz rollers develop near the riblet crest and promote earlier, stronger retransition through their interaction with the residual near-wall streaks. These findings provide a revised physical picture of riblet performance in non-equilibrium turbulent flows.

† Email address for correspondence: wu@olemiss.edu

**Abstract must not spill onto p.2**

## 1. Introduction

Streamwise-aligned riblets have been shown to reduce skin-friction drag by up to 10% in canonical turbulent flows, including zero-pressure-gradient (ZPG) turbulent boundary layers (TBLs) and channels (Walsh 1980; Choi 1989; García-Mayoral & Jiménez 2011a; Modesti *et al.* 2021). These gains have motivated applications in aeronautical (Kuntzagk 2024), maritime (Choi *et al.* 1989), and energy systems (Chamorro *et al.* 2013), making riblets a benchmark passive strategy for skin-friction-drag reduction.

In such ZPG wall-bounded flows, riblet performance is commonly characterized using viscous-scaled geometric parameters, most notably the peak-to-peak spacing, $s^+ = su_\tau/\nu$, and the square root of the groove cross-sectional area, $\ell_g^+ = \ell_g u_\tau/\nu$. This parameterization is attractive not only from a design perspective, as it provides a compact basis for comparing different riblet geometries, but also because current understanding indicates that drag modulation is closely tied to riblet-induced alterations of the near-wall flow. For $s^+ \lesssim 30$ and $\ell_g^+ \lesssim 20$, riblets reduce drag by impeding spanwise motion near the wall, thereby restricting turbulent momentum transfer to the riblet crests while maintaining viscous-dominated flow within the grooves (see, among others, Bechert *et al.* 1986; Choi 1989; Orlandi & Jiménez 1994; Goldstein *et al.* 1995; Lee & Lee 2001). This leads to an effective thickening of the viscous sublayer and an upward shift of the logarithmic region of the mean velocity profile, consistent with drag reduction (Luchini *et al.* 1991; Luchini 1996; Endrikat *et al.* 2021b).

As the riblet size increases beyond the optimal range, this mechanism deteriorates: turbulent structures of scale comparable to the riblet size penetrate the grooves, exposing the increased wetted area to enhanced momentum transfer and shear stress, and thereby producing drag penalty (Choi *et al.* 1993; Suzuki & Kasagi 1994; Goldstein & Tuan 1998; Modesti *et al.* 2021). For certain geometries, spanwise-coherent Kelvin–Helmholtz (KH)-type vortices may also arise near the riblet crest owing to inflectional mean velocity profiles (García-Mayoral & Jiménez 2011b; Endrikat *et al.* 2021a; Abu Rowin *et al.* 2025; Camobreco *et al.* 2025). The narrow drag-reducing range of riblet sizes, together with the dependence on riblet geometry, highlights the strong sensitivity of drag modulation in ZPG flows to the shear and turbulence near the riblet crest. This sensitivity poses a challenge both for the reliable application of riblets across different flow conditions and for predictive models developed on the basis of canonical ZPG behaviour. For example, the vertical displacement of turbulence has motivated protrusion-height-based approaches derived from affordable Stokes-flow calculations for extensive geometric studies (Luchini *et al.* 1991; Luchini 1996; Wong *et al.* 2024), while the shift in the logarithmic law has been incorporated into Reynolds-averaged Navier–Stokes models for engineering simulations (Wilcox 2008; Smith & Yagle 2025; Wang *et al.* 2025).

Riblet performance in non-equilibrium turbulent wall-bounded flows has received far less attention, despite the ubiquity of such flows. Among non-equilibrium effects, pressure gradients provide a particularly relevant setting in which to examine the robustness of canonical riblet behaviour, because even mild pressure gradients can substantially alter TBL structure. Under favourable pressure gradients (FPGs), the boundary layer thins and turbulence is suppressed, while sufficiently strong acceleration can induce relaminarization and lead to the loss of an identifiable logarithmic layer (Narasimha & Sreenivasan 1973; Blackwelder & Kovasznay 1972; De Prisco *et al.* 2007; Piomelli *et al.* 2000; Bourassa & Thomas 2009; Talamelli *et al.* 2002; Jiménez & Pinelli 1999; Schoppa & Hussain 2002). For finite-length FPGs, relaxation towards ZPG is accompanied by the reappearance of turbulent spots and a gradual recovery towards an equilibrium, higher-$Re$ TBL, a process commonly referred to as retransition (Sreenivasan 1982; Warnack & Fernholz 1998; Brandt *et al.* 2004; Piomelli & Yuan 2013; Falcone & He 2022). Conversely, adverse pressure gradients (APGs) thicken the boundary layer and enhance turbulence, particularly in the outer layer. Although APG TBLs can admit modified self-similar states (Monty *et al.* 2011; Kitsios *et al.* 2016, 2017; Bobke *et al.* 2017; Pozuelo *et al.* 2022), sufficiently strong or sustained APGs can destroy this quasi-similarity, leading to near-wall reversal and ultimately separation.

Given that riblet performance is closely tied to the near-wall flow, such pressure-gradient-induced modifications suggest that their drag-modulating behaviour may differ substantially from that observed in ZPG flows. The canonical picture of turbulence–riblet interaction established in ZPG flows may therefore not carry over directly to non-equilibrium conditions. This issue is especially relevant because pressure gradients act across the entire boundary layer, including the flow within the riblet grooves, and may therefore alter the coupling between the groove flow and the outer layer.

Existing studies of riblets in pressure-gradient flows remain limited. Most have focused on mild APGs, for which departures from ZPG behaviour are modest and drag reduction is largely retained (Choi 1990; Nieuwstadt *et al.* 1993; Debisschop & Nieuwstadt 1996; Klumpp *et al.* 2010; Boomsma & Sotiropoulos 2015). Likewise, under weak FPGs, riblets have been reported to maintain their drag-reducing performance (Choi 1990). Far fewer studies have considered strong or spatially varying pressure gradients that drive the flow well away from canonical ZPG behaviour or from quasi-self-similar states. Pargal *et al.* (2021) investigated riblets in an open-channel flow subjected to an impulse acceleration strong enough to induce relaminarization followed by retransition. In that case, the riblets had little influence on the relaminarization process, and the drag increase was attributed to their larger wetted area. During retransition, however, the riblets were found to stabilize near-wall streaks and delay the re-emergence of turbulence relative to the smooth-wall case. Our recent direct numerical simulation revealed an unprecedented drag reduction and a marked departure from equilibrium expectations when riblets were applied to a strong, yet attached, spatially developing APG TBL (Savino *et al.* 2026). The enhanced drag reduction was attributed to amplification by the APG of the reverse-flow portion of riblet-induced KH rollers. Similarly, a study of separating flow over riblets (Rouhi *et al.* 2025) reported KH rollers upstream of separation, together with enhanced separation reminiscent of the effect of sandgrain roughness (Wu & Piomelli 2018). Simulations of riblet-like grooves over a bump have likewise shown drag increase on the windward side under the FPG and enhanced separation in the leeward APG region (Hussain *et al.* 2024; García *et al.* 2025); in that configuration, the exchange of mass and momentum between the grooves and the outer flow, largely induced by the bump curvature, played a central role in the drag modulation.

Taken together, these studies indicate that riblet behaviour in strongly non-equilibrium flows may differ fundamentally from canonical expectations. However, it remains unclear whether viscous-scaled riblet-performance characterizations developed for ZPG flows retain predictive value under such conditions, and the mechanisms by which riblets modify drag, flow separation, relaminarization, and retransition in strongly developing pressure-gradient boundary layers remain insufficiently understood. Nor has this gap been resolved by the substantial body of application-oriented work in which riblets are applied directly to complex configurations, such as airfoils and compressor or turbine blades (Sareen *et al.* 2011; Lietmeyer *et al.* 2012; Kozul *et al.* 2025); these studies have largely assessed their effect through bulk metrics such as lift, drag, and loss coefficients, with comparatively limited attention to the detailed mechanisms governing drag modulation.

The present study investigates the interaction between riblets and a turbulent boundary layer subjected to a strong FPG that varies over an extended streamwise region, producing a ZPG region, a relaminarization region, and a retransition region. This arrangement allows the boundary-layer–riblet interaction to be examined under a range of mean-shear and turbulence states. Direct numerical simulations are performed for boundary layers over both a smooth wall and a riblet surface. As in Savino *et al.* (2026), the riblets are selected to lie in the optimal drag-reducing regime upstream of the pressure-gradient region, thereby providing a baseline for assessing subsequent changes in performance under downstream off-design conditions. The aims are therefore to determine whether riblets selected to be drag-reducing under upstream ZPG conditions remain beneficial under a strong, spatially developing FPG; to assess whether canonical ZPG performance characterizations based on viscous-scaled riblet geometry retain predictive value; to distinguish how riblets affect relaminarization and retransition, respectively; and to identify the mechanisms responsible for the resulting drag modulation and altered retransition dynamics.

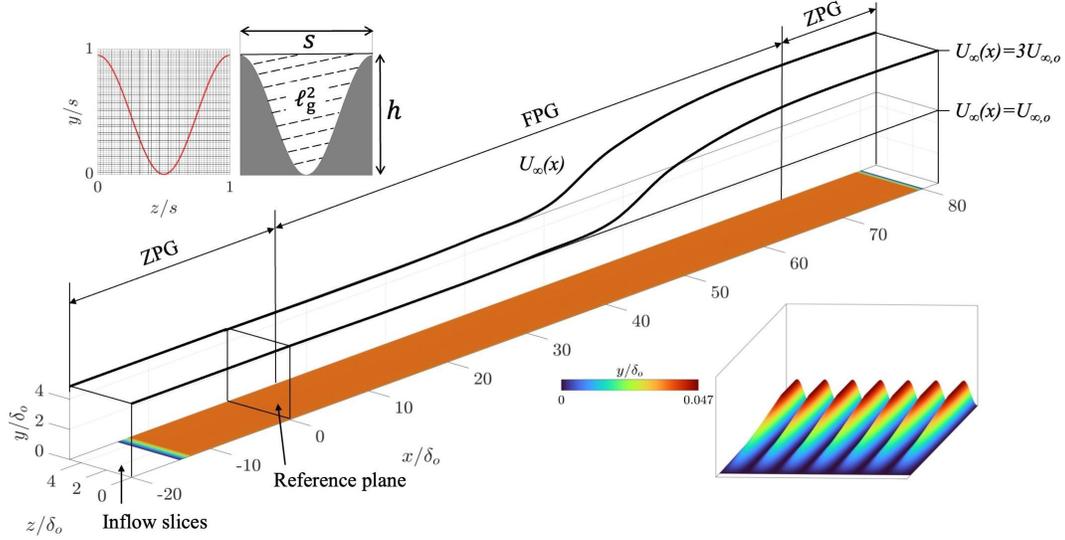

Figure 1: Schematic of the computational domain. The riblets used in Case RB are visualized by the isosurfaces of grid cell volume fraction of fluid equals 50% colored by wall-normal distance. The free-stream acceleration is sketched (not to scale) by the thick lines on the top boundary. The upper-left insets show the riblet profile with the grid superimposed, and the riblet sizing metrics. The lower-right inset shows the transition from smooth wall to the full riblets shortly downstream of the inflow.

The methodology is first summarized in §2. Instantaneous flow features are then illustrated through flow visualizations in §3. The development of the mean flow is examined in §4, followed by a detailed quantification and analysis of riblet-induced drag modulation in §5. Subsequently, statistical and structural changes throughout the evolution of the accelerating TBL are discussed in §6 and §7. Finally, §8 concludes the paper and outlines future work.

## 2. Methodology

### 2.1. *Problem formulation*

Turbulent boundary layers (TBLs) subjected to a favorable pressure gradient (FPG) are simulated by direct numerical simulation (DNS). Two cases are performed: one with a smooth wall, the second with a wall consisting of streamwise-aligned riblets. The smooth-wall case is denoted as 'SM', and the riblet case is denoted as 'RB'. A sketch of the computational domain is shown in figure 1 and case parameters are listed in Table 1.

The FPG is imposed by prescribing a freestream acceleration, $U_\infty(x)$, at the top boundary, while the mean wall-normal velocity at the top boundary, $V_\infty(x)$, is determined during the runtime by wall-normal integral mass conservation (Lund *et al.* 1998). We constructed a quasi-hyperbolic-tangent profile for $U_\infty(x)$ (figure 2a) so that the smooth-wall boundary layer closely reproduces the development of Case 2 reported by Warnack & Fernholz (1998).

| Cases | $K_{\max}$ | $Re_{\delta,o}$ | $Re_{\tau,o}$ | $Re_{\tau,\min}$ | $Re_{\tau,\max}$ | $N_i \times N_j \times N_k$ | $[L_x \times L_y \times L_z]/\delta_o$ |
|-------|------------|-----------------|---------------|------------------|------------------|------------------------------|-----------------------------------------|
| SM | $3.75 \times 10^{-6}$ | 6,800 | 317 | 312 | 475 | $7000 \times 270 \times 1400$ | $101 \times 5 \times 4.9$ |
| RB | $3.75 \times 10^{-6}$ | 6,800 | 309 | 309 | 583 | $7200 \times 306 \times 4800$ | $101 \times 5 \times 4.9$ |

Table 1: Case parameters.

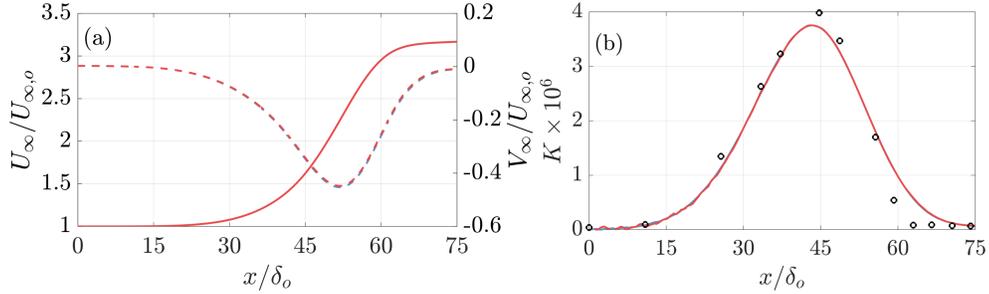

Figure 2: Profiles of (a) the prescribed freestream velocity $U_\infty(x)$ (solid, left axis) and resultant $V_\infty(x)$ (dashed, right axis), and (b) the induced acceleration parameter $K(x)$ (Eqn. (2.1)). ——— Case SM; ——— Case RB; ○ Case 2 from Warnack & Fernholz (1998).

This acceleration increases the freestream velocity by a factor of three over approximately 75 boundary layer thicknesses and produces the relaminarization–retransition scenario examined in the present study. Similar $U_\infty(x)$ profiles have been used in previous studies of TBLs over smooth walls with wall-resolved large-eddy simulation (De Prisco *et al.* 2007; Piomelli & Yuan 2013) and over sandgrain roughness with DNS (Yuan & Piomelli 2015). We adjusted it for better agreement with the experimental flow field. Following the experimental definition, the acceleration parameter reads as

$$K(x) = \frac{\nu}{U_e^2(x)} \frac{dU_e(x)}{dx}, \tag{2.1}$$

where $U_e(x)$ is the mean streamwise velocity at the edge of the boundary layer, determined following Griffin *et al.* (2021), and differs slightly from $U_\infty(x)$ imposed at the top boundary. The agreement in $K$ between the present results and the experimental measurements is shown in figure 2b. A comparison of the velocity profiles is provided in Appendix A.

At the inflow boundary, we prescribe a time series of velocity fields extracted from a transverse plane of an *a priori* DNS of a ZPG smooth-wall TBL. In each of the two cases considered, the incoming TBL is advanced under ZPG conditions until the two cases reach a common Reynolds number at a reference plane, which defines the streamwise coordinate in this study (figure 1). Quantities evaluated at the reference plane are denoted by the subscript 'o', and the common Reynolds number there is $Re_{\delta,o} = \delta_o U_{e,o}/\nu = 6,800$. The Reynolds number based on displacement thickness ($Re_{\delta,o}^* = \delta_o^* U_{e,o}/\nu = 1,270$) and momentum thickness ($Re_{\theta,o} = \theta_o U_{e,o}/\nu = 864$) on the smooth wall agree with those of Case 2 from Warnack & Fernholz (1998) to within 5%. While various reference length scales are used in the FPG-TBL literature, we match $Re_{\delta,o}$ between the two cases so that the incoming boundary layers have closely comparable mass fluxes before the onset of the FPG, so that riblet-induced drag changes can be compared directly. The remaining boundary conditions are periodicity in the spanwise direction, no slip on either the smooth wall or the riblet surface enforced through the immersed boundary method (see §2.3), and a convective outflow (Keating *et al.* 2004; Scotti 2006).

| Riblet | $Re_\tau$ | $\ell_g^+$ | $s^+$ | $h/\delta$ | Shape | Drag change % |
|---|---|---|---|---|---|---|
| Present Case RB, $x$=0 | 309 | 10.5 | 15.2 | 0.045 | 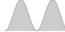 | -5.0 |
| Endrikat *et al.* (2021*b*) Case T615 | 395 | 9.7 | 14.7 | 0.032 | 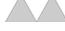 | -8.2 |
| García-Mayoral & Jiménez (2011*b*) 10S | 183 | 9.8 | 16.0 | 0.044 | 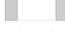 | -4.0 |
| Choi *et al.* (1993) C | 180 | 10.0 | 20.3 | 0.055 | 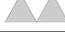 | -5.0 |
| Present Case RB, $x/\delta_o$=75 | 511 | 36 | 52 | 0.093 | 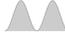 | +10.9 |
| Modesti *et al.* (2021) TA50 | 395 | 32.9 | 50.0 | 0.063 | 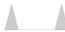 | +20.2 |
| García-Mayoral & Jiménez (2011*b*) 20S | 191 | 20.5 | 28.7 | 0.075 | 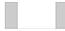 | +3.5 |

Table 2: Comparison between the riblet used in this study and selected ones in the literature in the ZPG regions. The upper entries correspond to drag-reducing cases, whereas the lower entries correspond to drag-augmenting cases.

### 2.2. *Riblet configuration*

On the bottom wall of case RB, streamwise-aligned riblets are applied. The riblet shape is defined by a sinusoidal function in the spanwise direction ($z$, see the top-left inset of figure 1): $h_l(z) = 0.5h\left[\sin\left(2\pi z/s\right) + 1.0\right]$. The parameters $s/\delta_o = 0.049$ and $h = 3s/\pi$ make the riblets small relative to the boundary-layer thickness (i.e., they are treated as surface roughness rather than large obstacles) and place them in the drag-reducing regime prior to the onset of acceleration. When normalized by $\nu/u_\tau$ (denoted by superscript +) at the reference plane the spanwise riblet spacing, $s_o^+ = 15.2$, and square root of riblet groove cross-sectional area, $\ell_g^+ = 10.5$, are close to the optimal drag-reducing values widely reported in the literature for ZPG channels and TBLs (Walsh & Lindemann (1984); Choi *et al.* (1993); García-Mayoral & Jiménez (2011*a*,*b*); Modesti *et al.* (2021); Endrikat *et al.* (2021*a*), among others). Here,

$$\ell_g^2 = \int_0^s (h - h_l(z))\,dz,$$
(2.2)

which gives $\ell_g = \sqrt{3/(2\pi)}\,s$ in this study. This setup allows us to examine how riblet performance changes when the spacing and size are initially selected to be drag-reducing, thereby characterizing deviations under downstream off-design conditions. As shown later, the streamwise variation of the local friction velocity during acceleration and retransition causes the riblet spacing and size in wall units to increase by more than a factor of three. Over the domain considered here, the maximum values of $s^+$ and $\ell_g^+$ are 60 and 41, respectively. As the boundary layer thickness varies during relaminarization and retransition, the riblet height varies between $0.0355\delta$ (at $x/\delta_o = 23$) and $0.121\delta$ (at $x/\delta_o = 63$).

Table 2 lists a comparison of riblet sizing metrics and drag modulation characteristics in the ZPG regions of the current study with selected benchmark cases. The present riblets are symmetric, with a sidewall inclination of approximately 65 degrees. Among the cases studied by Modesti *et al.* (2021), the triangular T6 series is geometrically the closest to the present riblet. Near the reference plane, and again in the downstream weak-pressure-gradient recovery region, the present riblet produces drag modulation comparable to that reported in previous ZPG studies.

Note that, to avoid an abrupt change from the smooth-wall inflow TBL to the riblet surface,

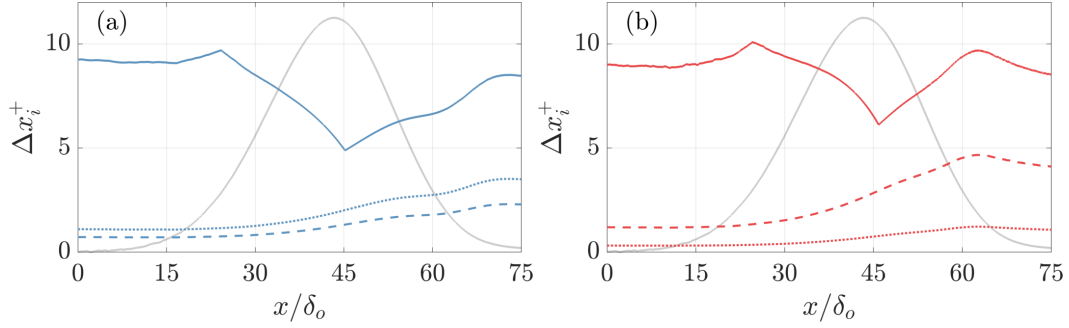

Figure 3: Grid spacing in wall units for (a) case SM, (b) case RB. ——— $\Delta x^+$; - - - $10\Delta y^+_{(1)}$; ······ $\Delta z^+$. The gray line shows $K(x)$.

the riblet height increases linearly over $x/\delta_o \in [-15, -12]$. This mild growth, together with the further development prior to the onset of the FPG, ensures that the incoming boundary layer is only weakly perturbed by the geometric transition and has recovered to the ZPG equilibrium state at the target $Re_{\delta,o}$.

Throughout this paper, the zero-plane displacement (virtual origin) caused by the riblet is denoted by $d$. Following Modesti *et al.* (2021), $d$ is obtained by vertically shifting the riblet turbulent shear stress profile at the reference location until it matches the smooth-wall profile, yielding $d = 0.9h$. This value is then held fixed throughout the present analysis, consistent with prior ZPG riblet studies and related FPG roughness/riblet studies (Yuan & Piomelli 2015; Pargal *et al.* 2021; Endrikat *et al.* 2021*a*; Modesti *et al.* 2021).

### 2.3. *Numerical methods*

We solve the non-dimensionalized incompressible Navier-Stokes equations, which read

$$\frac{\partial u_i}{\partial x_i} = 0, \text{ and } \frac{\partial u_i}{\partial t} + \frac{\partial (u_i u_j)}{\partial x_j} = -\frac{\partial p}{\partial x_i} + \frac{1}{Re}\frac{\partial^2 u_i}{\partial x_j^2} + f_i, \ (i, j = 1, 2, 3) \quad (2.3)$$

using a well-validated finite difference code with a staggered grid arrangement (Keating *et al.* 2004; Wu & Piomelli 2018; Savino & Wu 2024*a*,*b*). Here, $p$ is the modified pressure with the density factored out and $f_i$ is a body force that enforces the no-slip boundary condition on the riblet geometry by an immersed boundary method (IBM). All spatial derivatives are computed with second-order central differences. Time is advanced using a semi-implicit fractional step method in which the wall-normal diffusion terms are discretized with the Crank-Nicolson scheme, and all remaining terms with the Adams-Bashforth scheme. The Poisson equation is solved via a pseudo-spectral/direct method (Moin 2010) employing a Fourier transform in the spanwise direction, then a direct solve for each wavenumber using the blktri matrix solver from the FISH-PACK library (Sweet 1974; Swarztrauber & Sweet 1979). This approach eliminates iterative convergence errors and solves the discrete Poisson system to machine precision. The solver is parallelized using MPI.

The simulations span a domain of $[L_x \times L_y \times L_z]/\delta_o = [101 \times 5 \times 4.9]$ in the streamwise ($x$), wall-normal ($y$), and spanwise ($z$) directions. The streamwise length $L_x$ consists of the ZPG developing region ahead of the reference plane, the primary region of interest ($x/\delta_o \in [0, 75]$), and the last $6\delta_o$ to accommodate potential effects of the convective outflow condition. $L_z$ and $L_y$ reach their minimum of four times the local boundary layer thickness prior to strong acceleration, and increase to a maximum of 12-13 local boundary layer thickness by the end of acceleration. The chosen spanwise domain length also preserves the periodicity of the sinusoidal riblet geometry and accommodates 100 riblet realizations for ensemble averaging.

| Reference Case | Type/Region of Flow | $Re_\delta$ ($Re_\tau$) | $\Delta x^+$ | $\Delta y^+_{(1)}$ | $\Delta z^+$ |
|---|---|---|---|---|---|
| Present Case SM, $x = 0$ | smooth, ZPG TBL | 6,800 (317) | 9.2 | 0.07 | 1.1 |
| Present Case SM, $x/\delta_o = 75$ | smooth, ZPG TBL | 9,400 (446) | 8.5 | 0.22 | 3.5 |
| Schlatter & Örlü (2010) | smooth, ZPG TBL | (252–1,270) | 9 | 0.035 | 4 |
| Wu *et al.* (2017) | smooth, ZPG TBL | 6,800 (328) | 5.17 | 0.58 | 6.47 |
| Abe (2017) | smooth, ZPG TBL | 7,566 (350) | 8.24 | 0.11 | 4.40 |
| Present Case RB, $x = 0$ | riblet, ZPG TBL | 6,800 (309) | 8.5 | 0.11 | 0.3 |
| Present Case RB, $x/\delta_o = 75$ | riblet, ZPG TBL | 10,334 (511) | 8.3 | 0.40 | 1.1 |
| Endrikat *et al.* (2021a) T-series | riblet, ZPG chan. | (395) | 6.0 | 0.06 | 0.8 |
| García-Mayoral & Jiménez (2011b) | riblet, ZPG chan. | (180) | 6.0 | 0.3 | 0.7 |
| Present Case SM | smooth, FPG TBL | (317–475) | 4.8–9.5 | 0.07–0.2 | 1.1–3.5 |
| Piomelli & Yuan (2013) | smooth, FPG TBL | (200–405) | 3.7–8.8 | 0.5–1.2 | 1.6–3.8 |
| Falcone & He (2022) | smooth, FPG chan. | (178–324) | 3.3–6.0 | 0.24–0.44 | 2.0–3.6 |
| Present Case RB | riblet, FPG TBL | (309–583) | 6.0–10.0 | 0.11–0.43 | 0.3–1.2 |
| Pargal *et al.* (2021) | riblet, step-FPG chan. | (180–418) | 4.5–10.0 | 0.2–0.56 | 0.47–1.0 |
| Yuan & Piomelli (2015) | sandgrain, FPG TBL | (480–1,470) | 10–21 | 0.2–0.7 | 5–14 |
| Present Case SM, $x/\delta_o = 56$ | smooth, retransition | 8,000 (343) | 6.4 | 0.17 | 2.6 |
| Zaki (2013) | smooth, transition | 3,950 (170) | 10.2 | 0.13 | 4.1 |
| Wu *et al.* (2017) | smooth, transition | 4,700 (215) | 5.0 | 0.55 | 6.1 |
| Present Case RB, $x/\delta_o = 50$ | riblet, retransition | 8,540 (538) | 7.2 | 0.35 | 0.9 |
| Strand & Goldstein (2011) | riblet, transition | 1,093 | 20.0 | 0.08 | 4.0 |

Table 3: Comparison of grid resolution reported in various related DNS studies. Top two sections: ZPG flows; middle two sections: FPG flows; bottom two sections: transitional boundary layers.

We use an IBM based on a volume-of-fluid (VOF) approach to represent the riblet geometry embedded in the Cartesian grid (Peskin 1972; Scotti 2006). We first calculate the fraction of each computational cell volume that is occupied by fluid in pre-processing. During runtime, velocity in cells that contain the solid geometry is weighted by this fractional volume through the term $f_i$ in Eqn. (2.3). The total drag, and corresponding $u_\tau$, is obtained by taking the wall-normal integral of the time- and spanwise-average of $f_1$ (Yuan & Piomelli 2014a; Wu & Piomelli 2018). The riblet case uses $N_i \times N_j \times N_k = 7200 \times 306 \times 4800$ grid points, totaling 10.6 billion, in the $x$, $y$, and $z$ directions. The smooth wall has $7000 \times 270 \times 1400$ points, for a total of 2.6 billion. The $x$ grid is uniformly spaced between the inlet and $x/\delta_o = 25$ with spacing $\Delta x/\delta_o = 0.03$, and also between $x/\delta_o = 42$ and the domain outlet with spacing $\Delta x/\delta_o = 0.01$. In between, it is locally refined to ensure that the $\Delta x$ can sufficiently resolve all turbulent scales. The $y$ grid for case RB uses 61 uniformly spaced points below the riblet crest, followed by a hyperbolic-tangent stretched region above it. For case SM, the entire $y$ grid is hyperbolically stretched. The $z$ grid is uniform. In case SM, $N_k$ is set to be suitable for DNS, while in case RB, it is set finer to resolve each riblet period with 48 points. The computational grid is designed to resolve the turbulent motions, the sub-riblet-scale flow, and the retransition process, which may be sensitive to numerical accuracy. Our grid resolution is comparable to, or finer than, existing DNSs of TBLs and flow over riblets, as summarized in Table 3. Specifically, $\Delta x^+ < 10$, $\Delta y^+_{(1)} < 1$, and $\Delta z^+ < 1.2$ are maintained (see figure 3), despite $u_\tau$ increasing by a factor of 4.1 (3.3) in the riblet (smooth) case during flow acceleration. In addition, the ratio of the grid size ($\Delta h = \sqrt{\Delta x^2 + \Delta y^2 + \Delta z^2}$) to the Kolmogorov length scale ($\eta = (\nu^3/\varepsilon)^{1/4}$) is below 6 throughout the domain.

### 2.4. *Statistics collection and quantity decomposition*

Simulations are performed with a constant time step of $1.8 \times 10^{-4}\delta_o/U_{e,o}$. After each case reaches a statistically steady state, three-dimensional mean fields are obtained by time averaging during the simulation runtime. In addition, full-domain snapshots are saved at intervals of $1.0\delta_o/U_{e,o}$, while selected two-dimensional planes are extracted every $0.07\delta_o/U_{e,o}$, to support the flow analysis. Statistics were collected over a total sampling period of $150\delta_o/U_{e,o}$. Using half of this period produces differences of less than 1% in the mean velocity and drag, and less than 3% in the second-order statistics, indicating adequate convergence of the time- and ensemble-averaged quantities.

In the following analysis, an instantaneous flow quantity $\phi$ in case RB is first decomposed as

$$\phi(x, y, z, t) = \overline{\phi}(x, y, z_r) + \phi'(x, y, z, t). \tag{2.4}$$

Here, $\overline{\phi}(x, y, z_r)$ denotes the dispersive, or equivalently secondary, field that is averaged in time and over riblet periods. Thus, $z_r = \mathrm{mod}(z, s) \in [0, s)$ denotes the spanwise coordinate within a riblet period; points with the same $z_r$ in different riblet periods are treated as equivalent in the spanwise ensemble average. $\phi'(x, y, z, t)$ is the stochastic or turbulent fluctuation. The dispersive field is then decomposed as:

$$\overline{\phi}(x, y, z_r) = \left\langle \overline{\phi} \right\rangle(x, y) + \tilde{\phi}(x, y, z_r), \tag{2.5}$$

where $\left\langle \overline{\phi} \right\rangle(x, y)$ is the spanwise-average of the time- and riblet-period-averaged field. Here, operator $\langle \phi \rangle$ indicates the intrinsic average over the fluid domain only. The superficial average, including the solid and fluid, is indicated by $\langle \phi \rangle_s$ (Nikora *et al.* 2007). $\tilde{\phi}(x, y, z_r)$ is the spatial variation about $\langle \phi \rangle$. The full decomposition, therefore, becomes:

$$\phi(x, y, z, t) = \left\langle \overline{\phi} \right\rangle(x, y) + \tilde{\phi}(x, y, z_r) + \phi'(x, y, z, t). \tag{2.6}$$

This decomposition is consistent with the double-averaging framework widely used in studies of rough-wall (Raupach *et al.* 1991; Nikora *et al.* 2007; Mignot *et al.* 2009) and riblet flows (Modesti *et al.* 2021; Endrikat *et al.* 2021*a*). However, due to the streamwise inhomogeneity of the flow, quantities cannot be averaged over $x$ and thus it remains as an independent variable. In case SM, $\tilde{\phi} = 0$, thus $\overline{\phi} = \left\langle \overline{\phi} \right\rangle$ and the classical Reynolds decomposition is recovered. A list of operators and variables used in this paper is provided in Appendix B.

## 3. Instantaneous flow

To visualize the spatially developing flow and identify the regions discussed below, figure 4 shows instantaneous streamwise velocity in selected $x-z$ planes. The top panel corresponds to case SM at $y_o^+ \approx 10$, while the middle and bottom panels correspond to case RB at $y/h = 1.0$ and $y/h = 1.5$, respectively. Each panel spans the full streamwise domain of interest and includes three zoomed-in insets at representative locations. For reference, key streamwise positions based on the mean flow development are marked, including the maxima of $\delta$, $C_f$, and $K$, the mean retransition onset, minimum $\delta$, and the limits of quasi-ZPG regions marked by freestream velocity increase by $3\% U_{\infty,o}$ and $95\% \Delta U_\infty$. The full streamwise development of these mean quantities is discussed later in §4.

Over the smooth wall (figure 4a), the boundary layer initially evolves under near-ZPG conditions for approximately the first $25\delta_o$, during which the instantaneous near-wall streak pattern remains broadly similar to that upstream. As the imposed acceleration strengthens downstream, the near-wall streaks become progressively more elongated relative to the upstream quasi-ZPG region. This trend is especially clear when the inset at $x/\delta_o \in [43, 45]$ is compared with that at $x/\delta_o \in [8, 12]$. Over this same streamwise range, $C_f$ increases and reaches its maximum slightly upstream of the peak in $K$, at $x/\delta_o \approx 41$. The near-wall streaks continue to stretch downstream beyond these peaks, with streamwise-elongated quasi-laminar regions remaining visible until $x/\delta_o \approx 65$, even as both $C_f$ and $K$ decrease. In this downstream region, the freestream acceleration has entered its final 5%$\Delta U_\infty$ interval and $\delta$ reaches its minimum.

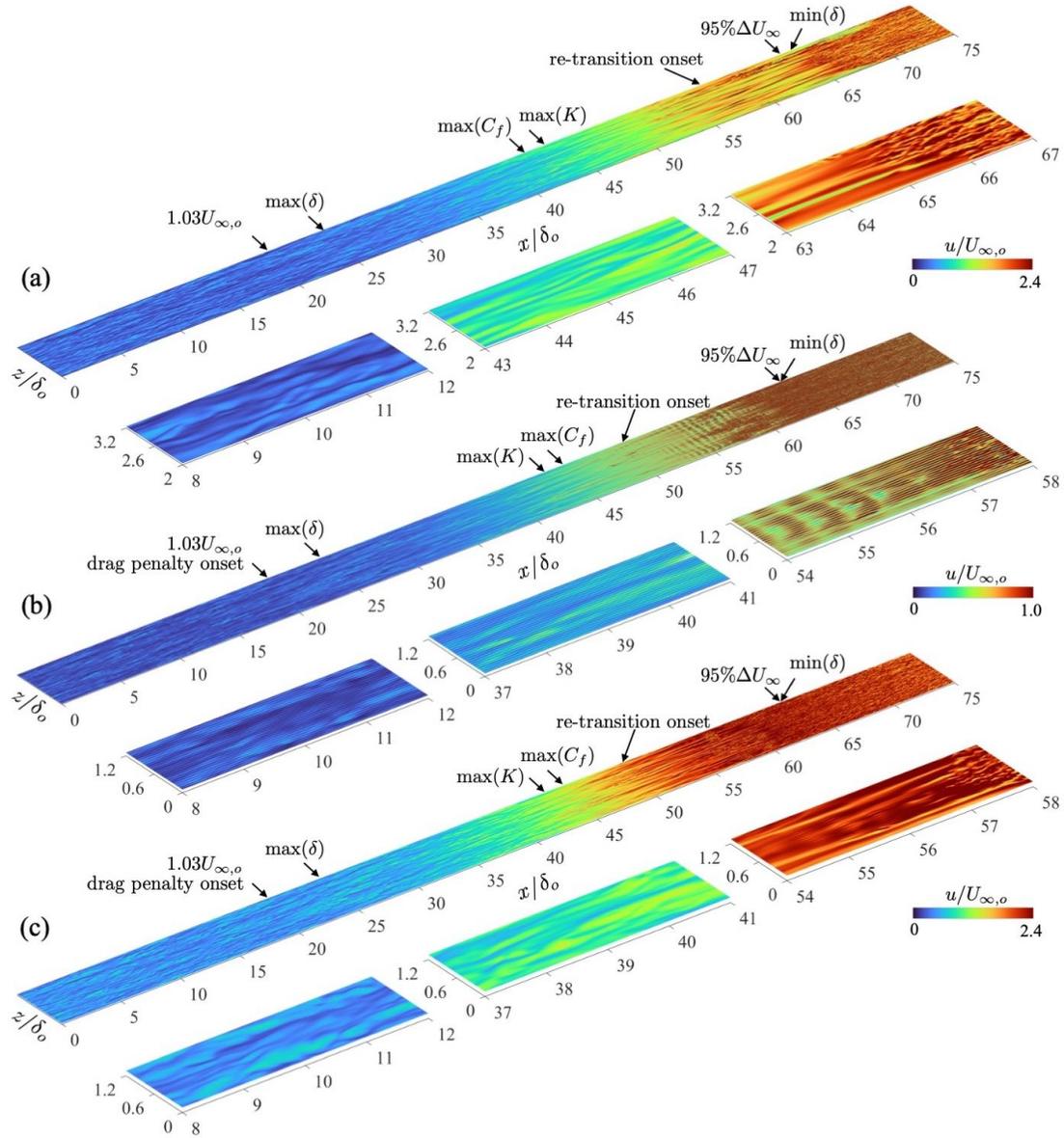

Figure 4: Instantaneous streamwise velocity in selected $x - z$ planes: (a) at $y^+$=10 for case SM; (b) at the riblet crest ($y = h$) for case RB; (c) at $y = 1.5h$ for case RB. Each panel includes three zoomed-in insets highlighting local flow features.

Also in this region, intermittent high-intensity turbulent spots emerge within the otherwise quasi-laminar flow. These spots are visible as early as $x/\delta_o \approx 56$ and rapidly expand across most of the span, with quasi-laminar flow becoming increasingly rare beyond $x/\delta_o \approx 72$.

The inset at $x/\delta_o \in [63, 67]$ clearly delineates the elongated near-wall streaks upstream from the high-intensity turbulent flow downstream. As shown later from the statistical analysis, retransition occurs over a relatively short streamwise interval, $x/\delta_o \in [56, 72]$, in contrast to the more gradual relaminarization upstream, despite the acceleration parameter profile being approximately symmetric about $x/\delta_o \approx 43$.

At $y = 1.5h$ ($7\nu/u_{\tau,o}$) above the riblet crest, figure 4c), the instantaneous flow over riblets initially resembles that over the smooth wall, with acceleration producing elongated near-wall streaks and reduced turbulence intensity. The maximum of $C_f$ is observed slightly downstream of the peak in $K$, in contrast to the smooth-wall case, although the two extrema remain close in the streamwise direction. This shift is relevant to the riblet-induced modulation of skin friction and is examined in detail in §5. Downstream of the peaks in $K$ and $C_f$, the near-wall streaks continue to stretch in a manner similar to the smooth-wall case. However, intermittent high-intensity turbulent spots are evident as early as $x/\delta_o \approx 54$ in the time instant shown (see inset at $x/\delta_o \in [54, 58]$), indicating that retransition is initiated substantially earlier over riblets; on average, its onset is about $6\delta_o$ upstream of that in case SM. Notably, this early onset of retransition occurs while the flow is still subject to substantial acceleration, in contrast to the smooth-wall case, where retransition occurs only after the acceleration has largely weakened.

Riblet-induced near-wall organization becomes more pronounced when the flow is examined at the riblet crest (figure 4b). Already by $x/\delta_o \approx 50$, spanwise-coherent bands of alternating high- and low-speed flow are observed near the wall. These structures, as will be examined in later sections, are consistent with Kelvin–Helmholtz (KH)-type rollers reported in ZPG flows over riblets (García-Mayoral & Jiménez 2011b; Endrikat et al. 2021a; Abu Rowin et al. 2025; Camobreco et al. 2025). They are much less apparent in the plane above the riblets shown in figure 4(c) and are absent in the smooth-wall case. As the flow develops downstream, the spanwise-coherent structures become progressively more pronounced, accompanied by a rapid emergence of intermittent high-intensity turbulent spots. Inspection of the inset at $x/\delta_o \in [54, 58]$ suggests that the first turbulent spots emerge locally from the spanwise-coherent structures, rather than through an immediate collapse across the full span. By $x/\delta_o \approx 60$, the flow appears to have largely retransitioned, approximately $12\delta_o$ upstream of that in the smooth-wall case, and the spanwise-coherent structures are no longer evident in the instantaneous velocity field.

Summarizing case RB, the qualitative features of relaminarization remain similar to those observed over the smooth wall, including the elongation of near-wall streaks and a reduction in turbulence intensity. However, the presence of riblets introduces additional spanwise-coherent near-wall organization that is absent in case SM. In particular, the downstream shift of the peak $C_f$ indicates that the riblets modulate the skin-friction response to the imposed acceleration. Moreover, spanwise-oriented KH rollers emerge near the riblet crest. Their subsequent local disruption is accompanied by the appearance of intermittent turbulent spots and a substantially earlier onset of retransition than in the smooth-wall case. These instantaneous observations provide the basis for the more detailed analyses presented in the following sections.

## 4. Mean flow

The mean evolution of the TBL under the imposed freestream FPG is illustrated by the time- and spanwise-averaged streamwise velocity contours in figure 5. These contours are used to establish a global view of the boundary-layer response to the FPG. The $y$-axis is plotted on a logarithmic scale to highlight the near-wall region. Over the smooth wall, the boundary layer initially evolves under weak acceleration, during which it continues to grow, and streamlines remain embedded within the TBL. As the acceleration strengthens downstream, the boundary layer undergoes pronounced thinning. At the same time, the streamlines gradually shift closer to the wall; however, this downward displacement occurs more slowly than the thinning of the boundary layer, so that the streamlines are positioned outward relative to the shrinking boundary-layer thickness. This behavior is consistent with previous observations in accelerating TBLs (Piomelli & Yuan 2013). As the acceleration diminishes, the boundary layer reaches its minimum thickness and subsequently thickens as the flow undergoes retransition.

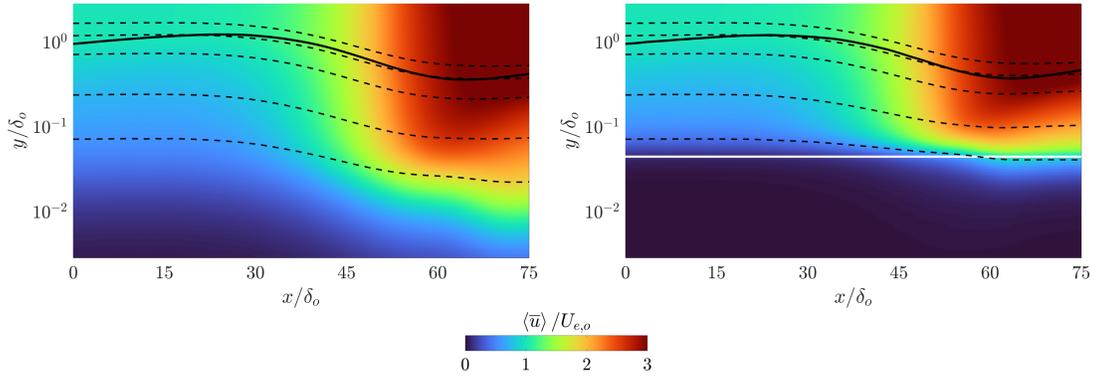

Figure 5: Mean streamwise velocity ($\langle \overline{u} \rangle /U_{e,o}$). Left: case SM; right: case RB. The solid black line shows $\delta(x)/\delta_o$, while the dashed black lines are streamlines beginning at $x = 0$, $y/\delta_o = 0.075, 0.25, 0.75, 1.25$, and $1.75$. For case RB, the riblet crest is indicated by the horizontal white line.

The overall boundary-layer development over riblets follows a similar trend, suggesting that the outer-layer response to the imposed FPG remains broadly comparable to that of the smooth-wall case. In the presence of riblets, the same downward migration of streamlines manifests differently near the riblet. Specifically, while the streamlines approach the wall in both cases, the lowest streamline shown here penetrates the riblet grooves. For example, the streamline initialized at $x = 0$ and $y/\delta_o = 0.075$ ($1.6h$), which is the closest to the wall among those shown, descends below the riblet crest at approximately $x \approx 57$. This behavior indicates that the groove flow becomes increasingly influenced by higher-momentum fluid from above the crest plane. The velocity contours further show that such penetration is primarily observed for $x/\delta_o \gtrsim 30$; upstream of this location, the streamlines remain nearly horizontal, indicating only weak penetration of the groove region by the accelerating outer flow. Even in the region where the streamline penetrates the grooves, the riblet geometry strongly limits the wall-normal extent of high-momentum intrusion: although the streamlines enter the grooves, the velocity contours show that the flow within the grooves remains relatively low-speed, unlike the smooth-wall case in which high-momentum fluid reaches the wall more directly. As a result, the wall-normal velocity gradient is preferentially concentrated in the vicinity of the riblet crest, giving rise to a locally enhanced shear compared with the smooth-wall case, while the flow within the riblet trough remains comparatively weakly sheared.

To quantify the mean-flow evolution illustrated in figure 5 and to assess how standard boundary-layer measures respond to the imposed FPG and riblets, profiles of several integral and near-wall parameters are compared in figure 6. These quantities provide a complementary description of the boundary-layer development, allowing the smooth-wall and riblet cases to be compared in terms of bulk thickness, near-wall state, and drag-related measures. In all panels, the bell-shaped distribution of the acceleration parameter $K$ is superimposed as a light gray line to facilitate comparison with the streamwise evolution of the flow. Selected quantities extracted from the figures and used to compare key events throughout this study are summarized in Table 4.

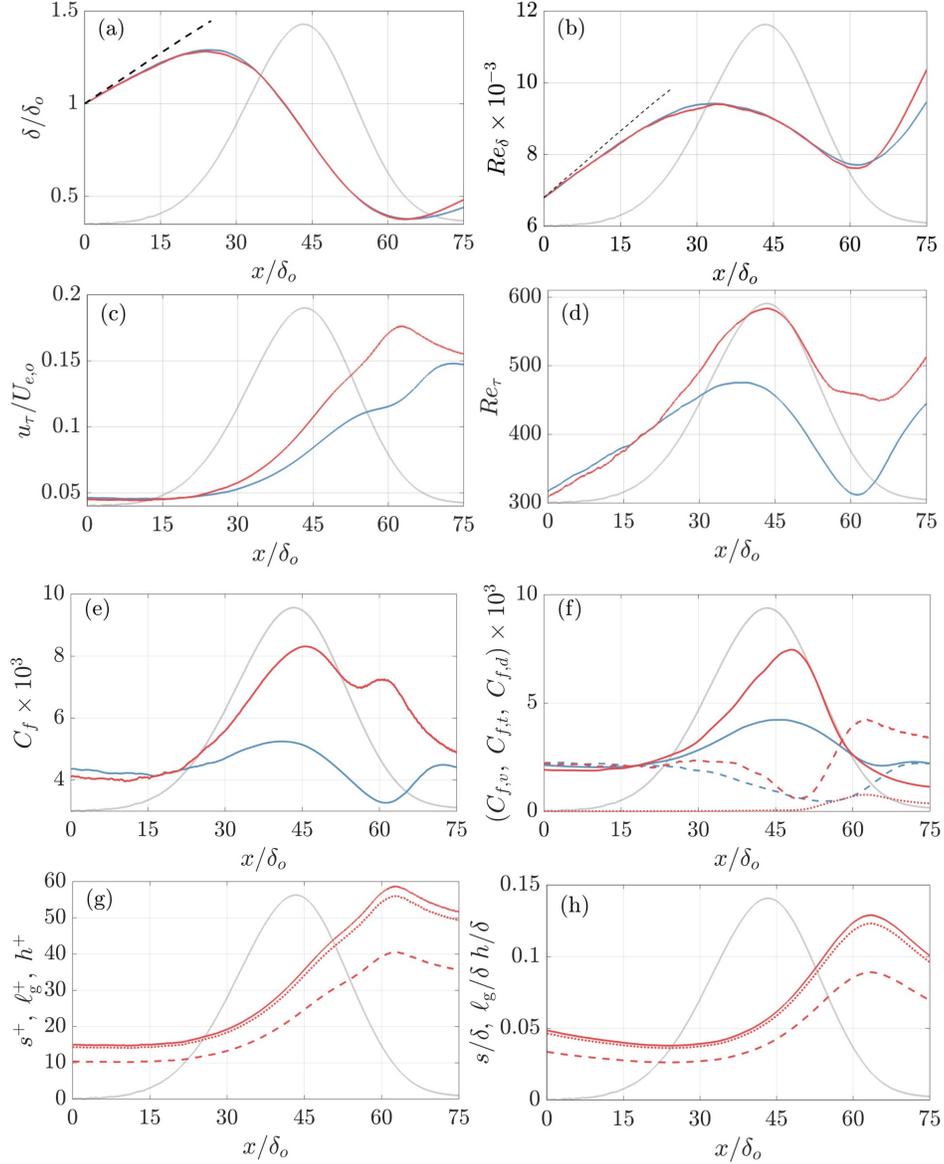

Figure 6: Streamwise profiles of (a) boundary layer thickness; (b) Reynolds number $Re_\delta$; (c) friction velocity; (d) friction Reynolds number $Re_\tau$; (e) skin friction coefficient $C_f = 2\tau_w/U_{e,o}^2$; (f) viscous contribution (——, $C_{f,v}$), turbulent contribution (– – –, $C_{f,t}$), and dispersive stress (······, $C_{f,d}$) contribution to $C_f$ (Zhang *et al.* 2024); (g) viscous-scaled riblet peak-to-peak spacing (——, $s^+$), square-root of groove cross-sectional area (– – –, $\ell_g^+$), and height (······ $h^+$); and (h) quantities in (g) normalized by local boundary layer thickness. —— case SM; —— case RB. The gray line shows $K(x)$. The dashed line in (a,b) shows ZPG TBL data of Wu *et al.* (2017) for reference.

The evolution of the boundary-layer thickness and the corresponding 'bulk' Reynolds number is shown in figure 6(a,b). The growth of the turbulent boundary layer is reduced relative to the ZPG reference, even under mild acceleration. Both the smooth-wall and riblet cases exhibit nearly identical behavior through the various stages of FPG. This close agreement indicates that the bulk development of the boundary layer may remain comparable between the two cases throughout the relaminarization phase. As the retransition begins, differences in $\delta$ become apparent: the boundary layer over the smooth wall becomes thinner than that over the riblets. Despite the small difference in $Re_\delta$ following retransition, these results indicate that the smooth-wall and riblet cases exhibit comparable bulk boundary-layer evolution under the imposed FPG. These results indicate that the smooth-wall and riblet cases undergo similar bulk boundary-layer development under the imposed FPG. This similarity supports a meaningful local comparison of wall stress and other drag-related quantities between the two cases at the same streamwise location.

| Criterion | Physical Significance | Smooth | Riblet |
|---|---|---|---|
| $C_{f,\text{RB}} > C_{f,\text{SM}}$ | Riblet becomes drag-increasing | - | 19 |
| $\delta_{\max}$ | Departure from quasi-ZPG TBL growth | 25 | 24 |
| $C_{f,\max}$ | Peak drag during acceleration | 41 | 45 |
| $K_{\max}$ | Peak acceleration | 43 | 43 |
| $C_{f,t,\min}$ | Onset of retransition | 56 | 50 |
| $C_{f,t,\max}$ | End of retransition; recovery toward quasi-ZPG | 72 | 62 |

Table 4: Streamwise locations ($x/\delta_o$) corresponding to key boundary-layer development events and definitive quantities for the smooth and riblet cases.

Given the comparability of the bulk boundary-layer evolution, the streamwise variation of wall-stress-related quantities is examined in Fig. 6(c–f), to characterize the turbulence state and drag response under the imposed FPG. At the reference plane, the riblet case exhibits a lower friction velocity than the smooth-wall case, consistent with the drag-reducing behavior expected for the selected riblet geometry under upstream ZPG conditions. As the flow accelerates downstream, $u_\tau$ increases in both configurations, reflecting the increasing wall stress induced by the FPG. Remarkably, the riblet becomes drag-augmenting as early as $x/\delta_o \approx 19$, where the freestream velocity has increased by only about 3%, and the acceleration remains weak.

While the friction velocity $u_\tau$ directly reflects the absolute wall stress, the corresponding evolutions of $Re_\tau$ and $C_f$ provide additional insight into the state of the near-wall turbulence and its relation to the outer flow. In particular, $Re_\tau$ incorporates the concurrent change in boundary-layer thickness, whereas $C_f$ normalizes the wall stress by the local edge velocity. In both cases, downstream of the peak acceleration at $x/\delta_o \approx 43$, $u_\tau$ continues to increase, whereas both $Re_\tau$ and $C_f$ decrease rapidly. This contrast suggests that the near-wall turbulence adjusts more slowly than the outer-layer mean flow to the evolving FPG: the outer layer adjusts rapidly through boundary-layer thinning and an increase in $U_e$, while the near-wall region experiences increasing wall stress without a corresponding enhancement of turbulence activity. As a result, $Re_\tau$ and $C_f$ decrease even as $u_\tau$ increases. Notably, in the smooth-wall case, both quantities drop to levels comparable to, or even lower than, those at the reference plane, whereas this level of reduction is not observed in the riblet case.

Further downstream, the retransition process is also reflected in the behavior of $Re_\tau$ and $C_f$. In both cases, these quantities subsequently increase during retransition, consistent with a near-wall turbulence response that strengthens more rapidly than the outer layer can adjust. However, over riblets, the retransition region exhibits a rapid and non-monotonic recovery that differs qualitatively from the smooth-wall case. Specifically, $Re_\tau$ exhibits a pronounced plateau and $C_f$ shows a secondary peak, indicating a renewed intensification of near-wall turbulence during retransition. No such secondary peak was not observed in the FPG TBL over sandgrain roughness reported by Yuan & Piomelli (2015). In the riblet case, this behavior is accompanied by a local maximum in $u_\tau$ near $x/\delta_o \approx 62$, suggesting a transient over-response of near-wall turbulence generation as the flow retransitions. This behavior is consistent with the earlier and more intense retransition observed in the instantaneous flow (figure 4b,c). By the end of the domain, the differences in $u_\tau$ and $C_f$ between the two cases diminish rapidly, while the difference in $Re_\tau$ remains significant, suggesting a more rapid thickening of the boundary layer featuring the retransition over the riblets.

The evolution of the TBL and the associated changes in $C_f$ can be further understood by decomposing $C_f$ into its viscous ($C_{f,v}$), turbulent ($C_{f,t}$), and, for case RB, dispersive ($C_{f,d}$) contributions, following the kinetic-energy-based decomposition of Zhang *et al.* (2024). For case SM, as the flow accelerates, $C_{f,v}$ increases and becomes the dominant contributor to the rise in $C_f$. Its peak approximately coincides with the peak $K$. In contrast, $C_{f,t}$ decreases more gradually, exhibiting a lagged response and reaching its minimum downstream of the peak $K$. These trends are consistent with previous studies of FPG TBLs, in which relaminarization has been interpreted as a viscous-dominated response accompanied by a delayed attenuation of turbulence relative to the mean shear. (Warnack & Fernholz 1998; Bourassa & Thomas 2009; Piomelli *et al.* 2000; Falcone & He 2022).

For case RB, $C_{f,v}$ also increases with the acceleration and broadly follows the trend of $K$, but its rise is stronger than in case SM, making it the primary contributor of riblet-induced drag penalty up to $x/\delta_o \approx 55$. $C_{f,t}$ also decreases similarly, indicating comparable turbulence suppression/relaminarization over the riblets. However, it reaches its minimum much earlier, near $x/\delta_o \approx 50$, demonstrating the early retransition. Further downstream, the rapid increase of $C_{f,t}$ makes it an important contributor to the total drag and gives rise to the secondary peak in $C_f$ at $x/\delta_o \approx 62$. Associated with the increase of $C_{f,t}$ is an increase of the third contributor, the dispersive component $C_{f,d}$, indicating riblet-scale wake stresses only appear once turbulence interacts with the flow inside the groove.

Because $C_{f,t}$ more closely tracks the relaminarization and retransition processes in both cases than total $C_f$, and exhibits close spatial correspondence with the qualitative flow changes observed in the instantaneous and mean fields, it is adopted here to define the onset and completion of these processes, as summarized in Table 4.

The change of the wall stress described above alters the local riblet length scales in wall units, which are commonly used in ZPG studies to categorize drag-modulation regimes and interpret riblet–turbulence interactions. In the present configuration, the riblet geometry is fixed in physical space; consequently, the streamwise evolution of all quantities shown in figure 6(f) reflects the variation of the friction velocity solely. As designed, the riblet length scales at the ZPG reference plane ($x = 0$) fall within the classical drag-reduction regime established for ZPG turbulent flows. Under canonical ZPG conditions, riblets operating within this range are understood to reduce drag by impeding spanwise motion near the wall, particularly in the near-wall region associated with the quasi-streamwise vortex legs. However, as the flow accelerates downstream in case RB, the viscous-scaled riblet dimensions no longer follow the canonical correlations between riblet size in wall units and drag modulation established for ZPG flows. Specifically, $\ell_g^+$ remains below 20 and $s^+$ remains below 30 up to $x/\delta_o \approx 38$; nevertheless, as shown in figure 6(c), the riblets become drag-augmenting as early as $x/\delta_o \approx 19$.

In addition, because the TBL thins during the acceleration, the riblet dimensions become progressively larger relative to the local boundary-layer thickness. For example, while the riblet height corresponds to approximately 4.5% of the boundary-layer thickness at the reference plane—placing them within the commonly used roughness classification—it increases to approximately 20% of the local boundary-layer thickness in the retransition region. At this stage, the riblets are large enough relative to the local boundary-layer thickness to potentially affect not only the immediate near-wall region but also a substantial fraction of the inner layer. A detailed discussion of the resulting drag modulation is given in the following section; the profiles shown here serve as a reference for the subsequent analyses of riblet performance in both viscous units and outer scaling.

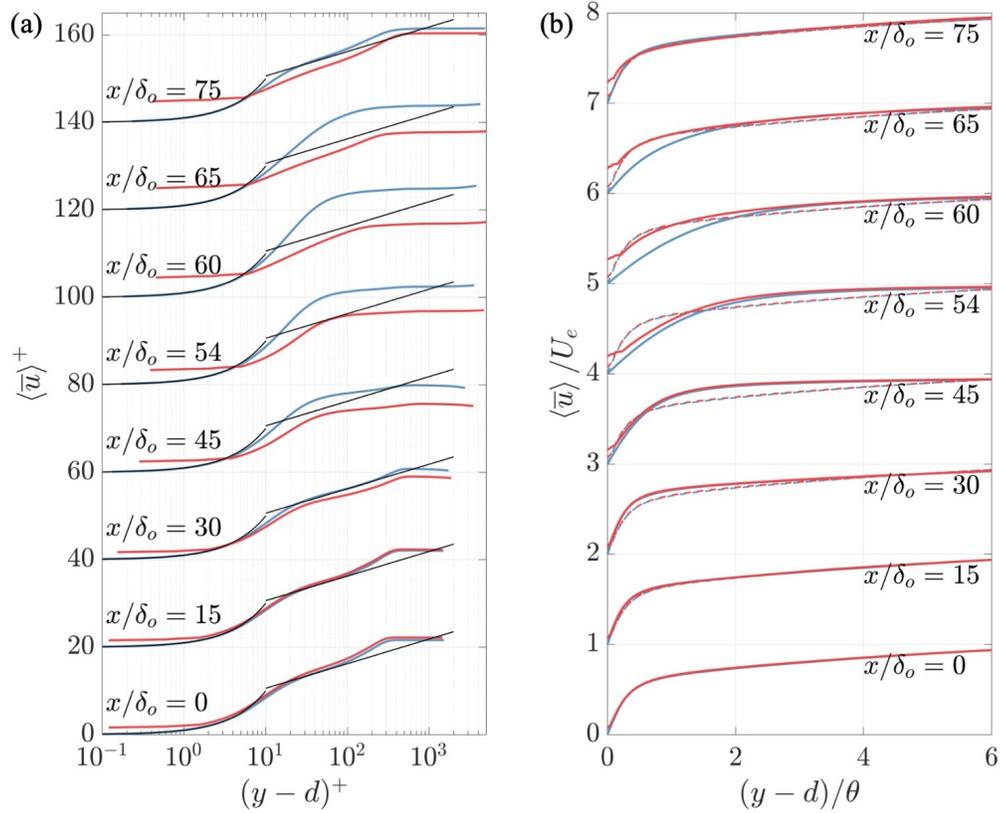

Figure 7: Development of mean streamwise velocity profiles scaled with (a) wall units and (b) outer units. The streamwise location of each profile is indicated by text. In both figures, —— case SM; —— case RB. In (a), the thin solid lines display $U^+ = y^+$ and $U^+ = 2.5\log(y^+) + 5.0$. For case RB, the wall-normal coordinate is shifted by the virtual origin $d = 0.9h$. $d = 0$ for case SM. In panel (b), the thin dashed lines represent the profiles at $x/\delta_o = 0$.

Mean streamwise velocity profiles extracted at selected streamwise locations are examined in figure 7. The profiles are presented using inner scaling (a) and outer scaling (b). For the riblet case, the profiles are shifted by the virtual origin $d = 0.9h$. At the ZPG reference plane ($x/\delta_o = 0$), the riblet case exhibits a modest upward shift of the logarithmic region relative to the smooth-wall case, consistent with its drag-reducing role. At $x/\delta_o = 15$ and 30, where the FPG remains mild, the velocity profiles in both cases retain near-equilibrium ZPG TBL characteristics, including an approximately logarithmic slope region. However, the upward shift of the riblet profiles diminishes progressively, eventually falling below the smooth-wall profile (i.e., drag-augmenting).

During the period of strong acceleration ($x/\delta_o = 45$ and 54), both cases exhibit a thickened viscous layer and a pronounced departure from the logarithmic law. For convenience, we refer to this region as the 'relaminarization' region, although this terminology is used to describe the observed *trend* toward laminarization rather than a fully laminar *state*. In this regime, the riblet profiles are shifted substantially downward relative to the smooth-wall case, continuing their drag-augmenting behavior identified earlier. Evidence of earlier retransition over riblets is observed at $x/\delta_o = 60$ and 65, where the riblet profiles recover toward the logarithmic slope, while the smooth-wall profiles remain quasi-laminar.

Additional evidence of relaminarization and earlier retransition over riblets is provided by the outer-scaled velocity profiles shown in figure 7(b). The ZPG profiles at $x/\delta_o = 0$ are repeated at each streamwise location for reference. During the acceleration phase, both cases exhibit quasi-laminar velocity profiles characterized by reduced wall-normal velocity gradients relative to the reference profiles. The earlier retransition of the riblet case is again evident from the near-collapse of the profiles at $x/\delta_o = 65$ onto the reference profile. Also noteworthy is the evolution of the mean velocity in the vicinity of the riblet crest, which increases markedly during the acceleration and reaches values of up to approximately 30% of the local boundary-layer edge velocity. A similar behavior was observed near the top of a sandgrain roughness layer in an FPG TBL by Yuan & Piomelli (2015), where it was considered as a contributing mechanism to drag increase by the sandgrains. A more detailed examination of the resulting drag modulation is presented in the next section.

## 5. Drag modulation

### 5.1. *Drag curve*

The previous section demonstrated that the riblets considered in the present study, which are designed to be drag-reducing at the reference plane under ZPG, produce a substantial increase in wall shear stress shortly after the onset of the FPG, even while the riblet scales expressed in wall units remain within the canonical drag-reducing range. Following previous works on riblets in ZPG flows (García-Mayoral & Jiménez 2011b; Gatti *et al.* 2020; Modesti *et al.* 2021; Endrikat *et al.* 2021b), we quantify drag modulation using the drag curve, which plots the change of wall stress with respect to the smooth wall, $\Delta\tau_w(x)$, as a function of $\ell_g^+(x)$. That is,

$$\Delta\tau_w(x) = 100 \times \frac{\tau_{w,RB}(x) - \tau_{w,SM}(x)}{\tau_{w,SM}(x)}\% . \qquad (5.1)$$

Drag is compared at corresponding streamwise locations between the two cases because the boundary-layer thickness Reynolds number $Re_\delta$, and therefore the bulk boundary-layer mass flux, remains nearly identical over most of the domain. In studies of surface roughness and riblets, drag modulation is also commonly quantified using the shift of the mean velocity profile in the logarithmic region. However, this approach is not applicable in the present flow, as the imposed FPG renders the log layer invalid over a significant streamwise extent of the domain (see figure 7).

Our results, compared with the ZPG drag curve from Gatti *et al.* (2020), are displayed in figure 8. The FPG drag curve shows a rapid departure from ZPG behavior with strong nonlinear streamwise variations. Starting from a $\sim 5\%$ drag reduction at the reference plane, where $\ell_g^+$=10.5, $\Delta\tau_w$ increases with $\ell_g^+$ far more rapidly than in the ZPG case. As noted earlier from the $C_f$ plot, the riblets become drag-increasing as early as $x/\delta_o = 19$, by which point the freestream velocity has increased by only 3%. Such an increase occurs even when $\ell_g^+$ remains within the canonical drag-reducing range defined for ZPG flows. For example, a 20% drag increase is observed at $\ell_g^+$=12.5, and a 45% increase at $\ell_g^+$=18, which is the (approximate) threshold for riblets to become drag-increasing in ZPG flows. The drag curve continues this initial concave increase before reaching an inflection point around $x/\delta_o = 50$, shortly downstream of the peak acceleration. This location coincides with the minimum of $C_{f,t}$ in case RB and marks the onset of retransition. Further downstream, $\Delta\tau_w$ rises more rapidly because, at the corresponding streamwise locations, the flow in case SM remains quasi-laminar, while case RB undergoes an earlier retransition. A maximum drag increase of $\Delta\tau_w \approx 124\%$ is reached when case RB completes retransition, while case SM only begins. Further downstream, the drag curve recovers toward the ZPG drag curve. By the end of the domain of interest, the drag penalty falls below the ZPG curve at $\ell_g^+ \approx 36$. This overshoot shows that the recovery stage is itself non-equilibrium, rather than a simple return to ZPG behaviour. The non-monotonic profile of the drag curve in the retransition and recovery regions shows that $\Delta\tau_w$ may not be a unique function of $\ell_g^+$, but depends on the local state of the boundary layer.

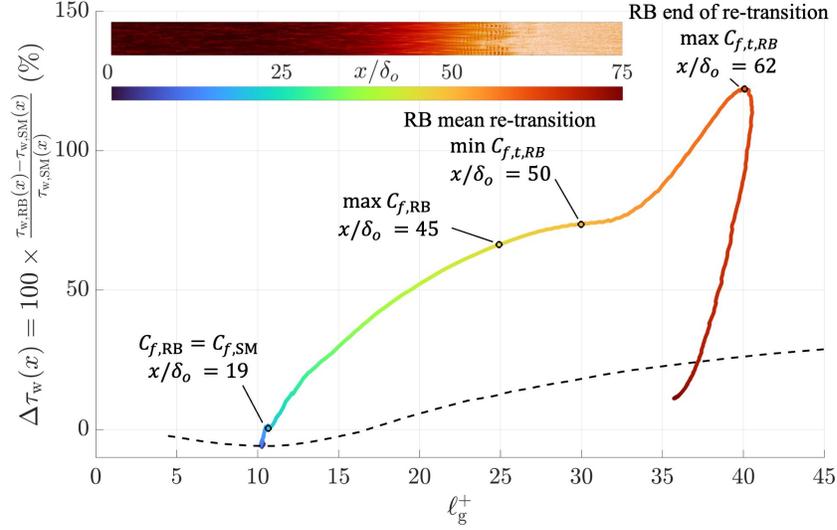

Figure 8: Drag curve for the current FPG TBL (rainbow-colored line) compared with the ZPG one adapted from Gatti *et al.* (2020) (dashed line). The downstream direction is indicated by a color transition from blue to red, as indicated by the color bar inset. For reference, the inset also includes an $x - z$ plane of instantaneous $u/U_{e,o}$ from case RB. Important streamwise locations are indicated by black circles and associated text.

Overall, the drag curve obtained in the present study departs significantly from that obtained in ZPG flows. In our separate work investigating the effects of riblets in APG TBLs (Savino *et al.* 2026), the drag curve also departs significantly from the ZPG behavior, exhibiting enhanced drag reduction as $\ell_g^+$ decreases in the flow. These deviations highlight the critical role of non-equilibrium effects in determining riblet performance.

## 5.2. *Sources of drag modulation*

The key question arised by figure 8 is what mechanisms are responsible for the drag increase in case RB relative to case SM. Several plausible candidates include: (1) an increase in the mean shear over the riblets, which imposes a larger driving force that must ultimately be balanced by skin friction drag on the riblet walls; (2) the development of secondary circulation within the riblet grooves; and (3) the generation of turbulent fluctuations inside the groove region. The latter two are commonly associated with drag increase in ZPG flows when riblets exceed their optimal size (Choi *et al.* 1993; Goldstein & Tuan 1998; Modesti *et al.* 2021). In the present spatially developing, non-equilibrium boundary layer, however, the relative importance of these mechanisms is not immediately clear. To clarify the underlying physics, we first examine the streamwise momentum budget within the riblet layer. This analysis allows us to identify the dominant terms in the mean momentum balance within the riblet region, and thereby to constrain the mechanisms most directly associated with the drag response.

Applying the double average to the streamwise momentum equation to isolate the spatial

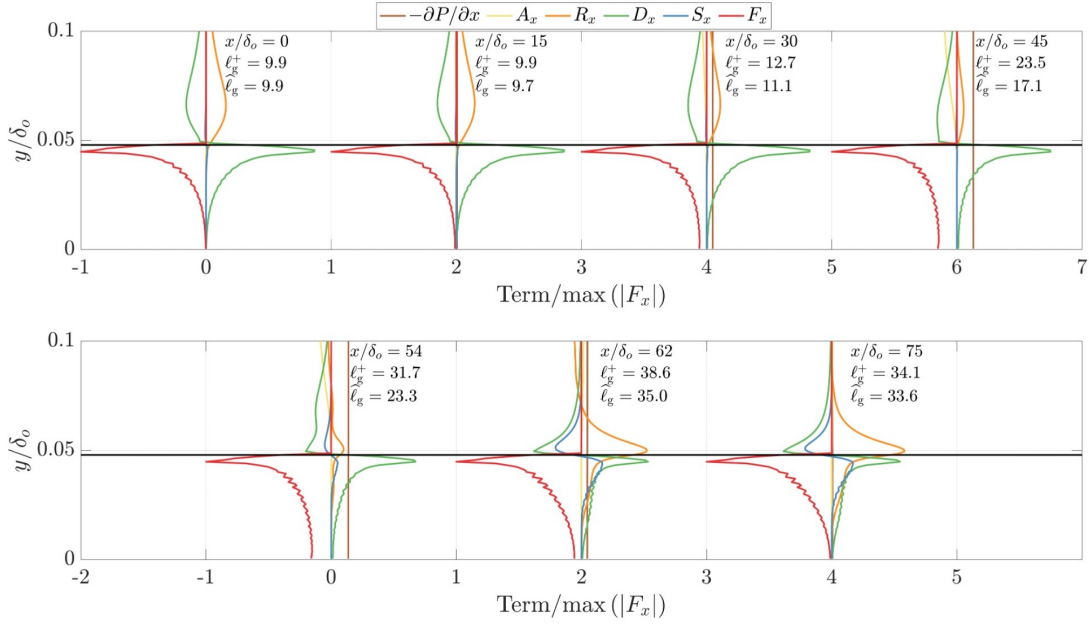

Figure 9: Mean streamwise momentum balance for case RB at selected streamwise locations. The balance terms at each location are normalized by its maximum absolute value of $F_x$ to demonstrate the contribution of each term to the total $\tau_{w,RB}$. The riblet crest is denoted by the horizontal solid black line.

dispersive fluctuations due to riblets, the balance reads (Yuan & Piomelli 2014*a,b*):

$$0 = -\frac{\partial \langle \overline{p} \rangle_s}{\partial x} - A_x + R_x + D_x + S_x + F_x. \tag{5.2}$$

The terms in the right-hand-side are the mean pressure gradient, mean convection ($A_x$), mean Reynolds stress divergence ($R_x$), mean viscous diffusion ($D_x$), mean divergence of the dispersive stress ($S_x$), and mean IBM force ($F_x$). Individual term definitions are given in Appendix C. The streamwise momentum balance is examined at selected locations in figure 9. Note that the terms at each location are normalized by the corresponding maximum magnitude of the IBM force to highlight the contribution of each component to $F_x$ and thus the total drag by the riblets. Moreover, the wall-normal coordinate is scaled with $\delta_o$ to allow direct comparison of the changes in distribution of balance terms in the riblet vicinity.

Across the majority of the domain, namely in the region upstream of retransition, both Reynolds and dispersive stress contributions within the groove are negligible. While this behavior is expected at $x = 0$, where the riblets operate in the drag-reducing regime, it contrasts with the drag-augmentation mechanisms often reported for large riblet sizes, where secondary flows and turbulence penetration into the groove are typically observed and attributed to the drag augmentation (Modesti *et al.* 2021). Instead, they do not contribute to the drag; the sole source of mean momentum is viscous transport associated with the mean shear. Consequently, $F_x$ arises entirely from viscous friction on the riblet walls, with negligible contribution from inertial transport between the groove and the outer boundary layer (i.e., Stokes-like, as in drag-reducing riblets in ZPG flows).

Note that zero $R_x$ and $S_x$ do not necessarily mean the Reynolds or dispersive stresses are zero. In the sections below, we will show that the Reynolds stresses are nearly zero in the groove until retransition, but the dispersive stresses caused by spanwise inhomogeneity has an appreciable streamwise component. The negligible $R_x$ and $S_x$ here, however, are a direct result of the mean momentum balance and show that the presence of turbulent and spanwise inhomogeneity in the groove does not imply a *direct* local contribution to drag. Instead, these motions, if present, may remain dynamically active and affect other terms in the mean momentum budget, for example, by modifying the mean flow and mean shear, while contributing negligibly to the mean streamwise momentum balance through their own terms. Ultimately, it is the mean shear-related viscous effects that lead to the drag.

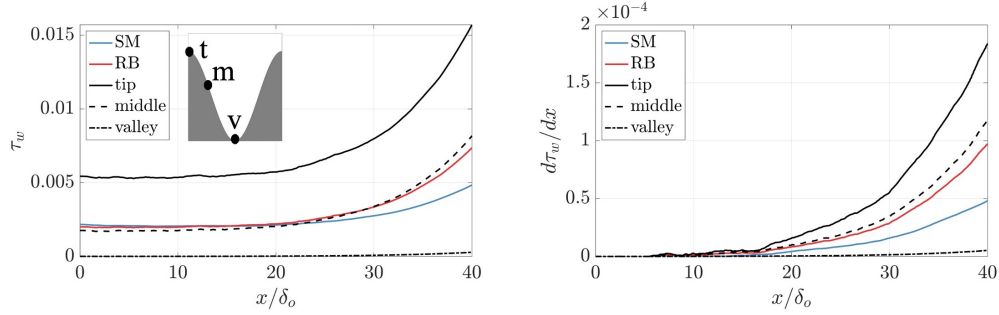

Figure 10: Left: spanwise-averaged wall shear stress produced by the smooth wall (———) and riblet wall (———), as well as local wall shear stress produced at the riblet tip (———); midway between the tip and valley (- - -); and riblet valley (-·-·-). Right: streamwise derivative of the stresses plotted on the left. The $x$-axis range is limited to the relaminarization region. The inset on the left provides a reference where the stress at the tip, middle, and valley is measured.

This scenario is altered downstream of the FPG peak as retransition occurs ($x/\delta_o$ = 54, 62, and 75 as shown here). Reynolds and dispersive stress contributions begin to appear near the riblet crest and progressively spread deeper within the groove. This marks the onset of inertial momentum transport and increasing interaction with the overlying turbulence. Besides, the dispersive-stress contribution extends into the overlying flow (i.e., forming a roughness sublayer) and, acting as a sink of mean streamwise momentum, reflects cross-stream transport generated by spatial variations of the mean flow induced by the riblet geometry. The Reynolds-stress contribution, meanwhile, supplies momentum to the mean flow in the vicinity of the riblet crest.

Note that the FPG acts as a body force and thus influences the flow throughout the groove in the wall-normal direction. Under strong FPG conditions (e.g., $x/\delta_o$ = 45 and 54 in the present figure), the very bottom of the riblet trough may become locally pressure-driven. Even so, the shear near the groove opening remains the primary driver of the groove flow, and its downward diffusion into the groove follows a qualitatively similar trend at all locations.

Therefore, over most of the acceleration region in the present flow, the riblet-induced drag increase differs from the mechanisms commonly invoked in ZPG studies, in which drag increase is typically associated with oversized riblets or with geometries that promote inertial effects within the groove. This picture appears to become relevant only after retransition sets in. A different mechanism is therefore needed to explain the upstream region. As will be shown below, the drag increase there is associated with viscous transport within the riblet groove. Meanwhile, in the retransition region, drag modulation by inertial effects is found to be closely associated with the spanwise Kelvin–Helmholtz rollers; this will be discussed later in the paper.

### 5.3. *Mechanism of drag modulation: a shear-driven, sheltered groove*

It is intuitive that streamwise acceleration increases the overall shear at the riblet crest plane. However, the viscous-drag distribution over riblets is known to be spanwise inhomogeneous, and it remains unclear whether the picture established in ZPG flows still applies when the groove flow is also subjected to FPG. Specifically, ZPG studies have shown that drag generation is highly localized: the riblet tips produce higher shear stress than a smooth wall, whereas lower shear occurs deeper within the grooves. The net drag then depends on the balance between these opposing contributions (Choi *et al.* 1993). Here, we examine how this local shear distribution evolves under FPG. Figure 10(a) shows the shear stress at the height of the riblet tip, mid-groove, and trough, and (b) shows their corresponding streamwise growth rate. The shear stress at the riblet tip and mid-groove increases more rapidly compared to the smooth-wall case, while the valley shear grows slowly. It indicates that the drag penalty arises primarily in the upper region of the current riblet, and reflects the dominance of viscous transport near the crest in the mean momentum budget, where increasing crest shear directly amplifies local stresses. The low-shear and possibly pressure-driven flow deep in the trough, in contrast, adds minor additional drag.

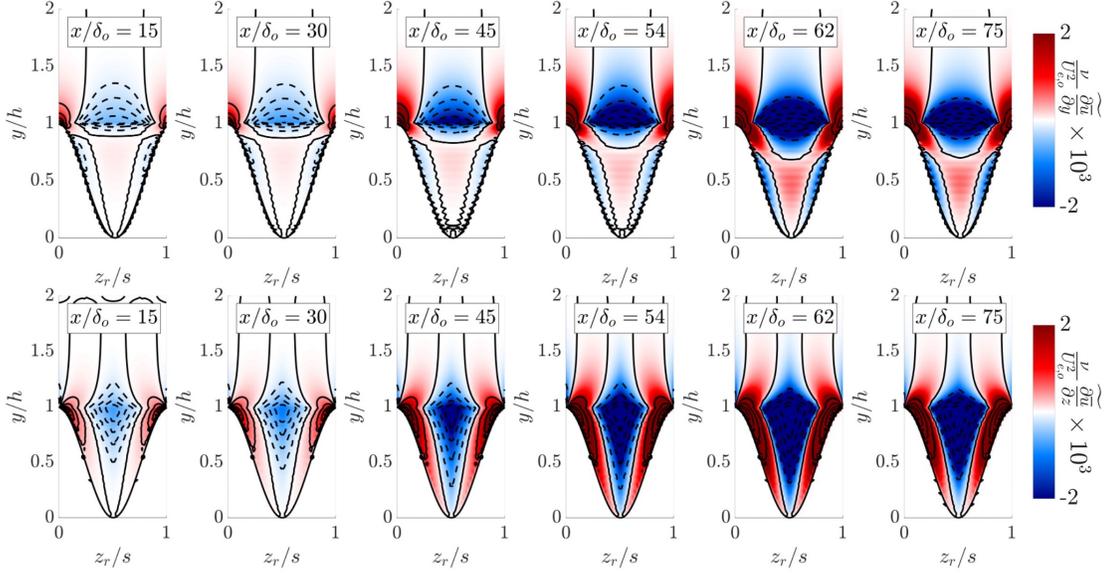

Figure 11: Spatial variation of the viscous shear-stress components relative to the spanwise average within the riblet groove for case RB. The filled contours may be viewed as a terrain map of the deviation from the spanwise average: red indicates positive "peaks" and blue negative "valleys", with darker color showing stronger spanwise inhomogeneity. Solid and dashed isocontours mark $[0, 0.2, 0.4, 0.6, 0.8] \times$ the local maximum and minimum variations, respectively. Their nearly invariant shapes indicate that the spatial pattern of the inhomogeneity remains almost the same. Thus, the groove shear-stress inhomogeneity undergoes mainly amplitude scaling rather than structural deformation.

The spatial distribution of the viscous shear is examined more closely in figure 11 through its spanwise deviation from the local spanwise average in the transverse ($z$–$y$) planes. At each wall-normal location, the spanwise-averaged shear is subtracted to isolate the across-groove variation, so that positive (negative) values indicate locally enhanced (reduced) shear relative to the spanwise mean. The filled contours, therefore, show the strength and sign of the spanwise variation, while the superposed isocontour lines mark fixed fractions of the local positive peak and negative trough, analogous to elevation lines on a terrain map that are used mainly to highlight the underlying spatial pattern. In this way, the color contours indicate how much the "peaks" rise and the "valleys" deepen, whereas the contour lines reveal whether the underlying spatial pattern changes shape.

At all streamwise locations, a preferential augmentation of wall-normal (spanwise) shear is observed near the riblet crest (along the sidewalls, particularly in the vicinity of the crest). In contrast, the groove mid-span consistently exhibits a relative shear deficit. This indicates a preferential distribution of shear toward those geometrically constrained regions, with the groove center remaining relatively weakly sheared. The result at $x/\delta_o = 15$ is nearly identical to the one at $x/\delta_o = 0$ (not shown) and can be considered as the baseline spanwise inhomogeneity caused by these riblets in the present TBL. As the FPG accelerates the flow, the magnitude of these variations intensifies, indicating a more rapid intensification of crest-generated shear than the spanwise average (i.e., increased inhomogeneity). Despite this amplification, the spatial pattern of the inhomogeneity remains largely unchanged, as evidenced by the consistent normalized isocontour lines across streamwise locations. Only near retransition ($x/\delta_o = 54$) does the spatial pattern near the riblet crest expand noticeably deeper into the groove, indicating enhanced interaction with the overlying TBL and the onset of riblet-scale turbulence. Nevertheless, the overall spatial pattern remains qualitatively similar. This persistence suggests that the non-uniform distribution of shear is governed by the riblet geometry, whereas the FPG mainly amplifies its magnitude, producing stronger shear concentration at the riblet crests and weaker shear in the groove opening.

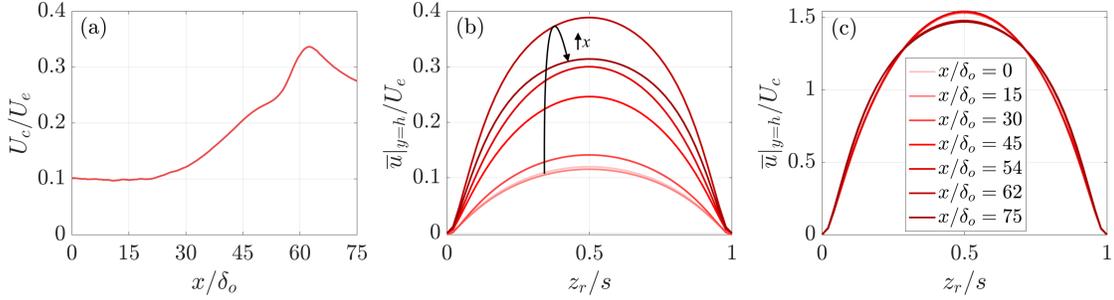

Figure 12: Profiles of (a) the time- and spanwise-averaged streamwise velocity at the riblet crest, $U_c$; (b) spanwise-dependent time-and ensemble-averaged streamwise velocity across the riblet opening, $\overline{u}|_{y=h}$; (c) $\overline{u}|_{y=h}/U_c$. In panels (b) and (c), darker shades of red correspond to increasing downstream locations.

Since the groove flow is driven primarily by the shear at the crest plane, and meanwhile, the spanwise inhomogeneity is most pronounced there, the velocity field at this location is examined in greater detail. Figure 12 shows the time- and spanwise-averaged streamwise velocity at the riblet crest ($U_c = \langle \overline{u} \rangle |_{y=h}$), together with the spanwise variation around it across the groove ($\overline{u}|_{y=h}$). Although both are expected to increase in absolute magnitude under streamwise acceleration, the figure further shows that they also increase relative to the velocity at the edge of the TBL. That is, the flow at the crest plane accelerates more than the outer region of the TBL. When the spanwise-varying velocity is normalized by its own spanwise average, the profiles collapse remarkably well at all streamwise locations (figure 12c), demonstrating a strong geometry-dependent self-similarity. This collapse persists even downstream of retransition ($x/\delta_o = 54$, 62, 75), with only minor departures in the form of slightly fuller profiles, indicating stronger interaction between the riblets and the overlying TBL that only happens during retransition in the current flow.

These observations suggest the following physical picture. Prior to retransition, the groove flow is largely decoupled from the overlying boundary layer and is driven by the shear associated with the mean velocity at the riblet crest plane. Negligible turbulence is present in the groove. As the FPG strengthens, this shear increases and is distributed by the riblet geometry, producing significantly enhanced stresses near the riblet crests and less intensified ones in the groove core. The upper part of the groove is most directly affected by this shear-driven mechanism and is responsible for most of the drag increase, whereas the lower part may behave more like a pressure-driven flow with only a limited contribution to drag. In this sense, the riblet geometry imposes a robust inhomogeneity of shear at the groove opening and preserves the overall flow and stress pattern within the groove despite the evolution of the overlying boundary layer. Thus, before retransition, the groove flow can be approximated as a geometry-dependent pattern whose magnitude is controlled by the FPG-altered mean velocity at the riblet crest plane. Following retransition, inertial momentum transport emerges within the groove, as evidenced by increased shear penetration and the appearance of Reynolds and dispersive stress contributions in the momentum balance, as well as a fuller crest velocity profile. Despite this shift in dynamics, the velocity and shear distributions retain their qualitative structure, again underscoring the persistent role of riblet geometry in organizing the near-riblet flow.

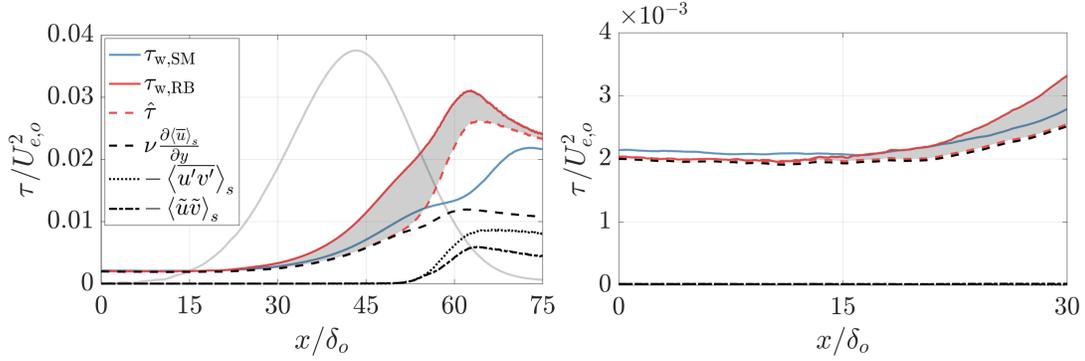

Figure 13: Comparison of various mean shear stresses. ——— $\tau_{\mathrm{w,SM}}$; ——— $\tau_{\mathrm{w,RB}}$; – – – $\hat{\tau}$; – – – viscous shear stress at riblet crest; ········· Reynolds shear stress at riblet crest; – · – · – dispersive shear stress at riblet crest. The $x$ range in the right plot is limited to $30\delta_o$ for clarity. The gray shaded region represents the difference between the total drag and $\hat{\tau}$.

### 5.4. Riblet crest plane as a partial-slip boundary

The isolation and geometric-dependency of the riblet groove flow suggest that the outer boundary layer is effectively decoupled from the motions within the riblet grooves, and that the outer flow does not directly experience the drag augmentation. In the wall-parallel plane at the riblet height, the boundary layer encounters narrow streamwise regions of elevated friction along the crests, while over most of the wall-parallel plane, it advects above the grooves with less resistance. This behavior may be interpreted as that of a partial-slip virtual boundary located at the riblet height.

The mean total shear stress across the groove opening (i.e., the effective overall shear stress exerted by this partial-slip plane on the flow above), denoted by $\hat{\tau}$, is defined as (Raupach *et al.* 1996; Nikora *et al.* 2007; Yuan & Piomelli 2014*a*,*b*):

$$\hat{\tau}(x) = -\left\langle \overline{u'v'} \right\rangle_s (x, h) + \nu \frac{\partial \langle \bar{u} \rangle_s}{\partial y}(x, h) - \langle \tilde{u}\tilde{v} \rangle_s (x, h). \qquad (5.3)$$

The quantity $\hat{\tau}$ and its three contributors are compared with $\tau_{\mathrm{w,RB}}$ and $\tau_{\mathrm{w,SM}}$ in figure 13. Prior to retransition ($x/\delta_o \approx 50$), $\hat{\tau}$ and $\tau_{\mathrm{w,SM}}$ evolve similarly while $\tau_{\mathrm{w,RB}}$ departs from their trend as early as $x/\delta_o = 13$. This is particularly evident over the initial $30\delta_o$, as shown in the zoom-in view in the right subplot, where the dashed red and solid blue lines are parallel, yet the red solid line shows a faster increase. The similarity between $\hat{\tau}$ and $\tau_{\mathrm{w,SM}}$ suggests that the overlying flow is subjected to an effective shear that remains close to that of the smooth-wall case, even though the total drag over the riblets increases substantially. Also note that up to $x/\delta_o = 50$ at retransition, $\hat{\tau}$ is entirely composed of viscous shear, with negligible contributions from the Reynolds and dispersive stresses. This offers further evidence, in addition to the mean momentum budget, for the absence of the latter two components.

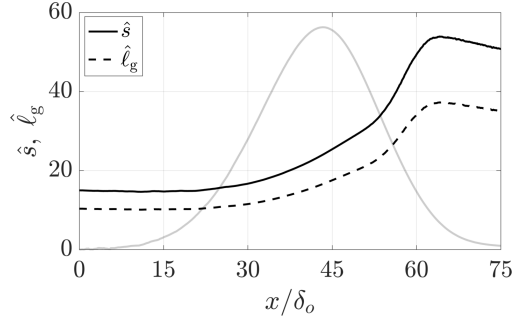

Figure 14: Riblet spacing ($\hat{s}$, ——) and square root of groove cross-sectional area ($\hat{\ell}_g$, – – –) normalized by the viscous length scale calculated with the total shear stress in the plane of the riblet crest.

These results show that the elevated drag generated near the riblet crests is largely confined below the crest plane such that it is not directly communicated to the outer flow. Instead, the boundary layer above behaves as if it were interacting with a partial-slip surface characterized by a total shear of $\hat{\tau}$, and a mean velocity of $U_c$. While this inference is drawn here from the mean-flow and stress distributions, it will be further examined using higher-order turbulence statistics in a later section.

A direct hypothesis from the groove-flow/TBL separation is that the characteristic size of the near-wall turbulent structures above the riblets is expected to remain governed by this effective friction scale, rather than by a viscous scale based on the total drag. Accordingly, although the riblets may generate additional drag within the grooves, they may still appear hydraulically drag-reducing to the overlying turbulent boundary layer. Indeed, when $s$ and $\ell_g$ are normalized by the effective viscous length scale $\nu/\hat{u}_\tau$, they remain in the conventional drag-reducing range throughout much of the FPG region, even as the total drag increases drastically (figure 14). Specifically, $s\hat{u}_\tau/\nu$ exceeds 30 and $\ell_g\hat{u}_\tau/\nu$ exceeds 20 at $x/\delta_o \approx 50$. Such a location is where the increasing trend of $\hat{\tau}$ deviates significantly from that of the smooth shear stress, retransition occurs, and turbulent and dispersive stresses are present in the groove; thus, the separation between the groove drag and the overlying TBL no longer persists. The applicability of the conventional drag-reduction criteria when $\hat{\tau}$ is used for scaling, together with the turbulence statistics presented in the next section, supports the hypothesis that, prior to retransition, the riblets modulate the overlying TBL at the crest plane in a manner broadly similar to that in ZPG flows, despite the augmented drag within the groove.

One way of characterizing a partial-slip surface is through a Robin-type slip condition. At the riblet crest plane, it can be written as

$$U_c - \lambda \left( \frac{\partial \langle \overline{u} \rangle}{\partial y} \right)_{y=h} = 0 \tag{5.4}$$

where $\lambda$ is the characteristic slip length. $\lambda = 0$ represents the no-slip limit, and $\lambda \to \infty$ corresponds to a free-slip condition. Figure 15 shows the profile of $\lambda$, normalized either by the effective viscous length scale or the riblet size. Evidently, $\lambda$ increases in response to the acceleration. Rather than indicating improved riblet efficiency, the rising slip length indicates that $U_c$ responds more rapidly to the acceleration than $\partial \langle \overline{u} \rangle / \partial y|_{y=h}$.

The slip length given in Eqn. (5.4) is often referred to as the streamwise, or longitudinal, protrusion height in ZPG riblet literature (Luchini *et al.* 1991; Luchini 1996; Endrikat *et al.* 2021*a*; Modesti *et al.* 2021). In conjunction with the transverse protrusion height, it is used to determine the log-law offset and corresponding drag modulation imposed by riblets. The ratio of it to riblet size has been claimed to be fixed for given riblet shapes in ZPG flows (Luchini *et al.* 1991). Figure 15 shows that $\lambda$ in case RB is initially constant in the ZPG and mild FPG regions, though from $x/\delta_o \approx 27$ it changes with the accelerating and later retransitioning flow, exhibiting flow dependence.

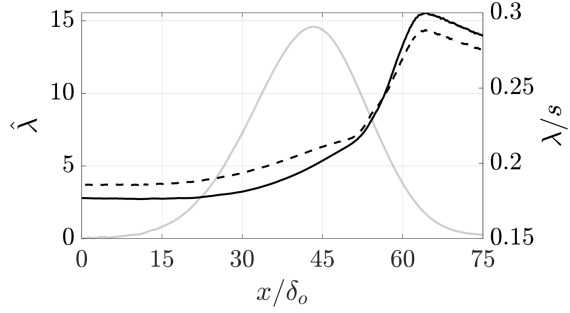

Figure 15: Slip length obtained by fitting the mean velocity and its gradient at the riblet crest to a Robin-type slip condition (Eqn. 5.4). The solid line (left axis) shows the slip length normalized by the crest-shear length scale ($\hat{\lambda} = \lambda \hat{u}_\tau / \nu$). The dashed line (right axis) shows the slip length normalized by the riblet spacing.

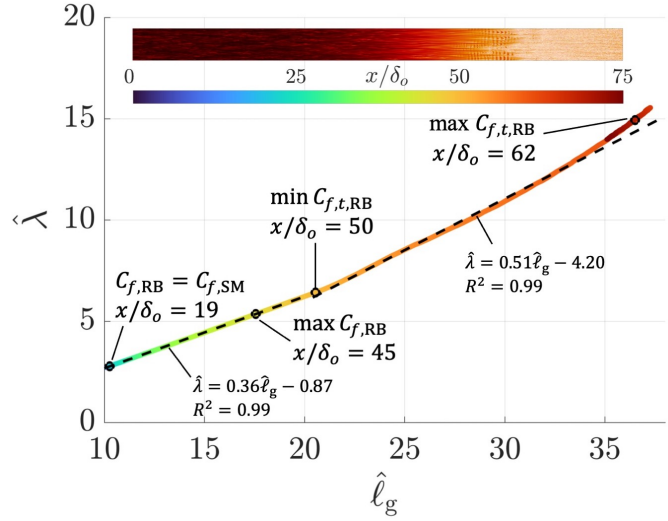

Figure 16: Slip length ($\hat{\lambda}$) as a function of $\hat{\ell}_g$. The color gradient from blue to red indicates the downstream direction, as shown by the legend. Dots mark selected streamwise locations of interest. The dashed lines represent linear fits from Eqn. 5.5 obtained in the regions $x/\delta_o \in [0,\ 50]$ and $x/\delta_o \in (50,\ 75]$.

Figure 16 presents the variation of slip length as a function of the groove-area length scale, normalized by the effective viscous length scale at riblet crest plane $\nu/\hat{u}_\tau$. The data collapse onto a well-defined curve that, except in the late retransition region, exhibits two linear regimes with distinct slopes. A switch between these two linear regimes occurs around $x/\delta_o \approx 50$, coinciding with the streamwise location where the qualitative change in the flow behavior from relaminarizing to retransition is observed.

Within each linear region, the relation between $\hat{\lambda}$ and $\hat{\ell}_g$ can be expressed as

$$\frac{\lambda(x)}{L^*(x)} = a\,\frac{\ell_g}{L^*(x)} + b, \tag{5.5}$$

where $a$ and $b$ are coefficients. It can be equivalently rewritten as

$$\lambda(x) = a\,\ell_g + b\,L^*(x). \tag{5.6}$$

Here, we use $L^*$ to denote a characteristic flow length scale in general, while in the plot, $L^* = \nu/\hat{u}_\tau$ is used. Eqn. (5.6) implies that, within a given linear region, the slip length can be decomposed into a geometry-dependent contribution proportional to the groove-area length $\ell_g$, and a flow-dependent contribution proportional to the local flow length scale $L^*$.

The existence of two distinct linear regions in our results indicates that the coefficients $a$ and $b$ are not universal, but depend on the flow regime. Since the riblet geometry and size remain unchanged throughout the domain, the observed variation in these coefficients must arise from changes in the underlying flow physics.

It is worth noting that, in previous studies of riblet flows under ZPG conditions, a linear relation between the slip length and the geometric scale has often been interpreted as evidence of a purely geometry-controlled regime, typically associated with Stokes-like flow behaviour (Luchini *et al.* 1991; Endrikat *et al.* 2021a). That is, $b$=0. In the present case, although linear (and piecewise linear) relations are also observed, the presence of a non-zero and negative coefficient $b$ indicates a persistent dependence on the flow scale $L^*$ in this flow. This demonstrates that linearity alone does not imply a purely geometry-dominated regime. Instead, the results here suggest that both geometric and flow effects remain active, with their sensitivities varying across different flow regions. Specifically, the first linear region—extending from the reference ZPG region to $x/\delta_o = 50$—demonstrates that the current riblet configuration, even when yielding drag reduction under ZPG, exhibits a slip length scale that remains dependent on the flow scale ($b$=-0.87). In other words, the pure Stokes scenario, corresponding to a complete decoupling of the viscous groove flow from the overlying turbulence and thus $b = 0$, represents an idealization rather than a physically accurate explanation for drag-reducing riblets in ZPG flows. Therefore, the observed linear scaling should be interpreted as an affine decomposition involving both geometry and flow contributions, rather than as evidence of a purely viscous or Stokes flow. It should be noted that the first linear region identified here need not remain valid below the lower bound of $\ell_g$ showed in this study, nor does it imply a non-zero $\lambda(x)$ at $\ell_g$=0.

A distinction should be made between the coefficients and the contributions themselves. The coefficients $a$ and $b$ quantify the sensitivity of the slip length to geometric and flow scales, respectively, while the corresponding contributions are $a\ell_g$ and $bL^*$. Both coefficients may depend on the flow, reflecting how the effect of the geometry and chosen flow scale varies with local conditions. In both linear regions observed here, $a > 0$, indicating that increasing $\ell_g$ enhances the slip length, as expected; while $b < 0$ indicating that an increase in the effective flow scale $L^*$ (i.e., a decrease in $\hat{u}_\tau$ here) reduces it which is also expected since decreasing $\hat{u}_\tau$ means less shear in Eqn. (5.4).

Furthermore, the increase in both $a$ and $|b|$ when switching to the second linear region indicates that the sensitivities to both geometry and flow are enhanced during retransition. In particular, the more positive coefficient $a$ implies that the slip length increases more strongly with the geometric scale $\ell_g$, while the more negative value of $b$ indicates a stronger dependence on the effective flow scale $L^*$, such that increases in the mean velocity at the riblet crest plane lead to a larger inferred slip.

Finally, and most importantly, the fact that the data collapse onto a piecewise linear relation when normalized by $L^* = \nu/\hat{u}_\tau$, together with the observation that the switch between the two linear regions coincides with the onset of flow retransition, suggests that this quantity serves as a dynamically relevant flow length scale governing the slip behaviour. Under this scaling, the problem is effectively reduced to the interaction between the geometric scale $\ell_g$ and a single flow scale $\nu/\hat{u}_\tau$, providing a reduced-order description of the complex near-wall flow dynamics.

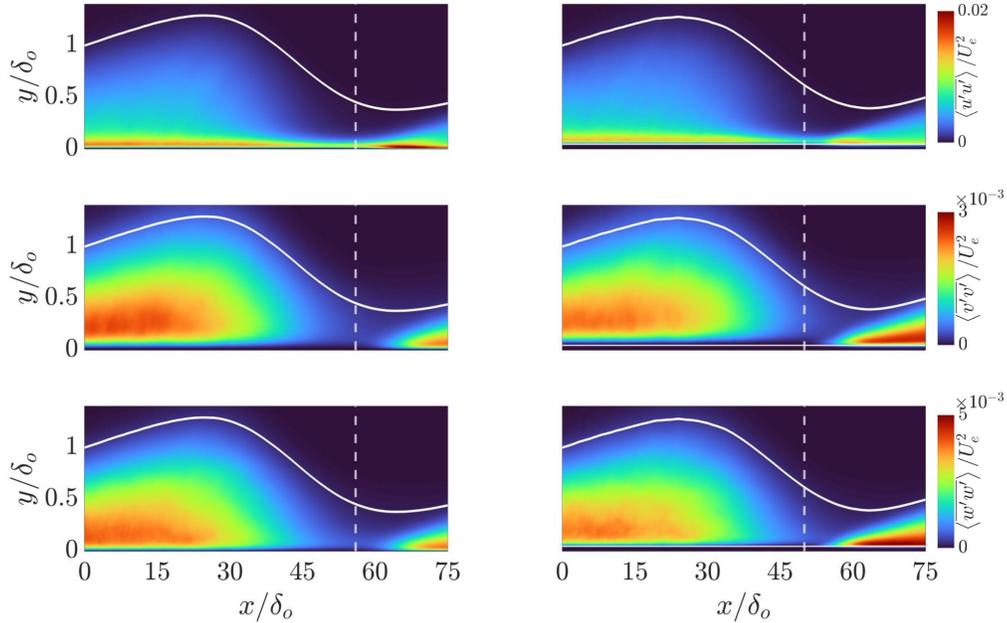

Figure 17: Contours of the mean Reynolds normal stresses. Left: case SM; right: case RB. In all plots, the thick white line represents $\delta/\delta_o$, while the thin horizontal white line in the right column shows the riblet crest. The vertical dashed line is located at $x/\delta_o = 56$ and 50 in cases SM and RB, respectively, corresponding to the streamwise position of the $C_{f,t,\min}$ and retransition onset in each case. The $y$ axis is stretched for clarity.

## 6. Statistical and structural changes of turbulence: relaminarization

Discussions in §4 show that the riblets used in this study have little influence on the TBL during the relaminarization process, while §5 reveals the associated viscous-dominated drag-penalty mechanism prior to retransition. In contrast, the promotion of earlier and more energetic retransition significantly alters both the boundary layer and the wall shear stress. The associated coherent structures and their statistical signatures are therefore essential for distinguishing the dynamics of relaminarization from those of retransition.

Figure 17 shows the mean Reynolds normal stresses, normalized by the local $U_e^2$, in order to highlight their relative streamwise evolution. In both cases, two high-turbulence regions are separated by an intermediate low-stress zone that clearly marks the strongly relaminarized part of the flow. Within the high-TKE incoming TBL, turbulence in the outer (wake) region appears to respond earlier to the acceleration than that in the near-wall region, exhibiting an earlier decrease, but diminishing more gradually than in the inner region. In the FPG literature, this behaviour is termed 'frozen turbulence' to distinguish it from the actively sustained near-wall turbulence (Narasimha & Sreenivasan 1973; Sreenivasan 1982; Bourassa & Thomas 2009; Piomelli & Yuan 2013; Falcone & He 2022). Comparing the three Reynolds normal stresses, $\langle \overline{u'u'} \rangle$ retains a noticeable magnitude near the wall even during strong relaminarization, whereas $\langle \overline{v'v'} \rangle$ and $\langle \overline{w'w'} \rangle$ are reduced to nearly zero. This is consistent with structural observations reported in the literature, where near-wall streaks are stretched and elongated under acceleration (Warnack & Fernholz 1998; Piomelli *et al.* 2000; Bourassa & Thomas 2009; Falcone & He 2022). In the second high-TKE region after the flow retransitions, all three stresses rapidly grow in magnitude. A notable difference between the smooth and riblet cases is that, while $\overline{\langle u'u' \rangle}$ during retransition is lower in magnitude over riblets than the smooth wall, the $\overline{\langle v'v' \rangle}$ and $\overline{\langle w'w' \rangle}$ are considerably augmented. This indicates that the two cases undergo retransition through different mechanisms. This section further characterizes the relaminarization process, with the retransition region examined in §7.

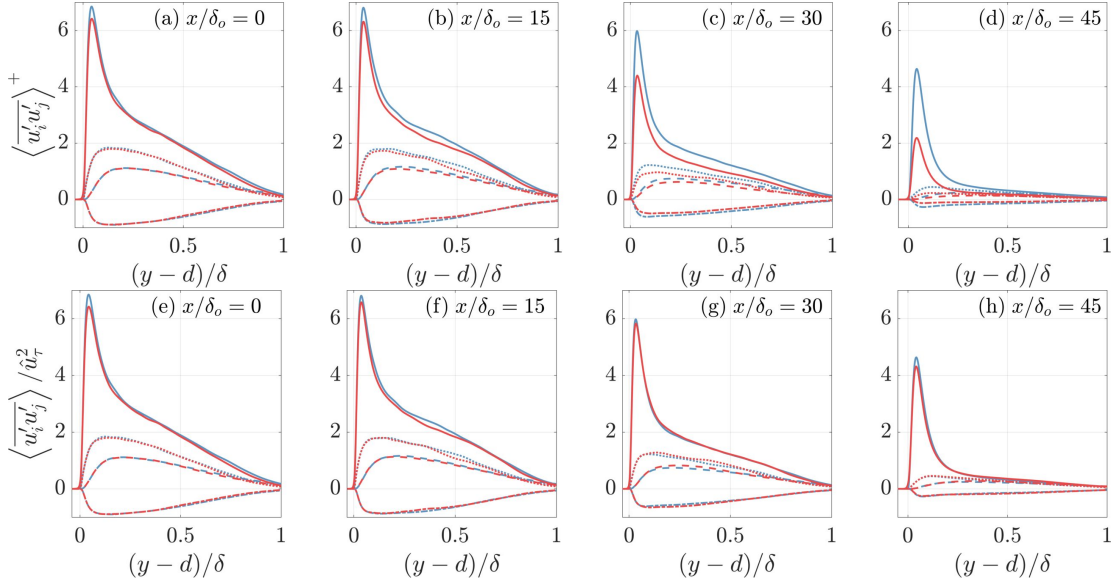

Figure 18: Wall-normal profiles of Reynolds stresses: —— $\langle u'u' \rangle$; – – – $\langle v'v' \rangle$; ⋯⋯ $\langle w'w' \rangle$; –·–· $\langle u'v' \rangle$. Blue lines are Case SM, red lines are case RB. In the top row, stresses are normalized by local $u_\tau^2$. In the bottom row, Case RB stresses are normalized by local $\hat{u}_\tau^2 = \hat{\tau}$ while Case SM stresses are normalized by local $u_\tau^2$.

### 6.1. *Reynolds stresses and turbulent structures during relaminarization*

Figure 18 presents Reynolds stress profiles extracted at selected streamwise locations up to $x/\delta_o = 45$. The two cases exhibit similar stress distributions at $x/\delta_o = 0$, except for a slight reduction in $\overline{\langle u'u' \rangle}^+$. This is consistent with findings in the literature that drag-reducing riblets do not substantially modify the turbulence structure, but instead primarily act to lift near-wall vortices (Choi *et al.* 1993; Suzuki & Kasagi 1994; Goldstein *et al.* 1995).

Moving downstream, relaminarization is evident in both cases through the reduction of all stress components, with a more pronounced decrease over riblets than over the smooth wall when normalized by the local $u_\tau^2$ (i.e., in viscous units accounting for the total drag; top row). However, as discussed earlier, the additional drag generated beneath the crest plane is not directly transmitted to the overlying TBL. Accordingly, when the stresses are normalized by $\hat{u}_\tau^2$ in the bottom row (see definition in Eqn. 5.3 and figure 13), the riblet and smooth-wall cases collapse at each streamwise location. This collapse shows that, relative to the effective stress experienced by the TBL, the two cases undergo the same turbulence attenuation during relaminarization.

This result shows that the turbulence dynamics are governed by the mean shear at the crest plane, $\hat{\tau}$, rather than the total drag, $\tau_{\mathrm{w,RB}}$. The stronger apparent attenuation of Reynolds stresses when scaled with the friction velocity based on the total drag is therefore not physical, but an artifact of an inappropriate viscous scaling. When scaled using the effective shear, the turbulence evolution remains consistent between the two cases.

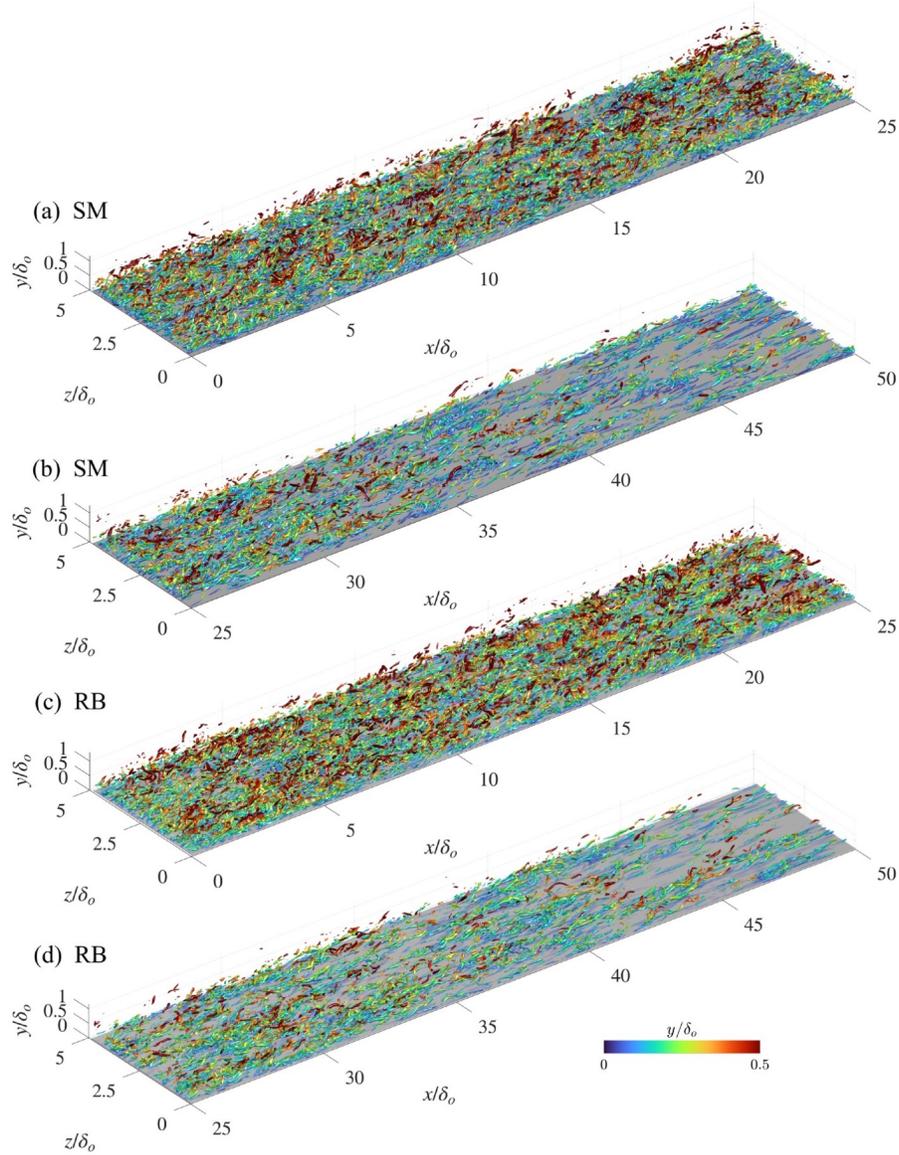

Figure 19: Turbulent structures in the accelerating boundary layer shown by isosurfaces of the second invariant of the velocity gradient tensor. Only the relaminarization region is shown. Isosurfaces are visualized at a level of $Q = 2.3U_{e,o}^2/\delta_o^2$ and are colored by distance from the wall or riblet crest. (a,b): case SM; (c,d) case RB. The wall or riblet surface is visualized in gray.

The underlying turbulent structures are likewise very similar in the two cases, as confirmed by the instantaneous fields and higher-order statistics. Figure 19 visualizes isosurfaces of the second invariant of the velocity gradient tensor ($Q = -\frac{1}{2}(\partial u_j/\partial x_i)(\partial u_i/\partial x_j)$). In both cases, the structural response to the imposed acceleration is qualitatively similar, with reduced small-scale turbulence and pronounced streamwise elongation of the streaks for $x/\delta_o \gtrsim 30$. This spatial evolution is consistent with previous instantaneous visualizations of FPG boundary layers (Piomelli *et al.* 2000; De Prisco *et al.* 2007; Piomelli & Yuan 2013; Falcone & He 2022).

Figure 20 presents the joint probability density functions (JPDFs) of the streamwise and wall-normal velocity fluctuations at selected streamwise locations. The riblet (lines) and smooth-wall (filled contours) cases are nearly identical. At the wall-normal location corresponding to the peak of $\langle u'u' \rangle$ (top row), where the legs of hairpin vortices and the associated streaks reside, both cases exhibit a broadening of the $u'/U_{e,o}$ distribution together with a reduced $v'/U_{e,o}$ component under the FPG. This is consistent with interpretations of FPG-driven relaminarization (Warnack & Fernholz 1998; McEligot & Eckelmann 2006; Bourassa & Thomas 2009; Piomelli *et al.* 2000), where reduced Q2 and Q4 activity reflects elongated, more stable streaks and suppressed bursting, ultimately weakening vertical momentum transport and the turbulence regeneration cycle.

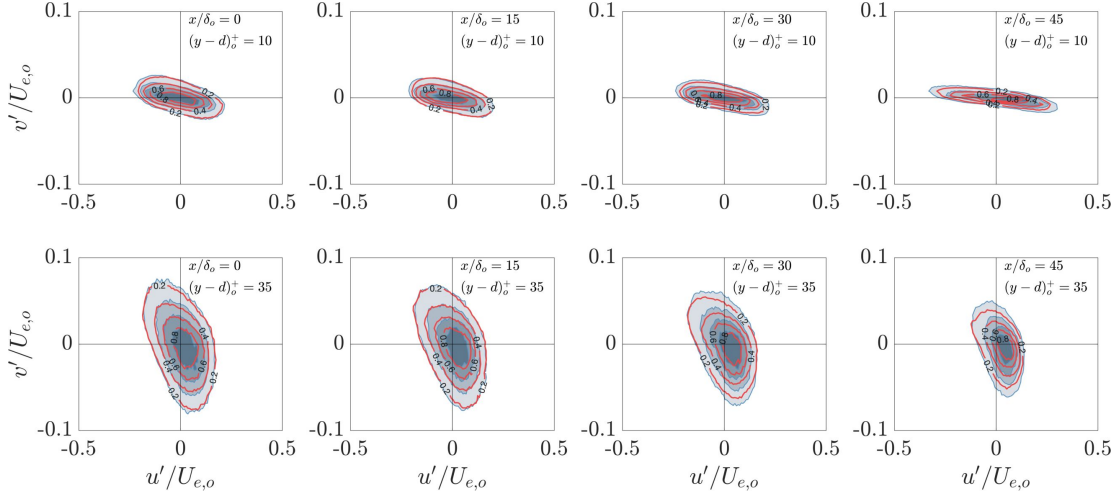

Figure 20: Joint probability density function (JPDF) of the streamwise ($u'/U_{e,o}$) and wall-normal ($v'/U_{e,o}$) velocity fluctuations at selected streamwise locations in the relaminarization region. Top row: at $(y - d)_o^+ = 10$ or $(y - d)/\delta_o = 0.03$ (wall-normal height of the near-wall streaks). Bottom row: at $(y - d)_o^+ = 35$ or $(y - d)/\delta_o = 0.12$. Each JPDF is normalized by its maximum value and shown at 0.2, 0.4, 0.6, and 0.8. The filled contours show case SM, and the red isocontour lines show case RB.

Further from the wall at $(y - d)_o^+ = 35$ $((y - d)/\delta_o = 0.12)$, the JPDF retains much of its upstream shape and is therefore less responsive to the FPG. However, high-amplitude $u'/U_{e,o}$ and $v'/U_{e,o}$ events gradually diminish, indicating slowly decaying, or "frozen," turbulence with a structure similar to the upstream ZPG TBL (Piomelli & Yuan 2013; Falcone & He 2022). As in the near-wall region, the two cases exhibit nearly identical behavior, reinforcing that riblets have minimal influence on turbulence structure across the boundary layer during relaminarization.

Taken together, the close agreement of turbulence statistics and structures across all diagnostics demonstrates that the increase in drag in case RB is decoupled from boundary-layer dynamics. The overlying TBL responds to the effective shear at the groove crest, which acts as a partial-slip plane, rather than to the total drag generated within the grooves. When scaled with this effective shear, the turbulence evolution under FPG is identical for the smooth and riblet cases, supporting the validity of outer-layer similarity.

### 6.2. *Dispersive stress: difference between riblets and heterogeneous roughness*

The riblets do not modulate turbulence during relaminarization because they are aligned with both the bulk-flow direction and the direction of acceleration. In the absence of streamwise obstruction, the riblet geometry does not directly generate form drag, local flow separation, or recirculation. In contrast, more isotropic surface roughness has been shown to significantly alter turbulence in an FPG TBL (Yuan & Piomelli 2015). In that study, sandgrain roughness operating in the transitionally rough regime at the reference location had a characteristic height comparable to the present riblets ($k_o^+ \approx 20$ versus $h_o^+ = 15$ here). Under the influence of the FPG, the skin-friction coefficient, turbulent stresses, and turbulence isotropy increased, rather than exhibiting relaminarization trends. In other words, relaminarization was hindered by such roughness. They further showed that dispersive stresses ($\langle \tilde{u}_i \tilde{u}_j \rangle$), arising from wakes behind roughness elements, increased more rapidly than the friction velocity. The sustained wall-normal and spanwise components of these dispersive stresses generated by the sandgrain roughness enhanced turbulence relative to the smooth-wall case through pressure work and the conversion of wake kinetic energy into turbulence kinetic energy. This process ultimately maintained the turbulence production cycle during the acceleration (Piomelli & Yuan 2013; Yuan & Piomelli 2015).

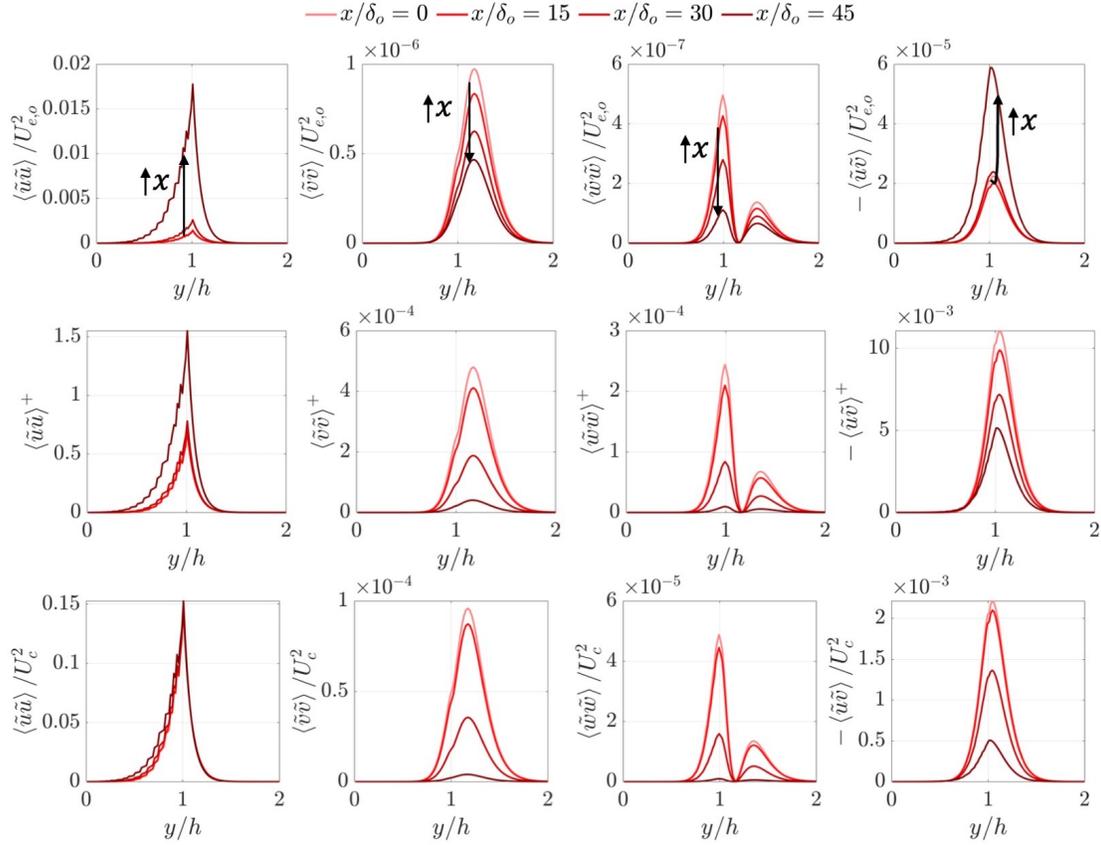

Figure 21: Dispersive stresses normalized by $U_{e,o}^2$ (top row), $u_\tau^2$ (middle row), and $U_c^2$ (bottom row). Darkening shade of red indicates the downstream direction.

Given their role in roughness-induced modulation of FPG turbulence, the dispersive stresses over riblets are also examined here, even though Figure 11 already indicates that the groove-flow pattern remains largely unchanged relative to the overlying TBL. Figure 21 presents the dispersive stresses throughout the relaminarization region for case RB, normalized using several velocity scales. The top row shows the absolute values normalized by the constant $U_{e,o}^2$, the middle row uses the local friction velocity associated with the total drag, and the bottom row employs the local mean velocity at the crest plane, $U_c^2$. Notably, the streamwise component is at least three orders of magnitude larger than the wall-normal and spanwise components, reflecting the suppression of transverse motion by the riblets. In contrast, the three components attain comparable magnitudes for the sandgrain roughness case of Yuan & Piomelli (2015). The reduction in transverse dispersive stresses during the FPG is consistent with Piomelli & Yuan (2013) and Yuan & Piomelli (2015), who showed that reductions in wake stresses promote relaminarization.

Although the qualitative trends with respect to $x$ are consistent across all three normalizations, scaling with $U_c$ yields the best collapse of the dominant streamwise dispersive stress. This collapse is consistent with the observation that, in case RB, the groove-flow pattern remains nearly unchanged throughout the domain. Since the mean velocity at the groove opening provides the effective driving scale for the groove flow, and turbulence does not significantly penetrate into the grooves before retransition, the mean flow within the grooves remains primarily geometry-controlled and therefore exhibits approximate self-similarity in the streamwise direction. The other dispersive stress components, by contrast, are much weaker and decrease as relaminarization proceeds. Their peak locations remain tied to the wall-normal region at or just above the riblet crest, again indicating the dominant role of the riblet geometry in setting their structure.

### 6.3. *Elongation of instantaneous groove flow under FPG*

The observed suppressed transverse motion and enhanced anisotropy indicate increasing alignment of the flow in the streamwise direction within the riblet grooves. Although the flow pattern within the riblet grooves is often described as a secondary flow, the dispersive velocity is a consequence of spatial inhomogeneity in the time-averaged flow and does not correspond to a realizable instantaneous flow structure. To characterize the instantaneous groove dynamics, we therefore examine the deviation of the instantaneous field from the local time-averaged velocity, focusing on its response to the imposed FPG.

The instantaneous streamwise velocity fluctuations ($u'/U_{e,o}$) are displayed in figure 22 for case RB at selected streamwise and wall-normal locations. Relative to the quasi-ZPG region (figure 22a,c), the near-wall flow becomes more streamwise-coherent, while the streaky structures exhibit markedly reduced spanwise undulation under the strong FPG (figure 22b,d). This trend is observed at both wall-normal locations and is consistent with the well-established stretching and stabilization of near-wall streaks under freestream acceleration (Warnack & Fernholz 1998; Bourassa & Thomas 2009; Piomelli & Yuan 2013; Falcone & He 2022). Under strong FPG, the flow at the riblet crest plane exhibits greater streamwise coherence and weaker spanwise waviness than the flow at the streak height, showing that the groove flow becomes even more streamwise-elongated than the overlying streaks.

The streamwise coherence is quantified with the streamwise auto-correlation of the fluctuating velocity, which reads:

$$R_{u'u'}(x, y, \Delta x) = \frac{\left\langle \overline{u'(x, y, z)u'(x + \Delta x, y, z)} \right\rangle}{\left\langle \overline{u'u'} \right\rangle (x, y)}. \tag{6.1}$$

From this, an integral streamwise length scale can be defined to measure the spatial persistence of the coherence:

$$L_{u'u'}(x, y) = \int_0^{\Delta x_0} R_{u'u'}(x, y, \Delta x) d\Delta x, \tag{6.2}$$

where $\Delta x_0$ is the first zero-crossing of the auto-correlation coefficient. The auto-correlations and coherence length scale are shown in figure 23(a) and (b). The auto-correlations widen due to the FPG, and correspondingly $L_{u'u'}$ increases with $x$, following a trend similar to the streamwise integral length scale reported by Warnack & Fernholz (1998). The growth of $L_{u'u'}$ appears to be associated with the strength of the FPG and is most pronounced at the riblet crest, where $L_{u'u'}$ increases by approximately a factor of three by peak acceleration. At positions slightly above the crest (e.g., $(y - d)/\delta_o = 0.03$ and $0.05$), this growth is attenuated, proving that the coherence within the grooves exceeds that in the flow above the crest as visualized in figure 22. These two locations are around 10 and 15 viscous units (based on $\hat{\tau}$) above the riblet crests and are associated with the typical position of hairpin legs. While it is well established that hairpin structures are elongated in FPG flows, this result shows that the streamwise uniformity imposed by the riblets further organizes the flow near the riblet crest plane during acceleration.

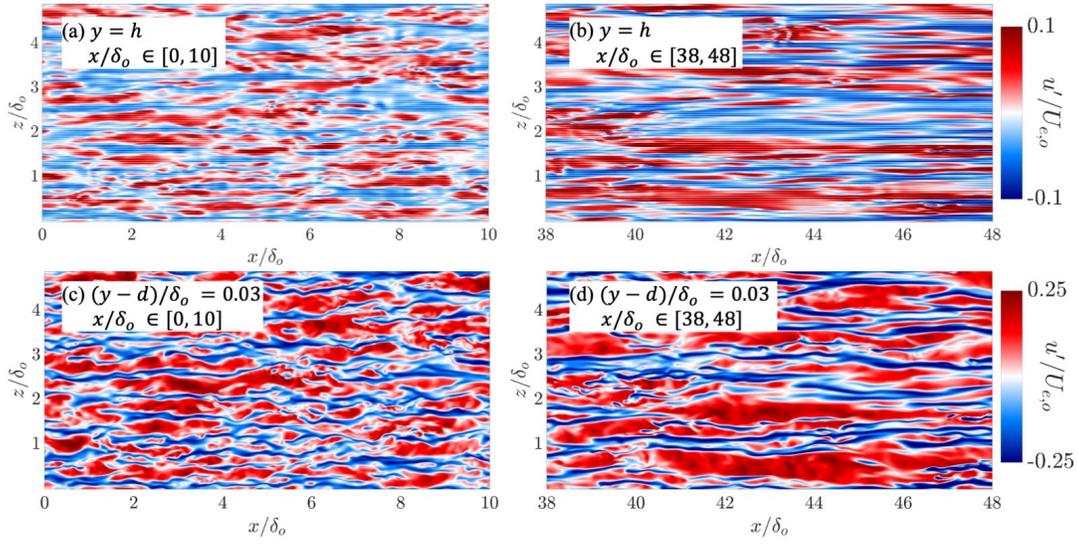

Figure 22: $x - z$ planes of instantaneous streamwise velocity fluctuations ($u'/U_{e,o}$) at (a,b) the riblet crest ($y = h$), and (c,d) the height of the near-wall streaks (($y - d)/\delta_o = 0.03$). (a,c) shows the ZPG region, while (b,d) shows the quasi-laminar high acceleration region.

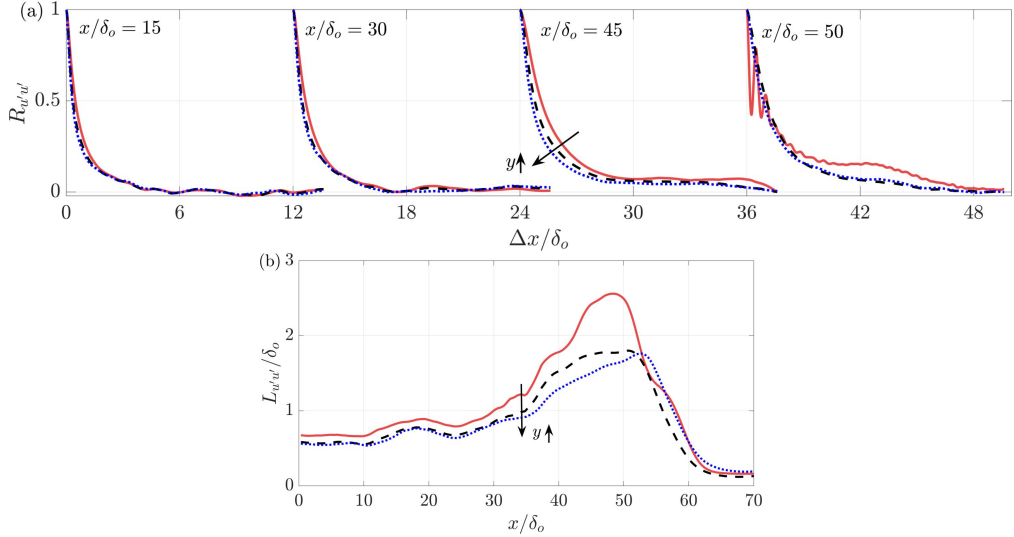

Figure 23: Profiles of (a): Streamwise auto-correlation of the streamwise velocity fluctuations at various $x$ and $y$ locations. Wall-normal position is indicated by line style: —— $y = h$; – – –: ($y - d)/\delta_o = 0.03$; ⋯⋯ ($y - d)/\delta_o = 0.05$. Profiles are shifted by 12 units for clarity. (b) The streamwise integral length scale at the three wall-normal locations where the correlations were measured.

In addition to providing insight into the relaminarization process, the auto-correlations and the streamwise integral length scale offer a statistical characterization of the subsequent retransition. At $x/\delta_o = 50$, the near-riblet auto-correlation develops distinct secondary peaks at relatively small streamwise separations ($\Delta x/\delta_o \approx 0.4$, and its harmonics). These features have been verified to be robust and not attributable to statistical noise. As discussed in the following sections, these peaks correspond to the characteristic wavelength of riblet-induced spanwise roller vortices that emerge prior to retransition. Beginning near this streamwise location, the streamwise integral length scale starts to decrease, indicating a breakdown of the previously elongated structures and the emergence of smaller-scale turbulence within the riblet grooves. Notably, this decrease is first observed near the riblet crest, suggesting that retransition in case RB initiates in the vicinity of the riblet grooves.

## 7. Statistical and structural changes of turbulence: retransition

Figure 24 displays the instantaneous flow field in the retransition region. Regions of intense, small-scale turbulence are identified with isosurfaces of $Q/(U_{e,o}/\delta_o)^2 = 20$, colored by wall-normal distance. Contours of the instantaneous wall-normal fluctuations ($v'/U_{e,o}$) in $x-z$ planes at $y/\delta_o = 0.01$ and $y = h$ in case SM and RB, respectively, are also shown. In both cases, residual turbulence is embedded within the incoming, still-accelerating boundary layer at the onset of retransition. These structures are shown in figure 19, but are not visible here because the higher $Q$ threshold used for the isosurfaces emphasizes turbulent spots during retransition. Nevertheless, their signatures remain evident in the instantaneous wall-normal fluctuations in the wall-parallel planes.

A clear difference between the two cases is the streamwise location at which clustered small-scale eddies appear. At the time instant shown, this occurs at $x/\delta_o \approx 62$ in case SM, whereas in case RB such structures are already evident by $x/\delta_o \approx 52$. The increasing turbulent activity coincides with the onset of an increase in the turbulent contribution to $C_f$ (figure 6f), a sharp rise in the drag penalty (figure 8), and a departure of the total shear stress at the riblet crest from the smooth-wall trend (figure 13). This indicates that the observed turbulent structures are closely linked to both drag modulation and the flow dynamics at the riblet scale. Despite the difference in retransition location, the intermittent turbulent patches exhibit similar growth and merging dynamics. By $x/\delta_o \approx 70$ and 60 for case SM and RB, respectively, turbulence has spread across most of the span.

However, the mechanisms responsible for retransition differ fundamentally between the two cases. In case SM, retransition proceeds through a classical bypass-like pathway, in which turbulent spots originate from the breakdown of streamwise-elongated near-wall streaks. Two canonical instability modes are observed in figure 24(a,b): varicose (spanwise-symmetric) and sinuous (spanwise-asymmetric) streak instabilities (Andersson *et al.* 2001; Brandt & Henningson 2002; Asai *et al.* 2002; Brandt *et al.* 2004), both leading to the generation of small-scale turbulence. These features closely resemble bypass transition in ZPG boundary layers (Jacobs & Durbin 2001; Wu & Moin 2009; Schlatter *et al.* 2008), and are consistent with previous observations in FPG retransition (De Prisco *et al.* 2007; Piomelli & Yuan 2013; Falcone & He 2022). The evolution of these breakdown mechanisms is investigated in more detail in §7.1.

In contrast, case RB exhibits a distinct riblet-induced retransition pathway involving spanwise-coherent structures generated near the riblet crests; such behaviour has not been reported for FPG turbulent boundary layers over sandgrain roughness (Yuan & Piomelli 2015). As shown in figure 24, the near-wall flow is dominated by the presence of spanwise-coherent regions of alternating positive and negative $v'/U_{e,o}$, indicative of spanwise-oriented KH roller vortices similar to observations in ZPG riblet flows (García-Mayoral & Jiménez 2011a,b; Endrikat *et al.* 2021a,b; Rouhi *et al.* 2022; Abu Rowin *et al.* 2025; Camobreco *et al.* 2025). The formation of small-scale turbulent spots is associated with the deformation and lifting of these rollers, as shown in figure 24c. As the rollers evolve, they undergo arched deformation and lifting, while streamwise-oriented vortices form near the riblet crests. Further downstream, the arched rollers develop into hairpin-like vortices, as illustrated in figure 24d, leading to rapid breakdown into turbulence.

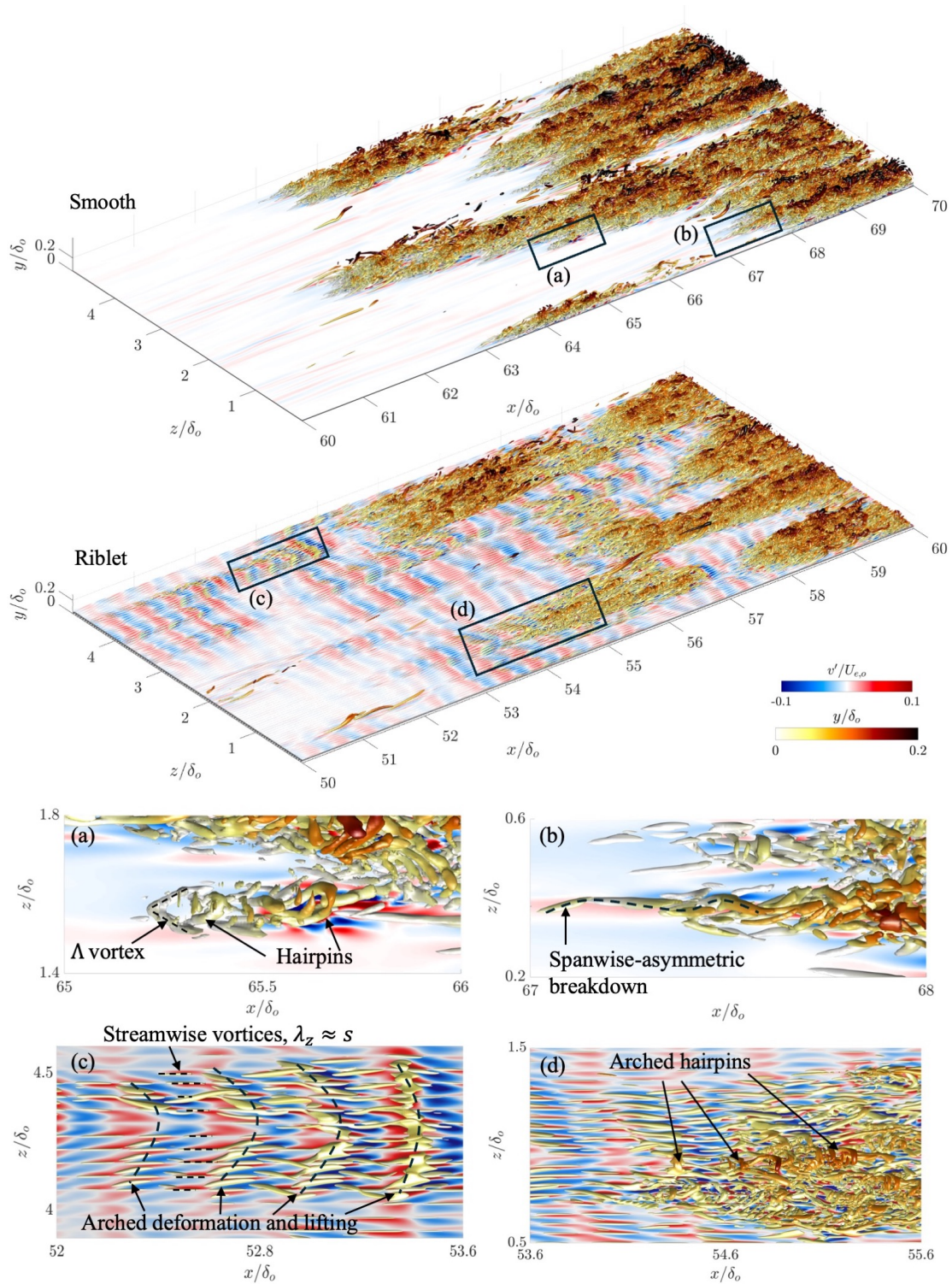

Figure 24: Instantaneous flow visualization in the retransition region. Three-dimensional isosurfaces of $Q/(U_{e,o}/\delta_o)^2 = 20$ (see text for definition) are colored by distance to the $y = 0$, while $x - z$ planes display the instantaneous wall-normal velocity fluctuation at $y/\delta_o = 0.01$ and $y = h$ for the smooth and riblet case, respectively. The regions of interest marked (a), (b), (c), and (d) are displayed in greater detail in the four subplots.

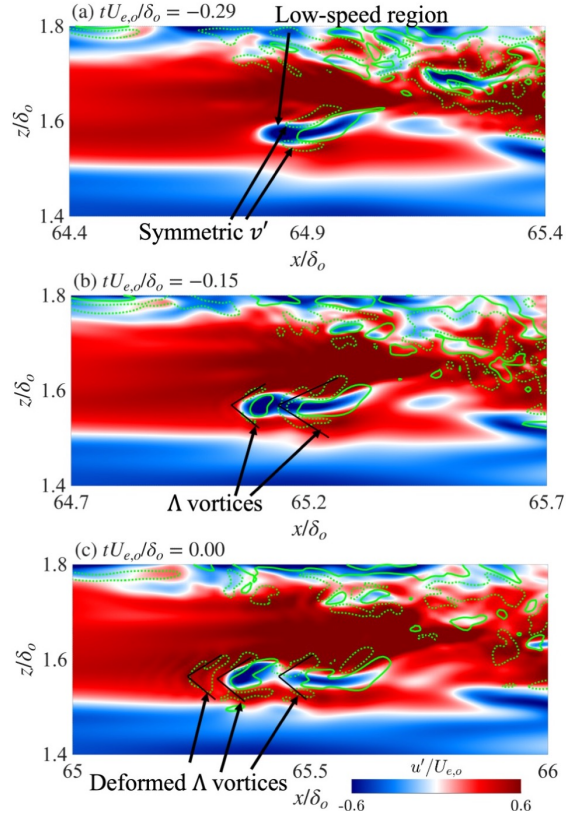

Figure 25: Evolution of a varicose instability event leading to a turbulent spot corresponding to figure 24(a). The contour shows $u'/U_{e,o}$, the green iso-contour lines show $v'/U_{e,o} = \pm 0.1$ (dashed: negative, solid: positive).

In case RB, turbulent spot formation is driven primarily by the breakdown of spanwise roller vortices, even though streak-instability pathways analogous to those in case SM cannot be ruled out completely. Closer inspection reveals weak streamwise streak signatures superposed on the spanwise rollers. However, the flow in case RB becomes fully turbulent by $x/\delta_o \approx 60$, whereas the streak-breakdown mechanisms are only beginning to emerge in the smooth case. This spatial discrepancy shows that the roller-breakdown mechanism injects small-scale perturbations that disrupt the quasi-laminar streaks and trigger retransition before the natural streak instability becomes dominant. Sections 7.2 and 7.3 will further quantify this process.

### 7.1. *Evolution of turbulent spots: over smooth wall*

Examples of varicose (spanwise-symmetric) instability during the smooth-wall retransition are shown in figure 25, while figure 26 shows sinuous streak instabilities. For both, the residual near-wall streaks in the boundary layer play a critical role. The incoming streaks are visualized using $u'/U_{e,o}$ in $x - z$ planes at $y/\delta_o = 0.03$. The growth of small-scale perturbations is highlighted by isocontours $v'/U_{e,o} = \pm 0.1$. In figure 26, the in-plane flow is indicated by the superposed in-plane fluctuating velocity vectors. The non-dimensional time shown in each frame is relative to the instant in figure 24. Although a quantitative assessment of the occurrence and lifetime of each instability event is difficult, we examined approximately 1000 instantaneous planes and selected representative examples to illustrate their common evolution. The sinuous breakdown is found to be the dominant mechanism for turbulent spot generation in the present configuration, accounting for the majority of events (approximately 95%, based on a bulk estimate) at the onset of turbulence in the smooth-wall case. This observation is consistent with previous studies that identified sinuous instability as the primary transition mechanism in ZPG boundary layers (Andersson *et al.* 2001; Brandt *et al.* 2004; Schlatter *et al.* 2008), and with Falcone & He (2022), who reported similar behaviour in FPG retransition.

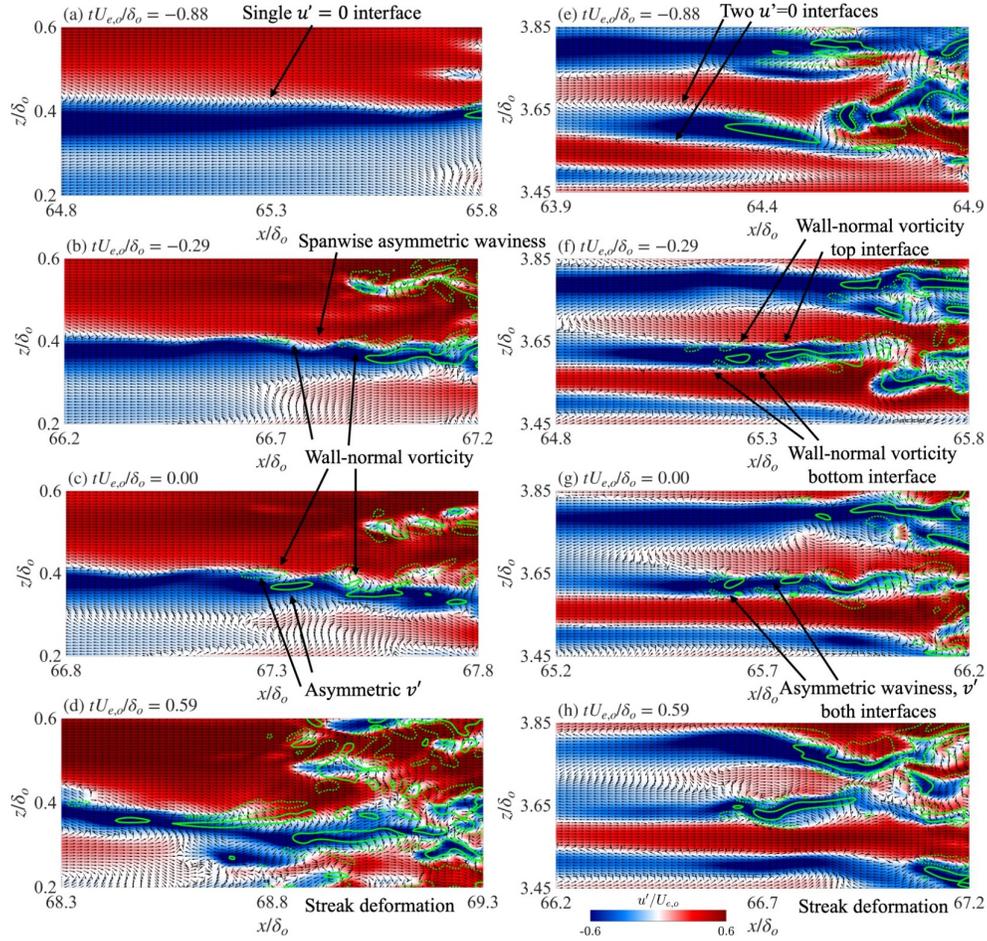

Figure 26: Left column: evolution of a one-sided sinuous instability corresponding to figure 24(b). Right column: evolution of a two-sided sinuous instability. The contours show $u'/U_{e,o}$, while the green contour lines show $v'/U_{e,o} = \pm 0.1$ (dashed: negative, solid: positive). The vectors show the in-plane velocity vector normalized by its local magnitude, and are shown at every second grid point. The time instance each plane is extracted, relative to the time instant shown in figure 24, is provided.

For the varicose (spanwise-symmetric) instability, which arises when a high-speed streak encounters a downstream low-speed streak, we observe strong $v'$ perturbations develop quasi-symmetrically about a confined low-speed region, followed by the formation of a $\Lambda$-vortex and subsequent hairpin packet (figure 24c). This evolution is consistent with varicose-type breakdown mechanisms reported in previous studies of bypass transition (e.g., Wu & Moin (2009); Brandt *et al.* (2004)).

For the dominant sinuous instability, a *one-sided* scenario corresponding to the turbulent spot identified in figure 24(b) is shown in the left column of figure 26. It occurs where initially, a high- and low-speed streak are nearly parallel, with a sharp spanwise gradient at their interface ($u' = 0$), corresponding to a locally inflectional velocity profile. At later times, the interface develops a spanwise-asymmetric waviness, characteristic of sinuous instability. Wall-normal vorticity emerges along the streak interface, accompanied by growth of asymmetric $v'$ perturbations. As the instability evolves, the perturbations amplify and spread, leading to increased distortion of the streak. The streak coherence is ultimately lost, resulting in breakdown to small-scale turbulence. The observed asymmetric distortion and associated vorticity are consistent with the sinuous instability mechanism reported in various works (Swearingen & Blackwelder 1987; Brandt & Henningson 2002; Brandt *et al.* 2004).

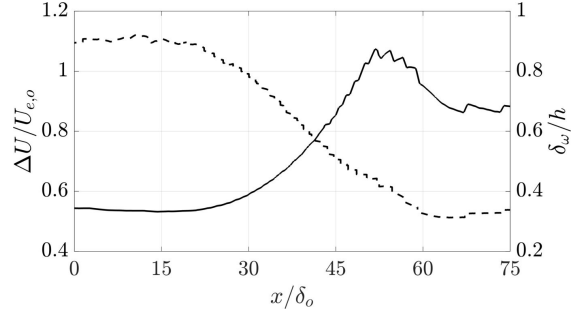

Figure 27: Profiles of the velocity difference across the riblet-induced shear layer ( ——— $\Delta U/U_{e,o}$, left axis) and the vorticity thickness ( – – – $\delta_\omega/h$, right axis).

The right column of figure 26 shows another type of evolution of a sinuous instability: a *two-sided* instability which develops on both sides of the low-speed streak. In this scenario, the low-speed streak is bounded by high-speed regions on either side, producing spanwise-inflectional velocity profiles at both interfaces. These inflectional regions give rise to wall-normal vorticity and asymmetric $v'$ perturbations about each $u' = 0$ interface. Because the instability develops simultaneously on both sides of the streak, the resulting deformation appears spanwise symmetric. However, the underlying mechanism remains sinuous, driven by local spanwise velocity gradients at each interface.

Overall, the smooth-wall results are consistent with previous studies of FPG TBLs and confirm that retransition proceeds through a bypass-like mechanism (Warnack & Fernholz 1998; Piomelli & Yuan 2013; Falcone & He 2022). The extent to which these dynamics mirror ZPG transition, and the factors governing the relative prevalence of sinuous and varicose streak breakdown, remain open questions and are not pursued here. Instead, we use these smooth-wall mechanisms as references to highlight the distinct retransition mechanisms induced by riblets, and to assess their impact on the downstream evolution of the boundary layer.

### 7.2. *Evolution of turbulent spots: formation of KH rollers over riblets*

The two retransition processes identified in the smooth case are absent in case RB. Instead, the onset of the earlier and more intense retransition in case RB is governed by interactions between the coherent structures present prior to breakdown, specifically the KH rollers and the residual streamwise streaks that persist after quasi-relaminarization. Figure 24 shows KH rollers that remain coherent across the span. These structures have been reported in previous studies of ZPG channel flows over riblets. While they are theoretically expected to possess large spanwise wavelengths, those observed in turbulent channels are typically distorted and confined by the overlying hairpin vortices (e.g., see figure 9 in Endrikat *et al.* (2021*b*) and figure 1 in Camobreco *et al.* (2025)). Here, the suppression of turbulence by the acceleration allows the KH rollers to remain more coherent across the span, making their spanwise organization much clearer than in canonical turbulent-channel conditions.

It is worth noting that previous studies have suggested that blunt riblets, such as those considered here, are less likely to generate strong KH rollers under ZPG conditions than steeper, more slender geometries (Endrikat *et al.* 2021*a*; Camobreco *et al.* 2025). KH rollers were likewise not reported by Pargal *et al.* (2021) in their study of impulsively accelerated open-channel flow over sawtooth riblets. Since that configuration differs from the present one in several respects, including riblet size, spanwise domain extent and flow development history, the reason for the absence of reported rollers remains unclear. By contrast, the present results demonstrate the formation of rollers with substantial strength and clear spanwise coherence, suggesting that the strong, spatially developing FPG considered here, in combination with the present riblet configuration, favours their development.

The KH rollers originate from the inner shear layer near the riblet crest and the instability of that shear layer. The fact that these rollers become appreciable only beyond $x/\delta_o = 50$, while remaining weak or absent further upstream, shows that the FPG progressively amplifies the instability until the flow enters a regime that supports coherent KH rollers. Figure 27 shows the streamwise evolution of two key quantities characterizing the internal shear layer: the velocity difference, $\Delta U$, and the shear-layer thickness,

$$\delta_\omega = \frac{\Delta U}{(\partial \langle \overline{u} \rangle / \partial y)_{\max}}. \tag{7.1}$$

Here, the velocity difference is obtained between the upper and lower bounds of the roller, which are taken at the critical wall-normal positions $(y-d)^+ = 20$ and $(y-d)^+ = -4$. The choice of these locations is guided by the premultiplied spectral analysis (figure 28) and is consistent with previous studies on KH rollers over riblets (García-Mayoral & Jiménez 2011*a*; Endrikat *et al.* 2021*a*; Camobreco *et al.* 2025). The denominator in the definition of $\delta_\omega$ is evaluated between these two heights and consistently attains its maximum at the riblet crest. The wall-normal location of the minimum second derivative of the mean velocity, commonly used to identify the shear layer and KH rollers (Jiménez *et al.* 2001; García-Mayoral & Jiménez 2011*b*; Camobreco *et al.* 2025), is also found to be fixed throughout the domain, occurring at $1.122h$, or $0.0055\delta_o$ above the riblet crest (roughly 3-10 viscous units depending on $u_\tau$ or $\hat{u}_\tau$). It can be seen that $\Delta U$ increases during relaminarization, particularly downstream of $x/\delta_o = 25$, where a strong FPG is present. In contrast, $\delta_\omega$ decreases over the same region. This indicates that the internal shear layer becomes progressively thinner while sustaining a larger velocity difference, leading to intensified shear and a stronger concentration of vorticity. Such conditions are conducive to the amplification of KH instability and promote the formation of spanwise rollers. The discussion regarding the increased sensitivity of $\lambda$ from $x/\delta_o = 50$ could be another factor contributing to the roller onset at this location. In summary, the FPG 1) increases the magnitude of shear at the crest and 2) reduces the viscous length scale, thereby confining the shear layer thickness. Combined with the geometry-dependent shear distribution fixing the shear layer location relative to the riblet height, these conditions favor KH vortex roll-up. That said, the present observations are based on a single case, and therefore do not allow for a definitive identification of universal threshold values for KH roller occurrence in general non-equilibrium TBLs.

In addition, while KH rollers originate near the riblet crest, the streaks in the overlying boundary layer have been reported to be present within $50\nu/u_\tau$ above the crest (Smith & Metzler 1983; Robinson 1991; Warnack & Fernholz 1998). This is supported by our data. Figure 28 shows the premultiplied spanwise energy spectra of streamwise velocity fluctuations shortly after retransition onset, where two distinct spectral peaks correspond to the two coherent structures. The streaks are generally situated above the rollers, however the lack of distinct wall-normal separation suggests the two structures are likely to interact.

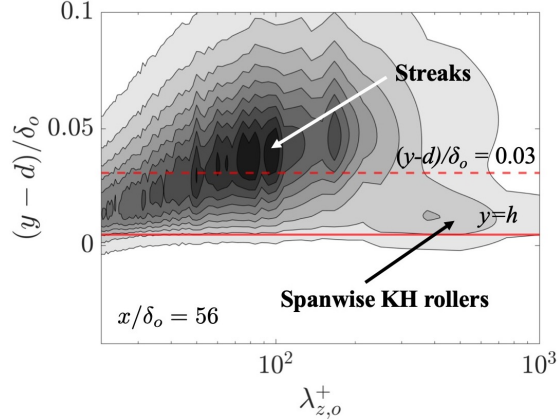

Figure 28: Premultiplied spanwise energy spectra of the streamwise velocity fluctuations for case RB, evaluated at $x/\delta_o = 56$.

### 7.3. *Evolution of turbulent spots: KH roller break down over riblets*

We now examine a typical turbulent spot formation over riblets using instantaneous flow fields near the two peaks shown in the premultiplied spectra, together with the 3D visualization in figure 24. In figure 29, riblet-induced rollers are visualized using $x-z$ planes of $v'/U_{e,o}$ at the riblet crest ($y = h$). The overlying streamwise-elongated streaks are indicated by superposed iso-contour lines of $u'/U_{e,o} = 0.2$, extracted at the streak height ($(y-d)/\delta_o = 0.03$) and are shaded to highlight high-speed regions. The non-dimensional time here is referenced to the instant shown in figure 24(c). A supplementary movie is also provided to show multiple occurrences of similar events across the retransition region.

At the earliest time instant shown (figure 29a), the spanwise rollers are approximately quasi-two-dimensional, and each has a streamwise extent of about $0.4\delta_o$, consistent with the scale shown in figure 23(a) at $x/\delta_o = 50$. This corresponds to about 353 (217) total (effective) viscous length scales evaluated at $x/\delta_o = 50$. Both scalings are on the order of the total-viscous-scaled wavelengths obtained by spectral analyses in ZPG turbulent flows (García-Mayoral & Jiménez 2011b; Endrikat *et al.* 2021a,b) which range between 65 and 290 viscous units, although the wavelength scaled by the effective shear shows closer agreement. The riblet crests segment the rollers but do not significantly disrupt their overall spanwise coherence. As the interaction develops, the portions of the rollers beneath the high-speed streak intensify due to the locally enhanced wall-normal shear, as evidenced by the darkening contours of $v'$. Yet, their size does not seem to change. Meanwhile, owing to the spanwise velocity gradient across the high-speed streak, the rollers undergo progressive deformation as they are continuously impacted by the former (figure 29d,e): the central portions are transported farther downstream, while the segments near the streak edges (and thus adjacent low-speed streaks) lag behind, resulting in an arched roller. From a vorticity-dynamics perspective, the streamwise inclination of this arched structure is consistent with vortex tilting, i.e., conversion of spanwise vorticity into streamwise vorticity in the presence of spanwise velocity gradients. Notably, this acceleration and deformation extend across multiple rollers rather than being confined to a single one. In the example shown here, approximately four consecutive rollers are influenced by the same high-speed streak.

As the rollers continue to be modified by the high-speed streak, the arched portion of the rollers tends to disrupt the center portion of the streak. By the fourth time instant shown here (figure 29d; also corresponding to figure 24c), the high-speed streak is observed to carry localized reduced-speed regions, and this trend persists at later times (figure 29e). Figure 24(c) shows that at this time instant, eddies are lifted off from the riblet crest plane by the induced rotation, primarily from regions of positive $v'$. These structures initially consist of riblet-spacing-scale streamwise vortices induced at riblet tips, which are connected by spanwise roller segments once the roller core becomes sufficiently strong for visualization. Since this roller-induced ejection acts against the turbulent sweeping related to the overlying high-speed streak, it prompts the interaction between the near-wall turbulence and the lifted KH rollers. Such interaction between the overlying streaks, the lifted roller vortices, as well as the streamwise vortices, subsequently grows into hairpin-like structures and concludes with rapid breakdown to turbulence (figure 24e, figure 29f).

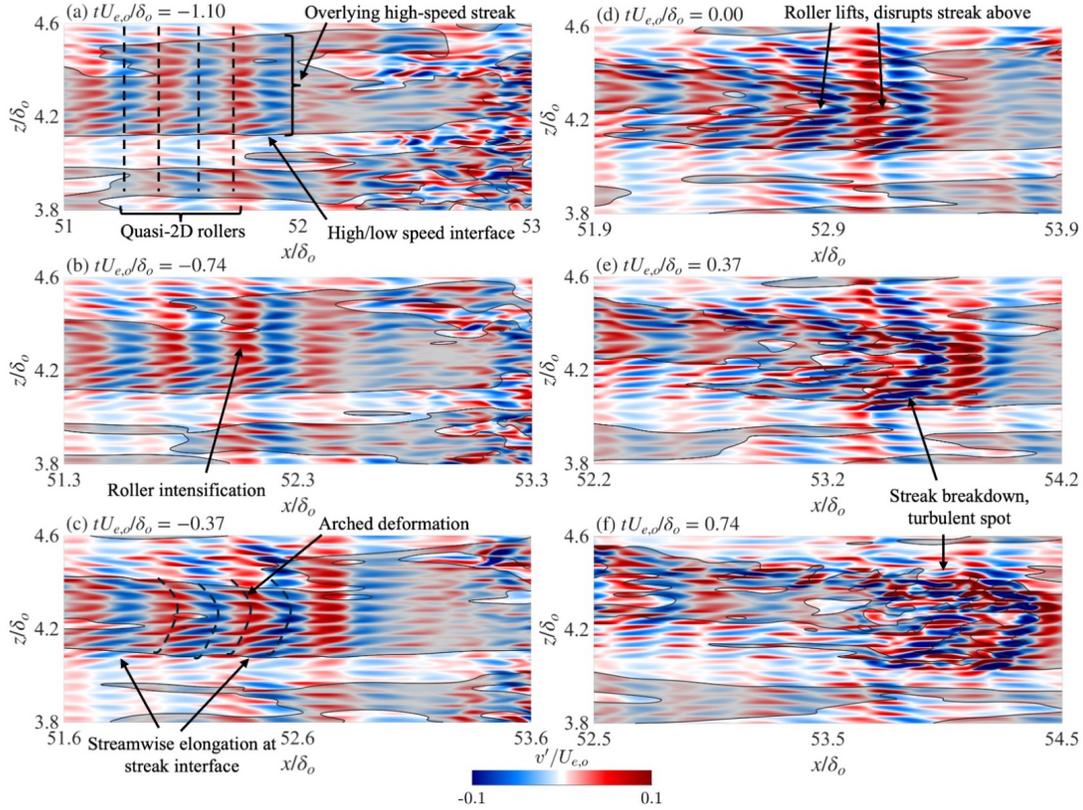

Figure 29: Evolution of spanwise-oriented roller structures in the ($x$–$z$) planes, shown by contours of velocity fluctuations. The blue-to-red contours represent the wall-normal fluctuation $v'/U_{e,o}$ at $y = h$ (the riblet crest). Superposed contours of $u'/U_{e,o} = 0.2$ at $(y - d)/\delta_o = 0.03$ mark the overlying high-speed streaks and illustrate their interaction with the rollers during retransition.

The above analysis shows that streak intermittency is strongly coupled to the occurrence of KH-induced retransition events. Both the streaks and the KH rollers convect downstream within the still-accelerating boundary layer while interacting continuously with one another. Their interaction is therefore inherently dynamic and spatially dependent. Previous studies report that near-wall streaks convect at $10$–$13u_\tau$ at $y^+ \sim 10$–$20$ (Liu & Gayme 2020; Zhang et al. 2025), whereas KH rollers travel more slowly, at approximately $5$–$6u_\tau$ (Endrikat et al. 2021a; Camobreco et al. 2025). Thus, the two structures move downstream at distinctly different speeds. Figure 30 presents space–time maps of the fluctuating streamwise velocity. The data are sampled at the spanwise midpoint between two riblet crests, at $y = h$ (riblet crest;

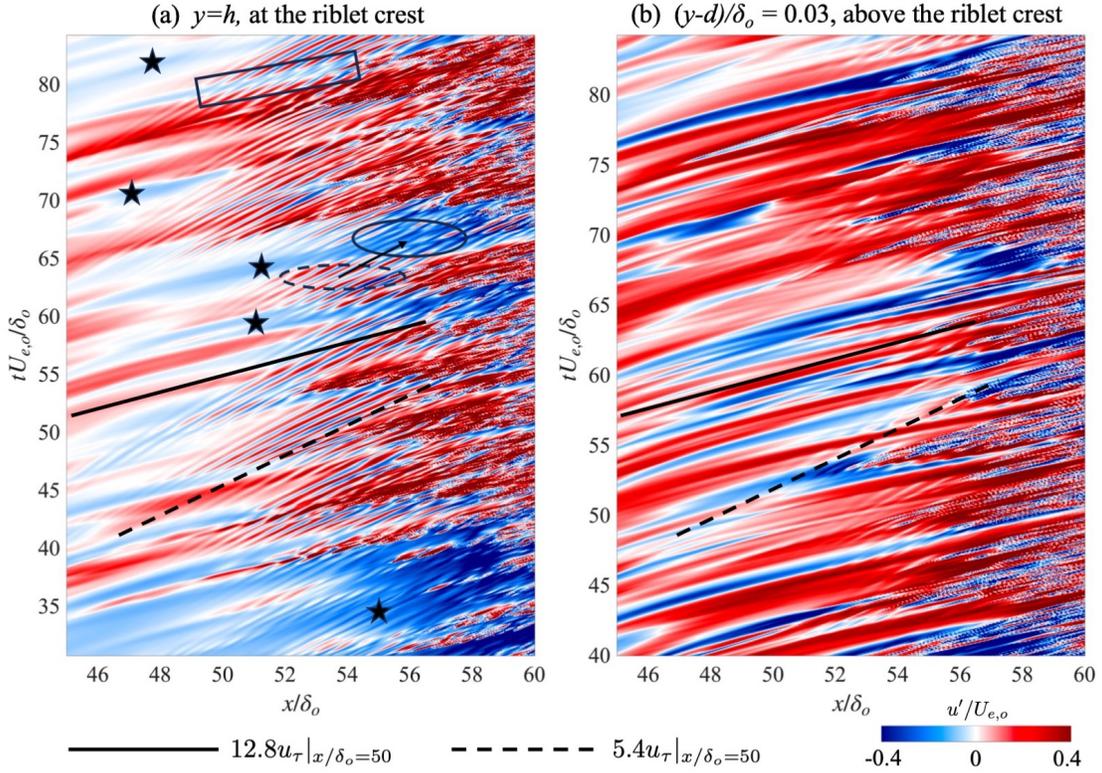

**(a)** $y=h$, at the riblet crest  **(b)** $(y\text{-}d)/\delta_o = 0.03$, above the riblet crest

— $12.8u_\tau|_{x/\delta_o=50}$    - - - $5.4u_\tau|_{x/\delta_o=50}$

Figure 30: Spatiotemporal maps of instantaneous streamwise velocity fluctuations at two wall-normal locations. The data are sampled at the spanwise midpoint between two riblet crests. (a) At $y = h$, the riblet crest; (b) at $(y - d)/\delta_o = 0.03$, located above the riblet crest. See text for description of the marked and enclosed regions.

left panel) and $(y - d)/\delta_o = 0.03$ (location of the residual streaks; right panel). The observed behaviour is robust and does not depend on the particular pair of riblet crests selected. The reference lines indicate measured convection velocities of $12.8u_\tau$ (or, $18.9\hat{u}_\tau$) and $5.4u_\tau$ (or, $8.0\hat{u}_\tau$) for the streaks and KH rollers, respectively, in the present case RB using the viscous velocity scale at $x/\delta_o = 50$. The map clearly demonstrates that high-speed streaks locally amplify the rollers, whereas low-speed streaks weaken or suppress them (examples are marked by stars in figure 30a).

Because the streaks overtake the rollers from above owing to their higher convection speed, each KH roller experiences the passage of multiple high- or low-speed streaks, leading to successive local amplification or suppression. Two representative examples are highlighted in Figure 30(a). The regions enclosed by ellipses show a group of KH rollers that are initially amplified by a high-speed streak, but are subsequently weakened and nearly extinguished as they are overtaken by a following low-speed streak. The region enclosed by the rectangle exhibits a similar sequence; however, the low-speed streak is not sufficiently strong to completely suppress the rollers. Instances of the opposite behavior—where rollers intensify following suppression—are also observed in the map (not marked).

### 7.4. *Changes in turbulent statistics during retransition*

Figure 31 shows that the different retransition mechanisms in the two cases produce distinct signatures in the turbulent-stress distributions. The smooth-wall profiles exhibit the expected signatures of bypass-like retransition. At $x/\delta_o = 50$, the streamwise normal stress $\langle \overline{u'u'} \rangle$ is the only Reynolds-stress component with appreciable magnitude. As the flow approaches the onset of retransition at $x/\delta_o = 56$, all stress components increase. The bypass retransition is marked by a rapid growth of $\langle \overline{u'u'} \rangle$ up to $x/\delta_o = 65$, indicating an abrupt transfer of kinetic energy from the mean flow to turbulent fluctuations. This breakdown of the incoming residual structures, together with the establishment of new near-wall turbulence, is reflected in the shift of the $\langle \overline{u'u'} \rangle$ peak from the dashed reference line toward the wall. Meanwhile, the transverse fluctuations grow in magnitude with their peak localizing near the original streak height, which is expected as they form due to the streak breakdown described above. Beyond $x/\delta_o = 65$, the newly developed turbulent boundary layer gradually recovers toward canonical TBL statistics as turbulent kinetic energy is redistributed among the Reynolds stress components. One signature of the gradual establishment of a canonical TBL is the tendency for an outer peak to emerge in the $\langle \overline{u'u'} \rangle$ profile.

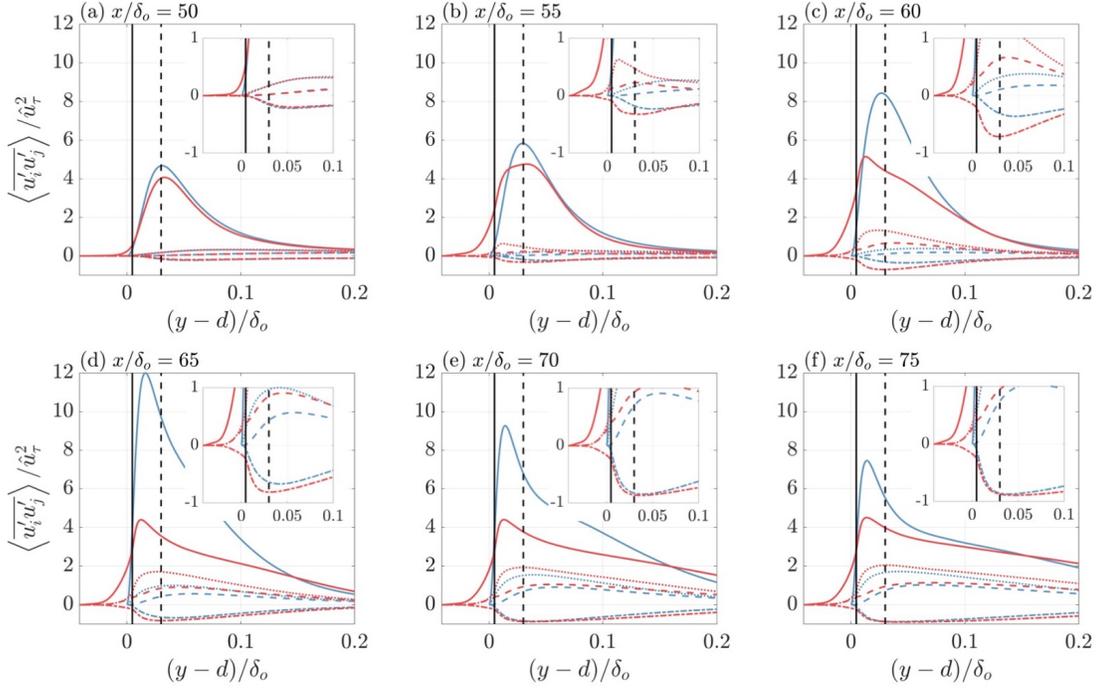

Figure 31: Reynolds stress profiles at various locations in the retransition region. The wall-normal coordinate is normalized by $\delta_o$ to visualize statistical changes with respect to the incoming streaks and riblet crest. The insets show the same profiles but zoomed to emphasize the near-riblet region. ——— $\langle \overline{u'u'} \rangle$; – – – $\langle \overline{v'v'} \rangle$; ······· $\langle \overline{w'w'} \rangle$; – · – · $\langle \overline{u'v'} \rangle$; ——— case SM; ——— case RB. The solid vertical line represents the riblet crest, and the dashed vertical line represents the location of the incoming streaks at $(y-d)/\delta_o = 0.03$.

Retransition and the establishment of a new TBL state begin about $6\delta_o$ earlier in the riblet case than the smooth-wall case. Several differences appear to be closely related to the observed KH roller–streak interaction. First, at $x/\delta_o = 55$, growth of $\langle \overline{u'u'} \rangle$, $\langle \overline{v'v'} \rangle$, and $\langle \overline{w'w'} \rangle$ appears near the riblet crest, beneath the outer peaks associated with the residual streaks. At the same location, the Reynolds shear stress $\langle \overline{u'v'} \rangle$ also increases sharply in magnitude. This specific feature is short-lived: by $x/\delta_o = 60$, retransition over the riblets appears to be close to complete, and the stress profiles mainly evolve toward the new turbulent boundary-layer state. Second, the Reynolds stresses penetrate beneath the riblet crest, in contrast to the nearly turbulence-free groove at $x/\delta_o = 50$. This process begins with a sharp increase in $\langle \overline{u'u'} \rangle$ throughout the groove, down to the trough, at $x/\delta_o = 55$. At that stage, the other stress components remain confined to the upper part of the groove, consistent with the local formation of KH rollers. Further downstream, the other stresses also grow throughout the groove as the flow fully retransitions.

## 8. Conclusions

In the present study, we examined how riblets designed to reduce drag under canonical conditions behave in a turbulent boundary layer subjected to freestream acceleration, relaminarization, and retransition. Direct numerical simulations (DNS) of flow over streamwise-aligned riblets, together with the corresponding smooth-wall case, show that the canonical ZPG picture of riblet drag reduction does not directly translate to this strongly non-equilibrium flow. Although the riblets were selected to lie in the optimal drag-reducing regime upstream of the favourable-pressure-gradient (FPG) region, they produce a substantial drag penalty downstream, while exerting little influence on the relaminarization process itself and promoting substantially earlier retransition.

The results reveal a clear distinction between drag generation within the riblet grooves and the effective shear experienced by the overlying boundary layer. The increase in total drag within the grooves is driven primarily by enhanced viscous shear, while contributions from turbulent fluctuations and secondary motions remain negligible there until retransition occurs. The riblet geometry sets the spatial distribution of the imposed mean shear at the groove opening over the riblet surface, and the FPG amplifies its magnitude, producing a concentration of elevated shear and a pronounced drag penalty near the riblet crest. At the same time, before the onset of retransition, the overlying turbulent boundary layer responds to the total shear stress at the groove opening, $\hat{\tau}$, rather than to the elevated total drag generated within the grooves themselves.

Before the onset of retransition, several observations show that the overlying boundary layer is governed by viscous scaling based on this effective total shear. These include the similar growth of $\hat{\tau}$ relative to the smooth-wall shear, the riblet size parameters remaining within the conventional drag-reduction range when scaled with the effective viscous length, and the collapse of higher-order turbulence statistics between the riblet and smooth-wall cases, consistent with Townsend's outer-layer similarity. In this sense, the riblet crest plane can be regarded as a partial-slip surface over which the outer boundary layer develops. Moreover, the piecewise linear relationship between the slip length and the groove-opening length scale, when both are normalized by the effective viscous length scale, identifies the groove opening as the dynamically relevant plane for defining the characteristic near-wall velocity and length scales in the present developing boundary layer. The breakpoint in this relationship occurs precisely at the onset of retransition.

From the perspective of coherent structures, freestream acceleration elongates the near-wall streaks in both the smooth-wall and riblet cases, and these streaks persist through relaminarization until the onset of bypass retransition. In the smooth-wall case, retransition occurs predominantly through sinuous instability of the streaks, with varicose instability appearing only intermittently. In the riblet case, by contrast, retransition is initiated substantially earlier by the emergence of spanwise Kelvin–Helmholtz rollers near the riblet crest and their interaction with the residual near-wall streaks. The interaction between the faster-convecting streaks and the more slowly travelling rollers drives a dynamic amplification process in which the rollers intensify intermittently beneath high-speed streaks before rolling up into intense turbulent spots. The drag penalty reaches its maximum during this stage, as turbulence begins to develop within the riblet grooves.

Taken together, these findings provide a revised physical picture of riblet performance in strongly non-equilibrium turbulent flows. Prior to retransition, drag modulation is governed by the effective shear at the groove opening so long as the outer turbulence remains largely decoupled from the groove flow. Once this regime breaks down, Kelvin–Helmholtz instability near the riblet crest becomes the dominant mechanism governing retransition and the accompanying drag amplification. The present results, therefore, clarify why viscous-scaled riblet-performance characterizations developed for canonical ZPG flows lose predictive value in strongly developing pressure-gradient boundary layers, while also suggesting how such characterizations might be reformulated in terms of dynamically relevant scales.

The quantitative changes in drag, turbulence modulation, and retransition location and rate reported here will depend on the particular riblet size and geometry considered in the present configuration. Owing to history effects in non-equilibrium turbulent flows, boundary layers subjected to different FPG magnitudes and streamwise development histories may interact differently with riblets. In addition, the scale separation between the wall turbulence and the riblets, and its role in the underlying mechanism, may introduce Reynolds-number dependence into the present observations. A broader parametric exploration is therefore required to establish the extent to which these findings persist across other flow and surface conditions. In particular, identifying the conditions under which Kelvin–Helmholtz rollers first emerge remains an important direction for future work.

**Funding** This material is based upon work supported by the Air Force Office of Scientific Research under award number FA9550-25-1-0033, monitored by Dr. Gregg Abate. BSS also appreciates the support of NSF GRFP Award No. 2235036.

**Acknowledgment** This work was supported by high-performance computer time and resources from the DoD High Performance Computing Modernization Program, as well as from the Expanse supercomputer at the San Diego Supercomputer Center under ACCESS. The authors gratefully acknowledge these sources of support. The authors gratefully acknowledge Dr. Junlin Yuan for providing the top-boundary condition used in Yuan & Piomelli (2015), which was refined for the present study.

**Data availability statement.** The data that support the findings of this study are available on request. In addition, one instantaneous three-dimensional flow-field snapshot from each case will be made publicly available upon acceptance of the manuscript on the University of Mississippi Fundamental and Applied Science of Turbulence Laboratory (FAST Lab) website (https://sites.google.com/view/fastum/fast-database).

**Declaration of interests.** The authors report no conflict of interest.

## Appendix A. Validation of the smooth-wall case

In addition to the agreement in the acceleration parameter shown in figure 2(b), the smooth-wall case also compares very well with the experimental data for Case 2 of Warnack & Fernholz (1998). Figure 32 shows the mean velocity in wall units for the present smooth-wall case together with the measurements of Warnack & Fernholz (1998). Excellent agreement is obtained throughout the boundary-layer development.

## Appendix B. Nomenclature

| | |
|---|---|
| $\overline{\phi}$ | Time average of quantity $\phi$ (or time-and-ensemble average over riblet realizations) |
| $\langle \phi \rangle$ | Intrinsic spanwise average of quantity $\phi$ |
| $\langle \phi \rangle_s$ | Superficial spanwise average of quantity $\phi$ |
| $\widetilde{\phi}$ | Spatial variation about $\langle \phi \rangle$, $\widetilde{\phi} = \phi - \langle \phi \rangle$ |
| $\phi'$ | Fluctuation from time/ensemble average, $\phi' = \phi - \overline{\phi}$ |
| $\phi_o$ | Quantity $\phi$ evaluated at $x = 0$ |

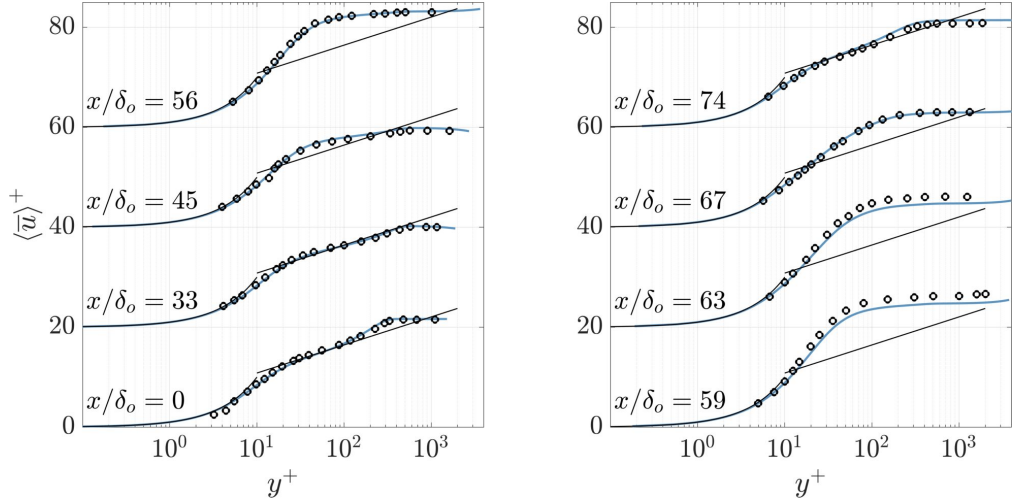

Figure 32: Mean velocity profiles in wall units. —— case SM; ○ Case 2 in Warnack & Fernholz (1998). Each profile is shifted by 20 units for clarity. The thin black lines indicate the viscous sublayer and logarithmic law of the wall.

| | |
|---|---|
| $\phi_{\text{SM}}$ | Quantity $\phi$ evaluated in the smooth wall case |
| $\phi_{\text{RB}}$ | Quantity $\phi$ evaluated in the riblet case |
| $\phi^+$ | Quantity $\phi$ normalized in local wall units |
| $\widehat{\phi}$ | Quantity $\phi$ normalized by units derived from total shear stress at riblet crest |
| $a$ | Linear coefficient for geometry-dependent contribution to slip length, Eqn. 5.6 |
| $b$ | Linear coefficient for flow-dependent contribution, Eqn. 5.6 |
| $C_f$ | Skin friction coefficient, $2\tau_w/U_e^2$ |
| $C_{f,v}$ | Viscous contribution to skin friction coefficient |
| $C_{f,t}$ | Turbulent contribution to skin friction coefficient |
| $C_{f,d}$ | Dispersive contribution to skin friction coefficient |
| $d$ | Zero-plane displacement (virtual origin), $0.9h$ |
| $\delta$ | Boundary layer thickness |
| $\delta^*$ | Displacement thickness |
| $\delta_\omega$ | Vorticity thickness, $\Delta U/(\partial \langle \overline{u} \rangle /\partial y)_{\text{max}}$ |
| $\varepsilon$ | Turbulent kinetic energy dissipation rate, $\overline{\nu(\partial u_i'/\partial x_j)(\partial u_j'/\partial x_i)}$ |
| $f$ | Immersed boundary method force |
| $h$ | Riblet height |
| $h_l$ | Spanwise riblet profile |
| $K$ | Acceleration parameter, $(\nu/U_e^2)(dU_e/dx)$ |
| $\ell_{\text{g}}$ | Square root of riblet groove cross-sectional area |
| $L_x$ | Streamwise domain length |
| $L_y$ | Wall-normal domain length |
| $L_z$ | Spanwise domain length |
| $L^*$ | Arbitrary characteristic flow length scale |
| $L_{u'u'}$ | Integral streamwise length scale obtained from streamwise auto-correlation |
| $\lambda$ | Robin slip length, $U_c/(\partial \langle \overline{u} \rangle /\partial y)_{y=h}$ |
| $N_i$ | Number of grid points in streamwise direction |
| $N_j$ | Number of grid points in wall-normal direction |
| $N_k$ | Number of grid points in spanwise direction |
| $\eta$ | Kolmogorov length scale, $(\nu^3/\varepsilon)^{1/4}$ |
| $p$ | Modified pressure |

| $Q$ | Second invariant of the velocity gradient tensor, $-(1/2)(\partial u_j/\partial x_i)(\partial u_i/\partial x_j)$ |
|---|---|
| $R_{u'u'}$ | Autocorrelation coefficient of streamwise velocity fluctuations |
| $Re_\theta$ | Momentum thickness Reynolds number, $U_e\theta/\nu$ |
| $Re_\tau$ | Friction Reynolds number, $u_\tau\delta/\nu$ |
| $Re_\delta$ | Boundary-layer thickness Reynolds number, $U_e\delta/\nu$ |
| $Re_{\delta^*}$ | Displacement thickness Reynolds number, $U_e\delta^*/\nu$ |
| $s$ | peak-to-peak riblet spacing |
| $t$ | Simulation time |
| $\tau_w$ | Wall shear stress, $\nu(\partial\langle\overline{u}\rangle/\partial y)_{y=0}$ for smooth wall, $-\int_0^h f_x\,dy$ for riblets |
| $\widehat{\tau}$ | Total shear stress evaluated at the riblet crest |
| $\Delta\tau_w$ | Drag change, $100\times(\tau_{w,RB}-\tau_{w,SM})/\tau_{w,SM}\%$ |
| $u$ | Streamwise velocity |
| $U_e$ | Mean streamwise velocity at the boundary layer edge |
| $U_\infty$ | Streamwise velocity applied at top of computational domain |
| $u_\tau$ | Friction velocity, $\sqrt{\tau_w}$ |
| $\widehat{u}_\tau$ | Velocity defined by total shear stress at the riblet crest, $\sqrt{\widehat{\tau}}$ |
| $\Delta U$ | Velocity difference across riblet shear layer |
| $\Delta U_\infty$ | Total freestream velocity increment |
| $v$ | Wall-normal velocity |
| $V_\infty$ | Wall-normal velocity at top of computational domain |
| $\nu$ | Kinematic viscosity |
| $w$ | Spanwise velocity |
| $x$ | Streamwise direction |
| $\Delta x_i$ | Grid spacing in the $i$-th direction |
| $y$ | Wall-normal direction |
| $z$ | Spanwise direction |
| $z_r$ | Spanwise location within a single riblet period, $z_r \in [0, s]$ |
| $\theta$ | Momentum thickness |

## Appendix C. Streamwise momentum budget

The double-averaged streamwise momentum budget for rough-wall flows, which we use in the analysis in §5, is derived by intrinsically-averaging the time-averaged streamwise momentum equation, applying the decomposition $\overline{\phi} = \langle\overline{\phi}\rangle + \widetilde{\phi}$, following the commutativity rules for the spanwise-averaging operator with spatial derivatives, and ultimately converting intrinsic averages to superficial averages. Detailed derivations, including those of the relationship between the IBM force and select terms resulting from non-commutativity of the spanwise average operator with spatial derivatives, are provided in various sources (Raupach & Shaw 1982; Nikora *et al.* 2007; Yuan & Piomelli 2014*a*,*b*), and thus one is not repeated here. The streamwise momentum budget reads:

$$0 = -\frac{\partial\langle\overline{p}\rangle_s}{\partial x} - A_x + R_x + D_x + S_x + F_x. \qquad (C\,1)$$

The terms on the RHS are the mean pressure gradient, mean convection ($A_x$), mean Reynolds stress divergence ($R_x$), mean viscous diffusion ($D_x$), mean divergence of the dispersive stress ($S_x$), and mean IBM force ($F_x$). Individual term definitions, excluding the pressure gradient, read as follows:

$$A_x = \frac{\partial\langle\overline{u}\rangle_s\langle\overline{u}\rangle_s}{\partial x} + \frac{\partial\langle\overline{u}\rangle_s\langle\overline{v}\rangle_s}{\partial y}, \qquad (C\,2)$$

$$R_x = -\frac{\partial \langle \overline{u'u'} \rangle_s}{\partial x} - \frac{\partial \langle \overline{u'v'} \rangle_s}{\partial y}, \tag{C 3}$$

$$D_x = \nu \left( \frac{\partial^2 \langle \overline{u} \rangle_s}{\partial x^2} + \frac{\partial^2 \langle \overline{u} \rangle_s}{\partial y^2} \right), \tag{C 4}$$

$$S_x = -\frac{\partial \langle \tilde{u}\tilde{u} \rangle_s}{\partial x} - \frac{\partial \langle \tilde{u}\tilde{v} \rangle_s}{\partial y}, \tag{C 5}$$

$$F_x = \left\langle \overline{f}_x \right\rangle_s. \tag{C 6}$$

## REFERENCES

ABE, H. 2017 Reynolds-number dependence of wall-pressure fluctuations in a pressure-induced turbulent separation bubble. *J. Fluid Mech.* **833**, 563–598.

ABU ROWIN, W., DESHPANDE, R., WANG, S., KOZUL, M., CHUNG, D., SANDBERG, R. D. & HUTCHINS, N. 2025 Experimental characterisation of Kelvin–Helmholtz rollers over riblet surfaces. *J. Fluid Mech.* **1009**, A65.

ANDERSSON, P., BRANDT, L., BOTTARO, A. & HENNINGSON, D. S. 2001 On the breakdown of boundary layer streaks. *J. Fluid Mech.* **428**, 29–60.

ASAI, M., MINAGAWA, M. & NISHIOKA, M. 2002 The instability and breakdown of a near-wall low-speed streak. *J. Fluid Mech.* **455**, 289–314.

BECHERT, D. W., BARTENWERFER, M., HOPPE, G. & REIF, W.-E. 1986 Drag reduction mechanisms derived from shark skin. In *Proceedings of the 15th Congress of the International Council of the Aeronautical Sciences*, pp. 1044–1068. ICAS.

BLACKWELDER, R. F. & KOVASZNAY, L. S. G. 1972 Large-scale motion of a turbulent boundary layer during relaminarization. *J. Fluid Mech.* **53** (1), 61–83.

BOBKE, A., VINUESA, R., ÖRLÜ, R. & SCHLATTER, P. 2017 History effects and near-equilibrium in adverse-pressure-gradient turbulent boundary layers. *J. Fluid Mech.* **820**, 667–692.

BOOMSMA, A. & SOTIROPOULOS, F. 2015 Riblet drag reduction in mild adverse pressure gradients: A numerical investigation. *Int. J. Heat Fluid Flow* **56**, 251–260.

BOURASSA, C. & THOMAS, F. O. 2009 An experimental investigation of a highly accelerated turbulent boundary layer. *J. Fluid Mech.* **634**, 359–404.

BRANDT, L. & HENNINGSON, D. S. 2002 Transition of streamwise streaks in zero-pressure-gradient boundary layers. *J. Fluid Mech.* **472**, 229–261.

BRANDT, L., SCHLATTER, P. & HENNINGSON, D. S. 2004 Transition in boundary layers subject to free-stream turbulence. *J. Fluid Mech.* **517**, 167–198.

CAMOBRECO, C. J., ENDRIKAT, S., GARCÍA-MAYORAL, R., LUHAR, M. & CHUNG, D. 2025 Why do only some riblets promote spanwise rollers? *J. Fluid Mech.* **1022**, A35.

CHAMORRO, L. P., ARNDT, R. E. A. & SOTIROPOULOS, F. 2013 Drag reduction of large wind turbine blades through riblets: Evaluation of riblet geometry and application strategies. *Renew. Energy* **50**, 1095–1105.

CHOI, H., MOIN, P. & KIM, J. 1993 Direct numerical simulation of turbulent flow over riblets. *J. Fluid Mech.* **255**, 503–539.

CHOI, K.-S. 1989 Near-wall structure of a turbulent boundary layer with riblets. *J. Fluid Mech.* **208**, 417–458.

CHOI, K.-S. 1990 Effects of longitudinal pressure gradients on turbulent drag reduction with riblets. In *Turbulence Control by Passive Means* (ed. E. Coustols), pp. 109–121.

CHOI, K.-S., GADD, G. E., PEARCEY, H. H., SAVILL, A. M. & SVENSSON, S. 1989 Tests of drag-reducing polymer coated on a riblet surface. *Flow Turbul. Combust.* **46**, 209–216.

DE PRISCO, G., KEATING, A. & PIOMELLI, U. 2007 Large-eddy simulation of accelerating boundary layers. In *45th AIAA Aerospace Sciences Meeting and Exhibit*, pp. 1–16. AIAA.

DEBISSCHOP, J. R. & NIEUWSTADT, F. T. M. 1996 Turbulent boundary layer in an adverse pressure gradient: Effectiveness of riblets. *AIAA. J.* **34** (5), 932–937.

ENDRIKAT, S., MODESTI, D., GARCÍA-MAYORAL, R., HUTCHINS, N. & CHUNG, D. 2021a Influence of riblet shapes on the occurrence of Kelvin-Helmholtz rollers. *J. Fluid Mech.* **913**, A37.

ENDRIKAT, S., MODESTI, D., MACDONALD, M., GARCÍA-MAYORAL, R., HUTCHINS, N. & CHUNG, D. 2021*b* Direct numerical simulations of turbulent flow over various riblet shapes in minimal-span channels. *Flow Turbul. Combust.* **107**, 1–29.

FALCONE, M. & HE, S. 2022 A spatially accelerating turbulent flow with longitudinally contracting walls. *J. Fluid Mech.* **945**, A23.

GARCÍA, E., HUSSAIN, F., YAO, J. & STOUT, E. 2025 Wall turbulence perturbed by a bump with organized small-scale roughness: structure dynamics. *J. Fluid Mech.* **1006**, A6.

GARCÍA-MAYORAL, R. & JIMÉNEZ, J. 2011*a* Drag reduction by riblets. *Phil. Trans. R. Soc. A* **369**, 1412–1427.

GARCÍA-MAYORAL, R. & JIMÉNEZ, J. 2011*b* Hydrodynamic stability and breakdown of the viscous regime over riblets. *J. Fluid Mech.* **678**, 317–347.

GATTI, DAVIDE, VON DEYN, LARS, FOROOGHI, POURYA & FROHNAPFEL, BETTINA 2020 Do riblets exhibit fully rough behaviour? *Experiments in Fluids* **61** (3), 81.

GOLDSTEIN, D., HANDLER, R. & SIROVICH, L. 1995 Direct numerical simulation of turbulent flow over a modelled riblet covered surface. *J. Fluid Mech.* **302**, 333–376.

GOLDSTEIN, D. & TUAN, T. -C. 1998 Secondary flow induced by riblets. *J. Fluid Mech.* **363**, 115–151.

GRIFFIN, K. P., FU, L. & MOIN, P. 2021 General method for determining the boundary layer thickness in nonequilibrium flows. *Phys. Ref. Fluids* **6**, 024608.

HUSSAIN, F., GARCÍA, E., YAO, J. & STOUT, E. 2024 Wall turbulence perturbed by a bump with organized small-scale roughness: flow statistics. *J. Fluid Mech.* **989**, A13.

JACOBS, R. G. & DURBIN, P. A. 2001 Simulations of bypass transition. *J. Fluid. Mech.* **428**, 185–212.

JIMÉNEZ, J. & PINELLI, A. 1999 The autonomous cycle of near-wall turbulence. *J. Fluid Mech.* **389**, 335–359.

JIMÉNEZ, J., UHLMANN, M., PINELLI, A. & KAWAHARA, G. 2001 Turbulent shear flow over active and passive porous surfaces. *J. Fluid Mech.* **442**, 89–117.

KEATING, A., PIOMELLI, U., BREMHORST, K. & NEŠIĆ, S. 2004 Large-eddy simulation of heat transfer downstream of a backward-facing step. *J. Turbul.* **5**, N20 1–27.

KITSIOS, V., ATKINSON, C., SILLERO, J. A., BORRELL, G., GUNGOR, A. G., JIMÉNEZ, J. & SORIA, J. 2016 Direct numerical simulation of a self-similar adverse pressure gradient turbulent boundary layer. *Intl J. Heat Fluid Flow* **61**.

KITSIOS, V., SEKIMOTO, A., ATKINSON, C., SILLERO, J. A., BORRELL, G., GUNGOR, A. G., JIMÉNEZ, J. & SORIA, J. 2017 Direct numerical simulation of a self-similar adverse pressure gradient turbulent boundary layer at the verge of separation. *J. Fluid Mech.* **829**.

KLUMPP, S., GULDNER, T., MEINKE, M. & SCHRÖDER, W. 2010 Riblets in a turbulent adverse-pressure gradient boundary layer. In *5th Flow Control Conference*, pp. 1–11. AIAA.

KOZUL, M., NARDINI, M., PRZYTARSKI, P. J., SOLOMON, W., SHABBIR, A. & SANDBERG, R. D. 2025 Optimal riblets applied to gas turbine compressor blades studied via direct numerical simulation (GT2024–122305). *Journal of Turbomachinery* **147** (8), 081020.

KUNTZAGK, S. 2024 Aeroshark – drag reduction using riblet film on commercial aircraft. In *Hamburg Aerospace Lecture Series*. AeroLectures.

LEE, S.-J. & LEE, S.-H. 2001 Flow field analysis of a turbulent boundary layer over a riblet surface. *Exp. Fluids* **30** (2), 153–166.

LIETMEYER, C., OEHLERT, K. & SEUME, J. R. 2012 Optimal application of riblets on compressor blades and their contamination behavior. *Journal of Turbomachinery* **135** (1), 011036.

LIU, C. & GAYME, D. F. 2020 An input–output based analysis of convective velocity in turbulent channels. *J. Fluid Mech.* **888**, A32.

LUCHINI, P. 1996 Reducing the turbulent skin friction. *Comp. Methods in App. Sc.* **3**, 466–470.

LUCHINI, P., MANZO, F. & POZZI, A. 1991 Resistance of a grooved surface to parallel flow and cross-flow. *J. Fluid Mech.* **228**, 87–109.

LUND, T. S., WU, X. & SQUIRES, K. D. 1998 Generation of inflow data for spatially-developing boundayr layer simulations. *J. Comput. Phys.* **140**, 233–258.

MCELIGOT, D. M. & ECKELMANN, H. 2006 Laterally converging duct flows. Part 3. Mean turbulence structure in the viscous layer. *J. Fluid Mech.* **549**, 25–59.

MIGNOT, E., BARTHELEMY, E. & HURTER, D. 2009 Double-averaging analysis and local flow characterization of near-bed turbulence in gravel-bed channel flows. *J. Fluid Mech.* **618**, 279–303.

MODESTI, D., ENDRIKAT, S., HUTCHINS, N. & CHUNG, D. 2021 Dispersive stresses in turbulent flow over riblets. *J. Fluid Mech.* **917**, A55.

MOIN, P. 2010 *Fundamentals of Engineering Numerical Analysis*. Cambridge University Press.

MONTY, J. P., HARUN, Z. & MARUSIC, I. 2011 A parametric study of adverse pressure gradient turbulent boundary layers. *Intl J. Heat Fluid Flow* **32** (3), 575–585.

NARASIMHA, R. & SREENIVASAN, K. R. 1973 Relaminarization in highly accelerated turbulent boundary layers. *J. Fluid Mech.* **61** (3), 417–447.

NIEUWSTADT, F. T. M., WOLTHERS, H., LEIJDENS, H., KRISHNA PRASAD, K. & SCHWARZ-VAN MANEN, A. 1993 The reduction of skin friction by riblets under the influence of an adverse pressure gradient. *Exp. Fluids.* **15**, 17–26.

NIKORA, V., MCEWAN, I., MCLEAN, S., , COLEMAN, S., POKRAJAC, D. & WALTERS, R. 2007 Double-averaging concept for rough-bed open-channel and overland flows: theoretical background. *J. Hydraul. Eng.* **133** (8), 873–883.

ORLANDI, P. & JIMÉNEZ, J. 1994 On the generation of turbulent wall friction. *Phys. Fluids* **6**, 634–641.

PARGAL, S., YUAN, J. & BRERETON, G. 2021 Impulse response of turbulent flow in smooth and riblet-walled channels to a sudden velocity increase. *J. Turbul.* **22** (6), R353–379.

PESKIN, C. S. 1972 Flow patterns around heart valves: a numerical method. *J. Comput. Phys.* **10**, 552–271.

PIOMELLI, U., BALARAS, E. & PASCARELLI, A. 2000 Turbulent structures in accelerating boundary layers. *J. Turbul.* **1** (1), 1–16.

PIOMELLI, U. & YUAN, J. 2013 Numerical simulation of spatially developing, accelerating boundary layers. *Phys. Fluids* **25**, 101304.

POZUELO, R., LI, Q., SCHLATTER, P. & VINUESA, R. 2022 An adverse-pressure-gradient turbulent boundary layer wiht nearly constant $\beta \simeq 1.4$ up to $Re_\theta \simeq 8700$. *J. Fluid Mech.* **939**, A34.

RAUPACH, M. R., ANTONIA, R. A. & RAJAGOPALAN, S. 1991 Rough-wall boundary layers. *App. Mech. Rev.* **44** (1), 1–25.

RAUPACH, M. R., FINNIGAN, J. J. & BRUNET, Y. 1996 Coherent eddies and turbulence in vegetation canopies: The mixing-layer analogy. *Bound.-Lay. Meteorol.* **25**, 351–382.

RAUPACH, M. R. & SHAW, R. H. 1982 Averaging procedures for flow within vegetation canopies. *Bound.-Lay. Meteorol.* **25**, 351–382.

ROBINSON, S. K. 1991 Coherent motions in the turbulent boundary layer. *Annu. Rev. Fluid Mech.* **23**, 601–639.

ROUHI, A., ENDRIKAT, S., MODESTI, D., SANDBERG, R. D., ODA, T., TANIMOTO, K., HUTCHINS, N. & CHUNG, D. 2022 Riblet-generated flow mechanisms that lead to local breaking of the reynolds analogy. *J. Fluid Mech.* **951**, A45.

ROUHI, A., KUMAR, V., LEHMKUHL, O., WU, W., KOZUL, M. & SMITS, A. J. 2025 Application of riblets to separating turbulent boundary layers. In *15th International ERCOFTAC Symposium on Engineering Turbulence Modelling and Measurements (ETMM-15)*, pp. 396–401.

SAREEN, A., DETERS, R. W., HENRY, S. P. & SELIE, M. S. 2011 Drag reduction using riblet film applied to airfoils for wind turbines. In *49th AIAA Aerospace Sciences Meeting*, pp. 2011–558. AIAA.

SAVINO, B. S., ROUHI, A. & WU, W. 2026 Attached decelerating turbulent boundary layers over riblets. In *AIAA SCITECH 2026 Forum*. AIAA.

SAVINO, B. S. & WU, W. 2024a Impact of spanwise rotation on flow separation and recovery behind a bulge in channel flows. *J. Fluid Mech.* **999**, A51.

SAVINO, B. S. & WU, W. 2024b Thrust generation by shark denticles. *J. Fluid Mech.* **1000**, A80.

SCHLATTER, P., BRANDT, L., DE LANGE, H. C. & HENNINGSON, D. S. 2008 On streak breakdown in bypass transition. *Phys. Fluids* **20** (10), 101505.

SCHLATTER, P. & ÖRLÜ, R. 2010 Assessment of direct numerical simulation of data of turbulent boundary layers. *J. Fluid Mech.* **659**, 116–126.

SCHOPPA, W. & HUSSAIN, F. 2002 Coherent structure generation in near-wall turbulence. *J. Fluid Mech.* **453**, 57–108.

SCOTTI, A. 2006 Direct numerical simulation of turbulent channel flows with boundary roughened with virtual sandpaper. *Phys. Fluids* **18** (3), 031701.

SMITH, B. R. & YAGLE, P. 2025 Rans turbulence model for drag reducing riblets and its predictions for aerodynamic applications. In *AIAA SCITECH 2025 Forum*, pp. 1–18. AIAA.

SMITH, C. R. & METZLER, S. P. 1983 The characteristics of low-speed streaks in the near-wall region of a turbulent boundary layer. *J. Fluid Mech.* **129**, 27–54.

SREENIVASAN, K. R. 1982 Laminarescent, relaminarizing and retransitional flows. *Acta. Mech.* **44**, 1–48.

STRAND, J. S. & GOLDSTEIN, D. B. 2011 Direct numerical simulations of riblets to constrain the growth of turbulent spots. *J. Fluid Mech.* **668**, 267–292.

Suzuki, Y. & Kasagi, N. 1994 Turbulent drag reduction mechanism above a riblet surface. *AIAA J.* **32** (9), 1781–1790.

Swarztrauber, P. N. & Sweet, R. A. 1979 Efficient fortran subprograms for the solution to separable elliptic partial differential equations. *ACM T. Math. Software* **5.3**, 352–364.

Swearingen, J. D. & Blackwelder, R. F. 1987 The growth and breakdown of streamwise vortices in the presence of a wall. *J. Fluid Mech.* **182**, 255–290.

Sweet, R. A. 1974 A generalized cyclic reduction algorithm. *SIAM J. Numer. Anal.* **11**, 506–520.

Talamelli, A., Fornaciari, N., Westin, K. J. A. & Alfredsson, P. H. 2002 Experimental investigation of streaky structures in a relaminarizing boundary layer. *J. Turbul.* **3** (18), 1–13.

Walsh, M. J. 1980 Drag characteristics of V-groove and transverse curvature riblets. In *Viscous Flow Drag Reduction* (ed. G. R. Hough), *Progress in Astronautics and Aeronautics*, vol. 72, pp. 168–184. AIAA.

Walsh, M. J. & Lindemann, A. M. 1984 Optimization and application of riblets for turbulent drag reduction. *AIAA Paper* pp. 84–0347.

Wang, Y., Sun, Z., Guo, D., Ju, S. & Yang, G. 2025 An improved modeling approach for riblet effects based on Reynolds-averaged Navier-Stokes turbulence model. *Phys. Fluids* **37**, 075188.

Warnack, D. & Fernholz, H. H. 1998 The effects of a favourable pressure gradient and of the reynolds number on an incompressible axisymmetric turbulent boundary layer. Part 2. The boundary layer with relaminarization. *J. Fluid Mech.* **359**, 357–381.

Wilcox, D. C. 2008 Formulation of the $k - \omega$ turbulence model revisited. *AIAA J.* **46**, 2823.

Wong, J., Camobreco, C. J., García-Mayoral, R., Hutchins, N. & Chung, D. 2024 A viscous vortex model for predicting the drag reduction of riblet surfaces. *J. Fluid Mech.* **978**, A18.

Wu, W. & Piomelli, U. 2018 Effects of surface roughness on a separating turbulent boundary layer. *J. Fluid Mech.* **841**, 552–580.

Wu, X. & Moin, P. 2009 Direct numerical simulation of turbulence in a nominally zero-pressure-gradient flat-plate boundary layer. *J. Fluid Mech.* **630**, 5–41.

Wu, X., Moin, P., Wallace, J. M., Skarda, J., Lozano-Durán, A. & Hickey, J. 2017 Transitional-turbulent spots and turbulent-turbulent spots in boundary layers. *Proc. Natl. Acad. Sci. U. S. A.* **114** (27), E5292–E5299.

Yuan, J. & Piomelli, U. 2014a Numerical simulations of sink-flow boundary layers over rough surfaces. *Phys. Fluids* **26** (1), 015113.

Yuan, J. & Piomelli, U. 2014b Roughness effects on the Reynolds stress budgets in near-wall turbulence. *J. Fluid Mech.* **760**, R1.

Yuan, J. & Piomelli, U. 2015 Numerical simulation of a spatially developing accelerating boundary layer over roughness. *J. Fluid Mech.* **178**, 192–214.

Zaki, T. A. 2013 From streaks to spots and on to turbulence: exploring the dynamics of boundary layer transition. *Flow Turb. Combust.* **91** (3), 451–473.

Zhang, H., Wu, T., Ning, D. & He, G. 2025 The approximate similarity between higher-order and second-order stress–velocity cross-spectra and its significance for the convection velocity of reynolds stress fluctuations. *J. Fluid Mech.* **1024**, R2.

Zhang, W., Yang, X. I. A., Chen, P. & Wan, M. 2024 Integral methods for friction decomposition and their extensions to rough-wall flows. *J. Fluid Mech.* **985**, A46.